%% file: main.tex
\documentclass[aps,prd,twocolumn,preprintnumbers,nofootinbib,superscriptaddress,amsmath]{revtex4-1}
\usepackage{graphicx}
\usepackage{amsmath}
\usepackage{amssymb}
\usepackage{amsfonts}
\usepackage{hyperref}
\usepackage{url}
\usepackage{color}
\usepackage{bm}
\usepackage{array}
\usepackage{placeins}
\usepackage{bbold}
\usepackage{algorithm}
\usepackage{algpseudocode}
\usepackage[normalem]{ulem}
\newcommand{\ssout}[1]{}
\newcommand{\sqz}{\medmuskip=0mu}

\begin{document}
\makeatother

\preprint{IFT-UAM/CSIC-23-100}

\title{Efficient Reduced Order Quadrature Construction Algorithms for Fast Gravitational Wave Inference}

\author{Gonzalo Morr\'as}
\email{gonzalo.morras@uam.es}
\affiliation{Instituto de F\'isica Te\'orica UAM/CSIC, Universidad Aut\'onoma de Madrid, Cantoblanco 28049 Madrid, Spain}
\author{Jose Francisco Nu\~no Siles}
\email{jose.nunno@uam.es}
\affiliation{Instituto de F\'isica Te\'orica UAM/CSIC, Universidad Aut\'onoma de Madrid, Cantoblanco 28049 Madrid, Spain}
\author{Juan Garc\'ia-Bellido}
\email{juan.garciabellido@uam.es}
\affiliation{Instituto de F\'isica Te\'orica UAM/CSIC, Universidad Aut\'onoma de Madrid, Cantoblanco 28049 Madrid, Spain}

\date{\today}

\begin{abstract}

Reduced Order Quadrature (ROQ) methods can greatly reduce the computational cost of Gravitational Wave (GW) likelihood evaluations, and therefore greatly speed up parameter estimation analyses, which is a vital part to maximize the science output of advanced GW detectors. In this paper, we do an in-depth study of ROQ techniques applied to GW data analysis and present novel algorithms to enhance different aspects of the ROQ bases construction. We improve upon previous ROQ construction algorithms allowing for more efficient bases in regions of parameter space that were previously challenging. In particular, we use singular value decomposition (SVD) methods to characterize the waveform space and choose a reduced order basis close to optimal and also propose improved methods for empirical interpolation node selection, greatly reducing the error added by the empirical interpolation model. To demonstrate the effectiveness of our algorithms, we construct multiple ROQ bases ranging in duration from 4s to 256s for compact binary coalescence (CBC) waveforms including precession and higher order modes. We validate these bases by performing likelihood error tests and P-P tests and explore the speed up they induce both theoretically and empirically with positive results. Furthermore, we conduct end-to-end parameter estimation analyses on several confirmed GW events, showing the validity of our approach in real GW data.

\end{abstract}
\maketitle

\section{Introduction}
\label{sec:intro}

Gravitational wave (GW) astronomy has been made possible in recent years by ground-based observatories like LIGO~\cite{AdvancedLigo}, Virgo~\cite{AdvancedVirgo}, and KAGRA~\cite{KAGRA:2018plz}, revolutionizing our understanding of the Universe by enabling the direct detection of GW signals emitted during extreme cosmic phenomena such as the mergers of binary black holes, binary neutron stars, and neutron star-black hole binaries. With the continuous improvement in sensitivity of current detectors~\cite{Abbott_2020} and the advent of next-generation detectors, including projects like the Einstein Telescope~\cite{Maggiore_2020}, Cosmic Explorer~\cite{evans2021horizon}, LISA~\cite{LISA_I,LISA_III,LISA_V}, we anticipate a dramatic increase in the number of GW candidates detected. For maximum science outputs, a parameter estimation (PE) for each candidate will have to be performed. With standard PE methods~\cite{Thrane_2019}, this can be prohibitively computationally expensive, especially as we reduce the frequency from which we can detect gravitational waves and the duration of the signals becomes much longer~\cite{Maggiore_Vol1}.

To fully exploit the enhanced sensitivity of these advanced detectors, it is essential to use accurate waveform models that incorporate important physical effects such as precession or higher-order modes~\cite{Mills:2020thr}. However, the computational challenge of calculating the likelihood of such signals poses a significant bottleneck in the analysis pipeline. Traditional likelihood calculations can be computationally intensive, particularly for long-duration waveforms. Several methods have been explored in the literature to reduce this computational burden, such as multi-banding~\cite{PhysRevD.104.044062}, heterodyned likelihood~\cite{Zackay:2018qdy, Cornish:2021lje}, likelihood-free approaches~\cite{Chua:2019wwt, Green:2020hst}, Reduced Order Quadrature methods~\cite{Antil:2012wf,Canizares:2013ywa,Smith:2016qas,Qi:2020lfr,Morisaki:2020oqk,Morisaki:2023kuq} and others~\cite{Pankow_2015,Lange:2018pyp,Pathak:2022ktt}.

In this work, we will focus on the ROQ method, which is one of the most promising approaches to fast GW likelihood evaluations, due to its ability to achieve very large speed-ups while maintaining high accuracy and being able to accommodate the effects of precession and higher-order modes~\cite{Smith:2016qas,Qi:2020lfr}. ROQ methods exploit the fact that for a given parameter range, the corresponding GW waveforms span only a small subspace of the vector space of all possible signals. By constructing reduced bases that capture the essential information of the templates, ROQ techniques provide an efficient representation that enables fast likelihood evaluations. The ROQ has a \textit{start-up} cost associated with the offline basis building stage, which needs to be performed in advance only once per waveform model and parameter space. However, since for typical PE analyses we have to compute more waveforms than what is needed to construct the ROQ and a basis can be used to perform multiple PEs, this start-up cost quickly pays off.

This paper presents several algorithms for ROQ construction, which offer some key advantages over existing methods. They are specifically designed to tackle the challenges of speed in the basis construction and accuracy in GW likelihood evaluation while maximizing the ROQ speedup. As we will see, these algorithms have the ability to handle complex waveform models in parameter ranges that were intractable with existing procedures.

The paper is organized as follows. In Sec.~\ref{sec:ThFramework}, we introduce the basic theoretical framework, including a discussion on GW inference as well as on the basics of ROQ. In Sec.~\ref{sec:Algorithm}, we describe the ROQ algorithms we introduce in depth, going through the construction of the reduced order basis, the choice of empirical interpolation model and how to construct a ROQ with a set tolerance for a given parameter space. In Sec.~\ref{sec:CodeValidation} we present several bases created for two phenomenological waveform models, \texttt{IMRPhenomPv2}~\cite{PhysRevLett.113.151101} and \texttt{IMRPhenomXPHM}~\cite{Pratten_2021}, and test their speed and accuracy. We further test the ROQ by performing parameter estimation analyses on three confirmed GW events.  In Sec.~\ref{sec:Conclusions} we finally conclude. We relegate some of the more convoluted numerical methods used by our algorithms to the Appendices.

The methods introduced in this paper have been implemented in a \texttt{python} code called \texttt{EigROQ}, which is publicly available at \url{https://github.com/gmorras/EigROQ}.

\section{Theoretical framework}
\label{sec:ThFramework}

In this section we will briefly describe the basic theoretical framework to contextualize the rest of the paper. In Sec.~\ref{sec:ThFramework:PrimerGWinference} we give a very brief overview on the basics of GW parameter estimation while on Sec.~\ref{sec:ThFramework:ROQbasics} we summarize the basics of the ROQ rule. For more details, we refer the reader to Refs.~\cite{Smith:2016qas,Thrane_2019}.

\subsection{A primer on gravitational wave inference}\label{sub:primer}
\label{sec:ThFramework:PrimerGWinference}

GW inference refers to the modern scientific discipline taking care among other things, of computing the posterior probability distribution of the GW model parameters $\vec{\theta}$ that best fit the data, using Bayes Theorem
\begin{equation}
    p(\vec{\theta}|d) = \frac{\mathcal{L}(d|\vec{\theta})\pi(\vec{\theta})}{\mathcal{Z}}.
    \label{eq:BayesTheorem}
\end{equation}
In this equation, there are several objects that enter the calculation. The first, $\pi(\vec{\theta})$ refers to the prior employed, from the nature of the event, which throughout this paper will always be a CBC to the distributions describing the parameters of the binary. Next, the likelihood function $\mathcal{L}(d|\vec{\theta})$ of the data given the parameters $\vec{\theta}$ and the evidence $\mathcal{Z}$ representing the probability of the data given the model. 

The likelihood is the most computationally expensive part of estimating the posterior. Given a CBC signal without eccentricity, there are 15 different parameters to fit that enter the likelihood computation. The typical gravitational-wave astronomy likelihood is based on the hypothesis that only Gaussian noise is present in the detector and deviations from it are the result of a GW signal. In such case, the likelihood can up to a normalization constant be expressed as~\cite{GWLikelihood_Finn}

\begin{align}
    \log\mathcal{L}(d|\vec{\theta}) & = -\frac{1}{2}(d-h(\vec{\theta}), d-h(\vec{\theta})) \nonumber \\ 
    & = -\frac{1}{2}(d,d) + (d, h(\vec{\theta})) - \frac{1}{2} (h(\vec{\theta}), h(\vec{\theta})) ,
\label{eq:Likelihood_def}    
\end{align}

\noindent where $h(\vec{\theta})$ represents, in this specific case, the CBC waveform with parameters $\vec{\theta}$ used to fit the data $d$. The overlap integral $( \cdot , \cdot)$ is defined as
\begin{equation}
    (d, h(\vec{\theta})) = 4 \Delta f \mathcal{R} \sum_{j=1}^L\frac{\tilde{d}^{*}(f_j)\tilde{h}(f_j; \vec{\theta})}{S(f_j)} \, ,
    \label{eq:overlap_integral}
\end{equation}

\noindent, where $S_n(f)$ is the detector's noise power spectral density (PSD) and $\tilde{a}(f)$, denotes the Fourier transform of $a(t)$. Since the data of GW detectors are discretely sampled, we will have discrete Fourier transforms having a frequency spacing $\Delta f = 1/T$, whith $T$ being the observation time. For a frequency window $(f_\mathrm{high} - f_\mathrm{low})$ there will be $L = \texttt{int}[(f_\mathrm{high} - f_\mathrm{low})T]$ terms in the sum of Eq.~\eqref{eq:overlap_integral}.\footnote{Here $\texttt{int}[x]$ refers to taking the integer part of $x$.} 
Repeatedly computing the overlap integrals in Eq.~\eqref{eq:Likelihood_def} is the bottleneck in gravitational waves inference, and the main part we aim to speed up in this paper. 

\subsection{Basics of Reduced Order Quadratures for Gravitational Wave inference}
\label{sec:ThFramework:ROQbasics}

The parameters $\Vec{\theta}$ of the GW signal $h(\Vec{\theta})$ we are fitting to the data (Eq.~\eqref{eq:Likelihood_def}) can be split on intrinsic and extrinsic parameters. The extrinsic parameters are common to all transient GW sources and they are the sky location, usually measured with right ascension $\alpha$ and declination $\delta$, the polarization $\psi$, luminosity distance $d_L$ and a reference time of arrival of the signal $t_c$.\footnote{We use $t_c$ because, for the CBC case, the reference time of arrival for the signal is usually given by the coalescence time at the geocenter.} The intrinsic parameters are related to the source of the GW and are generically referred to as $\vec{\lambda}$. For a quasi-circular CBC they are comprised of the 2 component masses $m_1$ and $m_2$, 3 components per BH spin vector $\vec{s}_i$, the inclination angle $\iota$ and the coalescence phase $\phi_c$. For CBCs with at least one neutron star (NS) $\vec{\lambda}$ can also contain a tidal deformability parameter $\Lambda$ per NS in the binary~\cite{Dietrich:2019kaq}, as well as any other matter effect information included in the model. If we break the assumption of quasi-circular orbits, the eccentricity $e$ would also have to be taken into account in the intrinsic parameters $\vec{\lambda}$~\cite{Chiaramello:2020ehz}.

We assume that the signal $h(t, \Vec{\theta})$ is short enough to ignore the dependence of the detector antenna patterns $F_{+, \times}$ with time and the time-varying Doppler shift due to motion of the detector with respect to the solar system barycenter~\cite{Jaranowski:1998qm}. In practice, the signal will have to last less than a few hours, to be able to ignore the effects of Earth's rotation. Then, in the frequency domain, the GW signal can be written as:

\begin{align}
    \tilde{h}(f, \Vec{\theta}) & = e^{-i 2 \pi f t_c} \frac{1}{d_L} \Big( F_+ (\alpha,\delta,\psi) \tilde{h}_{+}(f, \Vec{\lambda}) \nonumber \\
    &\quad\quad\quad\quad\quad\quad\;+ F_\times (\alpha,\delta,\psi) \tilde{h}_{\times}(f, \Vec{\lambda})\Big) \nonumber \\
    & \equiv e^{-i 2 \pi f t_c} \tilde{h}(f, \Vec{\Lambda}) 
    \label{eq:short_Signal_detector}
\end{align}

The main idea of the ROQ is to represent the GW waveform model $\tilde{h}(f_i;\Vec{\theta})$ and its modulus squared $|\tilde{h}(f_i;\Vec{\theta})|^2$ in terms of an empirical interpolant each, which is described in more detail in Sec.~\ref{sec:Algorithm}. For now, we assume that they can be approximated to arbitrary precision as:

\begin{subequations}
\label{eq:wf_ROQ}
\begin{align}
    \tilde{h}(f_i;\Vec{\Lambda}) & \approx \sum_{j=1}^{N_\mathrm{L}} B_j(f_i) \tilde{h}(F_j;\Vec{\Lambda}) \,
    \label{eq:wf_ROQ:L} \\
    |\tilde{h}(f_i;\Vec{\Lambda})|^2 &\approx \sum_{k=1}^{N_\mathrm{Q}} C_k(f_i) |\tilde{h}(\mathcal{F}_k;\Vec{\Lambda})|^2 \label{eq:wf_ROQ:Q} \, ,
\end{align}
\end{subequations}

\noindent where the main focus of this paper is to find the optimal values of the interpolation nodes $\{F_j\}_{j=1}^{N_\mathrm{L}}$ and $\{\mathcal{F}_k\}_{j=1}^{N_\mathrm{Q}}$ and of the ``bases'' $B_j(f_i)$ and $C_k(f_i)$ such that we minimize the required number of elements $(N_\mathrm{L}+N_\mathrm{Q})$ entering Eq.~\eqref{eq:wf_ROQ} while respecting a given specified precision.

If we input Eq.~\eqref{eq:short_Signal_detector} into Eq.~\eqref{eq:Likelihood_def} and use the approximation for the GW waveform $\tilde{h}(f_i;\Vec{\theta})$ and its modulus squared $|\tilde{h}(f_i;\Vec{\theta})|^2$ of Eq.~\eqref{eq:wf_ROQ}, we can represent the likelihood as

\begin{align}
    \log\mathcal{L}(d|\vec{\theta}) \approx -\frac{1}{2}(d,d) + (d, h(\vec{\theta}))_\mathrm{ROQ} - \frac{1}{2} (h(\vec{\theta}), h(\vec{\theta}))_\mathrm{ROQ} \, ,
\label{eq:ROQ_Likelihood_def}    
\end{align}

\noindent where the term $-\frac{1}{2}(d,d) \equiv \log\mathcal{L}_\mathrm{noise} $ is a constant that depends only on the data and cancels with the evidence $\mathcal{Z}$ when we compute the posterior probability distribution using Bayes theorem (Eq.~\eqref{eq:BayesTheorem}). In Eq.~\eqref{eq:ROQ_Likelihood_def} we have also implicitly defined the quantities:

\begin{subequations}
\label{eq:overlaps_ROQ}
\begin{align}
    (d, h(\vec{\theta}))_\mathrm{ROQ} & \equiv \mathcal{R} \sum_{j=1}^{N_\mathrm{L}} w_j(t_c) \tilde{h}(F_j;\Vec{\Lambda}) \,
    \label{eq:overlaps_ROQ:L} \\
    (h(\vec{\theta}), h(\vec{\theta}))_\mathrm{ROQ} & \equiv \sum_{k=1}^{N_\mathrm{Q}} \psi_k |\tilde{h}(\mathcal{F}_k;\Vec{\Lambda})|^2 \label{eq:overlaps_ROQ:Q} \, , 
\end{align}
\end{subequations}

\noindent which approximates the corresponding overlap integrals appearing in the Likelihood calculation of Eq~\eqref{eq:Likelihood_def}. In Eq.~\eqref{eq:overlaps_ROQ} we have introduced the linear and quadratic ROQ weights, $w_j(t_c)$ and $\psi_k$, defined as:

\begin{subequations}
\label{eq:weights_ROQ}
\begin{align}
    w_j(t_c) & \equiv  4 \Delta f \sum_{i=1}^L\frac{\tilde{d}^{*}(f_i) B_j(f_i)}{S(f_i)} e^{-i 2 \pi f_i t_c} \,
    \label{eq:weights_ROQ:L} \\
    \psi_k & \equiv 4 \Delta f \sum_{i=1}^L\frac{C_k(f_i)}{S(f_i)} \label{eq:weights_ROQ:Q} \, .
\end{align}
\end{subequations}

Before starting PE analysis on an event, the weights have to be computed for the observed data strain $\tilde{d}(f)$ and the corresponding PSD ($S(f)$). Since the linear weights are smooth functions of time, they are usually evaluated in a discrete set of times $N_t$ and are interpolated for the PE analysis~\cite{Smith:2016qas}. The spacing between time samples is usually of the order of the expected resolution in $t_c$, which for CBC signals can be as small as $0.1\mathrm{ms}$, and for the typical $t_c$ prior, which is uniform in $\pm 0.1\mathrm{s}$ around trigger time, this equates to $N_t \sim O(10^3)$. Therefore, at the beginning of the analysis, we have to perform $N_t N_\mathrm{L} + N_\mathrm{Q}$ full overlaps, as prescribed in Eq.~\eqref{eq:weights_ROQ}, and the startup cost of the ROQ is $O((N_t N_\mathrm{L} + N_\mathrm{Q})L)$.

Once the weights have been initialized, computing the ROQ likelihood will only require $N_\mathrm{L} + N_\mathrm{Q}$ terms to estimate the overlap integrals (Eq.~\eqref{eq:overlaps_ROQ}), compared to the $L$ terms in the full overlap integrals. We can therefore expect a speed-up in the likelihood computation of $O(L/(N_\mathrm{L} + N_\mathrm{Q}))$ when using the ROQ rule. In GW astronomy, typical CBC PE analyses require $O(10^8-10^9)$ likelihood evaluations, which dominate the computational cost required to sample the posterior of Eq.~\eqref{eq:BayesTheorem}. In most applications the startup cost of the ROQ is negligible compared to the sampling time and the ROQ will greatly speed up the whole analysis. The likelihood speedup is further explored in Sec.~\ref{sec:CodeValidation:Speeduptests}.

The biggest overhead when using the ROQ rule is in constructing the ROQ basis (Eq.~\eqref{eq:wf_ROQ}), since to explore typical CBC parameter spaces we need $O(10^6 - 10^7)$ random waveforms. With the methods outlined in this paper, we also aim to reduce the computational time of the basis generation, allowing us to handle complex waveform models in parameter ranges that were intractable with existing procedures.
In practice, for the CBC case, we train the ROQ on the $h_+$ polarization, varying only the values of the intrinsic parameters $\vec{\lambda}$, defined in Eq.~\eqref{eq:short_Signal_detector}. The same ROQ basis is valid for both polarizations since they can be jointly decomposed in spherical harmonics of spin weight $-2$,  ${}_{-2}Y_{l m}$ as~\cite{Mills:2020thr}
\begin{equation}
    h_{+} - i h_\times = \sum_{l=2}^\infty \sum_{m=-l}^l {}_{-2}Y_{l m}(\iota, \phi_c) h_{l m}
    \label{eq:}
\end{equation}
\noindent where the inclination $\iota$ and coalescence phase $\phi_c$ are also being sampled.

\section{Efficient algorithm for ROQ computation}
\label{sec:Algorithm}

\subsection{Reduced Order Basis}
\label{sec:Algorithm:ROB}
We generate $N$ templates from the waveform model we are trying to approximate:
\begin{equation}
    \{h_A(x), A=1,..,N\} \, ,
    \label{eq:WaveformSet}
\end{equation}
\noindent where, in GW astronomy, $x$ can be either frequency $f$ or time $t$. We can define the matrix of inner products between templates as
\begin{equation}
    M_{A B} = \langle h_A, h_B \rangle \, .
    \label{eq:MAB}
\end{equation}
In this context, the inner product is usually defined as:
\begin{equation}
    \langle h_A, h_B \rangle = \int_{f_\mathrm{low}}^{f_\mathrm{high}} \tilde{h}^{*}_A(f) \tilde{h}_B(f) df
    \label{eq:innerprod_hab}
\end{equation}
\noindent although we could also use a reference PSD $S_n(f)$ to give different weights at different frequencies to the integrand, as in Eq.~\eqref{eq:overlap_integral}. Since $M_{A B}$ is a matrix of inner products, it is hermitian and positive semi-definite, and therefore can always be diagonalized as
\begin{equation}
    M_{A B} = \sum_{C=1}^N E_{A C} \lambda_C E^{*}_{B C} \, ,
    \label{eq:MAB_diag}
\end{equation}
\noindent where $\lambda_C \geq 0$ are the eigenvalues and $E_{A B}$ is a unitary matrix whose columns are the orthonormal eigenvectors
\begin{equation}
    \sum_{C=1}^N E^{*}_{C A} E_{C B} = \delta_{A B}\, .
    \label{eq:EAB_def}
\end{equation}

In the waveform space we can then define the eigenvectors with $\lambda_A \neq 0$ as:
\begin{equation}
    e_A(x) = \frac{1}{\sqrt{\lambda_A}} \sum_{C=1}^N h_C(x) E_{C A} \, .
    \label{eq:eA_def}
\end{equation}

It can be proven that these are an orthonormal set of vectors under $\langle \cdot, \cdot \rangle$. That is:
\begin{align}
    \langle e_A, e_B \rangle & = \left\langle \frac{1}{\sqrt{\lambda_A}} \sum_{C=1}^N h_C(x) E_{C A}, \frac{1}{\sqrt{\lambda_B}} \sum_{D=1}^N h_D(x) E_{D B} \right\rangle \nonumber \\
    & = \frac{1}{\sqrt{\lambda_A \lambda_B}} \sum_{C=1}^N \sum_{D=1}^N E^{*}_{C A} E_{D B} \underbrace{\langle h_C, h_D \rangle}_{M_{C D}} \nonumber \\
    & = \frac{1}{\sqrt{\lambda_A \lambda_B}} \sum_{C=1}^N E^{*}_{C A} \underbrace{\sum_{D=1}^N M_{C D} E_{D B}}_{\lambda_B E_{C B}} \noindent \\
    & = \sqrt{\frac{\lambda_B}{\lambda_A}} \underbrace{\sum_{C=1}^N E^{*}_{C A} E_{C B}}_{\delta_{A B}} = \delta_{A B}
    \, . \label{eq:eAeB_proof}
\end{align}

We can also prove that the inner product between one of the waveforms used to compute $M_{A B}$ and a given eigenvector will be given by:
\begin{align}
    \langle h_A, e_B \rangle & = \left\langle h_A, \frac{1}{\sqrt{\lambda_B}} \sum_{C=1}^N h_C(x) E_{C B} \right\rangle  \nonumber \\ 
    &= \frac{1}{\sqrt{\lambda_B}} \sum_{C=1}^N \underbrace{\langle h_A, h_C \rangle}_{M_{A C}} E_{C D} \nonumber \\ 
    &= \frac{1}{\sqrt{\lambda_B}} \underbrace{\sum_{C=1}^N M_{A C} E_{C D}}_{\lambda_B E_{A B}} = \sqrt{\lambda_B} E_{A B} \, .
    \label{eq:hAeB_proof}    
\end{align}

We can define our reduced order basis (ROB) as a subset of $n<N$ elements of $\{e_A\}_{A=1}^N$, which we will learn how to optimally select later. To represent the waveform $h_A$ in terms of this ROB $\{e_a\}_{a=1}^n$, we project $h_A$ using the orthonormality property of the ROB:
\begin{equation}
    h_A^\mathrm{ROB}(x) = \sum_{b=1}^n \langle e_b, h_A \rangle e_b(x) = \sum_{b=1}^n \sqrt{\lambda_b} E^{*}_{A b} e_b(x) \, .
    \label{eq:hA_ROB}
\end{equation}

We can compute the representation error of projecting $h_A$ as:
\begin{align}
    & \sigma_{\mathrm{ROB},A}  = \Vert h_A - h_A^\mathrm{ROB}\Vert^2 = \langle h_A - h_A^\mathrm{ROB}, h_A - h_A^\mathrm{ROB}\rangle \nonumber \\
    & = \left \langle h_A - \sum_{b=1}^n \langle e_b, h_A \rangle e_b, \; h_A - \sum_{c=1}^n \langle e_c, h_A \rangle e_c  \right \rangle \nonumber \\
    & = \langle h_A, h_A \rangle - \sum_{b=1}^n |\langle e_b, h_A \rangle|^2 = \langle h_A, h_A \rangle - \sum_{b=1}^n \lambda_b |E_{A b}|^2 \, .
    \label{eq:sigma_ROB}
\end{align}

Ideally, to construct a ROB we would take a very large number of templates $\{h_A\}_{A=1}^N$, that capture most of the variability of the waveform in the parameter space of interest, compute the matrix $M_{A B}$ as in Eq.~\eqref{eq:MAB}, diagonalize it and, to construct our ROB, pick the minimum number of eigenvectors $\{e_a\}_{a=1}^n$ such that the ROB error of Eq.~\eqref{eq:sigma_ROB} is smaller than a specified tolerance. Unfortunately, this cannot be done in practice, since the number of random templates needed to fully span the typical parameter spaces for GW applications is of order $O(10^7)$. Using the fact that $M_{A B}$ is hermitian, we need $N(N-1)/2$ complex numbers to store the off-diagonal elements, and $N$ real numbers for the diagonal elements. Assuming that each real number is stored with $n_B$ Bytes, the memory required to store $M_{A B}$ is\footnote{$1 \mathrm{GB} = 10^9 \mathrm{Bytes} = 8 \cdot 10^9 \mathrm{bits}$}:
\begin{equation}
    \mathrm{Memory}(M_{AB}) = N^2 n_B = 80\,\mathrm{GB} \left(\frac{N}{10^5}\right)^2 \left(\frac{n_B}{8\,\mathrm{B}}\right) \, .
    \label{eq:Memory_MAB}
\end{equation}

Therefore, in current computers, examining more than a few tens of thousands of waveforms at a time is unfeasible, and we will not be able to analyze the entire parameter space at once. Motivated by this issue, we have developed a multi-step approach summarised in Algorithm~\ref{alg:create_ROB}. We construct a first ROB for a set tolerance with random waveforms. Then, we calculate its orthogonal space and obtain the corresponding ROB which we add to the original ROB. We repeat this process iteratively, reducing the tolerance at every step. The equivalent to the matrix $M_{A B}$ of Eq.~\eqref{eq:MAB} for the orthogonal space to the basis $\{e_a\}_{a=1}^n$ is:

\begin{align}
    M_{A B}^\mathrm{ROB} & = \left \langle h_A - h_A^\mathrm{ROB}, h_B - h_B^\mathrm{ROB}\right \rangle \nonumber \\
    & = \left \langle h_A  - \sum_{c=1}^n \langle e_c, h_A \rangle e_c, h_B - \sum_{d=1}^n \langle e_d, h_B \rangle e_d \right \rangle \nonumber \\
    & = \langle h_A, h_B \rangle - \sum_{c=1}^n \langle h_A, e_c \rangle \langle e_c, h_B \rangle \, .
    \label{eq:MAB^ROB}    
\end{align}

\begin{algorithm}[H]
\caption{Construction of reduced order basis}
\label{alg:create_ROB}
\begin{algorithmic}[1]
\State {\bf Input:} Maximum number of waveforms selected $N$, tolerances of each step $[\sigma_0,\dots,\sigma_s]$, maximum number of waveforms computed per step $[N_{\mathrm{lim},1},\dots,N_{\mathrm{lim},s}]$
\vskip 10pt
\State Generate $N$ waveforms $\{h_A \}_{A=1}^N$
\State Compute the matrix $M_{A B} = \langle h_A, h_B \rangle$
\State Diagonalize $M_{A B}$ to obtain eigenvalues $\lambda_A$ and eigenvectors $E_{A B}$
\State Input $\{\sigma_0, \{h_A\}_{A=1}^N, \lambda_A, E_{A B}\}$ in Algorithm~\ref{alg:select_lambda} to obtain inital ROQ basis $\{ e_i \}_{i=1}^{n_0}$
\vskip 10pt
\For{$j = 1 \to s$}
    \Repeat
        \State Generate $N_{\mathrm{lim},j}$ waveforms $\{h_A \}_{A=1}^{N_{\mathrm{lim},j}}$ and compute their ROB error $\sigma_{\mathrm{ROB},A}$
        \State Select the N waveforms $\{h_A\}_{A=1}^N$ with largest $\sigma_\mathrm{ROB}$
        \State Save the minimum value of $\sigma_\mathrm{ROB}$ for the selected waveforms: $\sigma_{\mathrm{ROB},\mathrm{min}}$
        \State $M^\mathrm{ROB}_{A B} = \langle h_A, h_B \rangle - \sum_{c=1}^{n_{j-1}} \langle h_A, e_c \rangle \langle e_c, h_B \rangle$
        \State Diagonalize $M^\mathrm{ROB}_{A B}$ and obtain eigenvalues $\lambda_A$ and eigenvectors $E_{A B}$
        \State Input $\{\sigma_j, \{h_A-h_A^\mathrm{ROB}\}_{A=1}^N, \lambda_A, E_{A B}\}$ in Algorithm~\ref{alg:select_lambda} to obtain next ROQ basis elements $\{ e_i \}_{i = n_{j-1}+1}^{n_j}$
    \Until{$\sigma_{\mathrm{ROB},\mathrm{min}}<\sigma_s$}
\EndFor
\vskip 10pt
\State {\bf Output:} ROB $\{ e_i \}_{i=1}^n$
\end{algorithmic}
\end{algorithm}

We observe that the same formulas and reasoning of Eqs.~(\ref{eq:MAB}-\ref{eq:Memory_MAB}) apply to the space orthogonal to the ROB if we make the identification $h_A \to h_A - h_A^\mathrm{ROB}$. To find the minimum number of elements that have to be added to the ROB to reduce the error below the set tolerance $\sigma$, we use Algorithm~\ref{alg:select_lambda}, where we iteratively subtract the contribution of the eigenvalue that produces the largest drop in any $\sigma_{\mathrm{ROB},A}$, according to Eq.~\eqref{eq:sigma_ROB}, until $\sigma_{\mathrm{ROB},A} < \sigma$ for all $A$.

\begin{algorithm}[H]
\caption{Selection of Eigenvectors}
\label{alg:select_lambda}
\begin{algorithmic}[1]
\State {\bf Input:} Tolerance $\sigma$, waveforms $\{h_A\}_{A=1}^N$, eigenvalues $\lambda_A$ and eigenvectors $E_{A B}$ of the matrix $M_{A B} = \langle h_A, h_B \rangle$
\vskip 10pt
\State Initialize $\sigma_A$: $\{\sigma_A = \langle h_A, h_A \rangle\}_{A=1}^{N}$
\State Compute the maximum contribution of each eigenvector \newline $\{\delta\sigma_{A,\mathrm{max}} = \lambda_A \underset{B}{\mathrm{max}} |E_{B A}|^2\}_{A=1}^{N}$ 
\State Find order of $\delta\sigma_{A,\mathrm{max}}$: $\{B_n\}_{n=1}^N =\mathrm{argsort}(\delta\sigma_{B,\mathrm{max}})$
\State $n = N$
\Repeat
    \State Compute current error $\{ \sigma_A \gets \sigma_A - \lambda_{B_n} |E_{A B_n}|^2\}_{A=1}^N$
    \State $n \gets n - 1$
\Until{$\sigma_A < \sigma \, \forall \, A=1,\dots,N$}
\vskip 10pt
\State {\bf Output:} Eigenvectors in waveform domain \newline
$\left\{e_k(x) = \frac{1}{\sqrt{\lambda_{B_k}}} \sum_{A=1}^N h_A(x) E_{A B_k}\right\}_{k=n}^N$
\end{algorithmic}
\end{algorithm}
 
The process of diagonalizing the matrix $M_{A B}$ of Eq.~\eqref{eq:MAB} and finding the eigenvalues in the waveform domain using Eq.~\eqref{eq:eA_def} is equivalent to performing Singular Value Decomposition (SVD) on a set of waveforms $\{h_A\}_{A=1}^N$, which has been previously used in the literature for the Reduced Order Modeling (ROM) of GW waveforms (See Refs.~\cite{Purrer:2014fza, Cotesta:2020qhw}). However, we follow the procedure outlined in this paper since it has a few numerical advantages. Namely, if we have waveforms with a number of sampling points $M$, storing them will require $2 M N n_B$ bytes, which in the usual case that $M \gg N$, will be much larger than the memory needed to store $M_{A B}$ (Eq.~\eqref{eq:Memory_MAB}) and we will be even more limited in the number of waveforms we can analyze at once. Moreover, if we are studying the ROB of the space orthogonal to $\{e_a\}_{a=1}^n$, our algorithm is equivalent to computing the SVD of the orthogonal part of the waveforms $\{h_A - h_A^\mathrm{ROB}\}_{A=1}^N$. Finding this orthogonal part is, in general, a computationally expensive process that can be avoided if $M_{AB}^\mathrm{ROB}$ is obtained using Eq.~\eqref{eq:MAB^ROB}. Since we are going to select $n_\mathrm{new} \ll N$ eigenvectors of $M_{AB}^\mathrm{ROB}$, we can just compute the orthogonal projection of their corresponding eigenvectors in the waveform domain at the end of the algorithm.

\subsection{Empirical Interpolation Model}
\label{sec:Algorithm:EIM}

Writing a given template in the form of Eq.~\eqref{eq:hA_ROB} will not save computational cost, since one needs the full waveform $h_A(x)$ to compute the inner product $\langle h_A, e_B \rangle$. To avoid this, we approximate the inner products $\langle h(\Vec{\lambda}), e_i \rangle$ by some coefficients $c_i(\Vec{\lambda})$ that will in general be functions of the parameters of the waveform $\Vec{\lambda}$ (e.g. for a CBC this would be masses, spins, inclination and coalescence phase). The approximate waveform can then be written as:
\begin{equation}
    I_n[h](x,\Vec{\lambda}) = \sum_{i=1}^n c_i(\Vec{\lambda}) e_i(x) \, .
    \label{eq:Inh_ci}
\end{equation}

We force the approximation to be exact at some interpolation nodes $\{X_j\}_{j=1}^m$
\begin{equation}
    I_n[h](X_j,\Vec{\lambda}) = h(X_j, \Vec{\lambda}) = \sum_{i=1}^{n} c_i(\Vec{\lambda}) e_i(X_j) \, .
    \label{eq:Interpolant_def}
\end{equation}
\noindent This is what we define as an interpolant. If we identify the matrix 
\begin{equation}
    A_{i j} = e_j(X_i) \, ,
    \label{eq:A_def_ejXi}
\end{equation}
\noindent and take the number of interpolation nodes $m$ to be equal to the number of basis elements $n$, then $\hat{A}$ is a square matrix which we construct by choosing the interpolation nodes $\{X_j\}_{j=1}^n$. Assuming that we construct $\hat{A}$ to be invertible, we can solve Eq.~\eqref{eq:Interpolant_def} for $c_i(\Vec{\lambda})$ in the following way:
\begin{equation}
    c_i(\Vec{\lambda}) = \sum_{j=1}^n (\hat{A}^{-1})_{i j} h(X_j, \Vec{\lambda}) \, .
    \label{eq:cEIM_value}
\end{equation}

We therefore observe that the value of $c_i(\Vec{\lambda})$ will just be a linear combination of the values of the waveform at the different interpolation nodes $\{X_j\}_{j=1}^n$. In practice, the functions $h(x)$ and the ROB elements $\{e_i(x)\}_{i=1}^{n}$ are discretely sampled in a set of points $\{x_i\}_{i=1}^M$, and we can define the matrix:
\begin{equation}
    \hat{V} \equiv [\Vec{e}_1, \dots, \Vec{e}_n] \in \mathbb{C}^{M \times n}\, ,
    \label{eq:Vmatrix_def}
\end{equation}
\noindent where $\Vec{e}_A = e_A(\Vec{x}) \in \mathbb{C}^{M} $. From Eq.~\eqref{eq:A_def_ejXi}, we observe that the matrix $\hat{A}$ can be written in terms of $\hat{V}$ as:
\begin{equation}
    \hat{A} = \hat{P}^\dagger \hat{V} \in \mathbb{C}^{n \times n}\, ,
    \label{eq:Amatrix_def}
\end{equation}
\noindent where the matrix $\hat{P} \in \mathbb{C}^{M \times n}$ is a projector that selects the rows of $\hat{V}$ corresponding to the interpolation nodes. That is:
\begin{equation}
    P_{\alpha j} = \delta_{\alpha \beta_j}
    \label{eq:Pij_def}
\end{equation}
\noindent with $\{\beta_j\}_{j=1}^n$ the indices of the interpolation nodes (i.e. $x_{\beta_j} = X_j$). In terms of these matrices, the empirical interpolation model (EIM) can be written as:
\begin{equation}
    I_n[\vec{h}] = \hat{V} (\hat{P}^\dagger \hat{V})^{-1} \hat{P}^\dagger \vec{h} \, .
    \label{eq:Interpolant_matrix_def}
\end{equation}
\noindent which is an interpolant because $\hat{P}^\dagger I_n[\vec{h}] = \hat{P}^\dagger \vec{h}$. In terms of the matrix $\hat{V}$, the ROB representation of $\vec{h}$ is given by
\begin{equation}
    \vec{h}^\mathrm{ROB} = \hat{V} \hat{V}^\dagger \vec{h} \, .
    \label{eq:ROB_matrix_def}
\end{equation}

Note that even though the basis elements $\Vec{e}_A$ are orthonormal, and therefore $\hat{V}^\dagger \hat{V} = \mathbb{1}_{n \times n}$, since the matrices are not square, we have that in general $\hat{V} \hat{V}^\dagger \neq \mathbb{1}_{M \times M}$. From Eqs.~(\ref{eq:Interpolant_matrix_def}, \ref{eq:ROB_matrix_def}) we can explicitly see that the EIM acting on a waveform in the ROB space will have no effect. That is:
\begin{align}
    I_n[\vec{h}^\mathrm{ROB}] & =  \hat{V} (\hat{P}^\dagger \hat{V})^{-1} \hat{P}^\dagger (\hat{V} \hat{V}^\dagger \vec{h}) = \hat{V} (\hat{P}^\dagger \hat{V})^{-1} (\hat{P}^\dagger \hat{V}) \hat{V}^\dagger \vec{h} \nonumber \\
    &= \hat{V} \hat{V}^\dagger \vec{h} = \vec{h}^\mathrm{ROB}
    \label{eq:EIM_ROB}    
\end{align}

This can be used to relate the representation error of the EIM with the representation error of the ROB. Computing the modulus of the difference between the exact waveform and its EIM representation we obtain:
\begin{align}
    \sigma_\mathrm{EIM}(\Vec{h}) & = \left\Vert \Vec{h} - I_n[\vec{h}] \right\Vert^2 = \left\Vert \left[\mathbb{1} - \hat{V} (\hat{P}^\dagger \hat{V})^{-1} \hat{P}^\dagger\right]\Vec{h} \right\Vert^2  \nonumber \\
    & = \left\Vert \left[\mathbb{1} - \hat{V} (\hat{P}^\dagger \hat{V})^{-1} \hat{P}^\dagger\right](\Vec{h} -\vec{h}^\mathrm{ROB})  \right\Vert^2 \nonumber \\
    & \leq \left\Vert \mathbb{1} - \hat{V} (\hat{P}^\dagger \hat{V})^{-1} \hat{P}^\dagger \right\Vert_2^2 \underbrace{\Vert \Vec{h} -\vec{h}^\mathrm{ROB}  \Vert^2}_{\sigma_\mathrm{ROB}(\Vec{h})} \, ,
    \label{eq:sigmaEIM}
\end{align}
\noindent where $\Vert \cdot \Vert_2$ denotes the matrix 2-norm, which is given by:
\begin{equation}
    \Vert \hat{M} \Vert_2 = \max_{\Vec{x} \neq 0} \frac{\Vert \hat{M} \Vec{x} \Vert}{\Vert \Vec{x} \Vert} = \sqrt{\lambda_\mathrm{max}(\hat{M}^\dagger \hat{M})} = \sqrt{\lambda_\mathrm{max}(\hat{M} \hat{M}^\dagger)} \, ,
    \label{eq:Matrix2normDef}
\end{equation}
\noindent where $\Vert \Vec{x} \Vert$ is the usual vector norm and $\lambda_\mathrm{max}(\hat{M}^\dagger \hat{M})$ denotes the maximum eigenvalue of $\hat{M}^\dagger \hat{M}$. Since $\hat{V} (\hat{P}^\dagger \hat{V})^{-1} \hat{P}^\dagger$ is idempotent, that is $(\hat{V} (\hat{P}^\dagger \hat{V})^{-1} \hat{P}^\dagger)^2 = \hat{V} (\hat{P}^\dagger \hat{V})^{-1} \hat{P}^\dagger$, and it is different from $\mathbb{0}$ or the identity $\mathbb{1}$, it follows that~\cite{Szyld2006}:
\begin{equation}
     \left\Vert \mathbb{1} - \hat{V} (\hat{P}^\dagger \hat{V})^{-1} \hat{P}^\dagger \right\Vert_2 =  \left\Vert\hat{V} (\hat{P}^\dagger \hat{V})^{-1} \hat{P}^\dagger \right\Vert_2 \, .
    \label{eq:Matrix2norm_idempotent}
\end{equation}

Furthermore, since $\hat{V}^\dagger \hat{V} = \mathbb{1}_{n \times n}$ and $\hat{P}^\dagger \hat{P}  = \mathbb{1}_{n \times n}$, from the definition in Eq.~\eqref{eq:Matrix2normDef} of the matrix 2-norm, we have that
\begin{equation}
     \left\Vert\hat{V} (\hat{P}^\dagger \hat{V})^{-1} \hat{P}^\dagger \right\Vert_2 = \left\Vert (\hat{P}^\dagger \hat{V})^{-1} \right\Vert_2 \, .
    \label{eq:Matrix2norm_eliminateV}
\end{equation}

Substituting in Eq.~\eqref{eq:sigmaEIM}
\begin{align}
    \sigma_\mathrm{EIM}(\Vec{h})  \leq & \Vert(\hat{P}^\dagger \hat{V})^{-1}\Vert_2^2 \sigma_\mathrm{ROB}(\Vec{h}) = \Vert\hat{A}^{-1}\Vert_2^2 \sigma_\mathrm{ROB}(\Vec{h}) \, .
    \label{eq:sigmaEIM_upperbound}
\end{align}

Using the definition of the matrix 2-norm of Eq.~\eqref{eq:Matrix2normDef}, we have that

\begin{align}
    \Vert\hat{A}^{-1}\Vert_2^2 &  = \lambda_\mathrm{max}\left((\hat{A}^{-1})^\dagger \hat{A}^{-1}\right) = \lambda_\mathrm{max}\left((\hat{A}^\dagger)^{-1} \hat{A}^{-1}\right) \nonumber \\
    & = \lambda_\mathrm{max}\left((\hat{A} \hat{A}^\dagger)^{-1}\right) = \frac{1}{\lambda_\mathrm{min}(\hat{A} \hat{A}^\dagger)} \, ,
    \label{eq:norm22invA_rewriten}
\end{align}

\noindent and we can rewrite Eq.~\eqref{eq:sigmaEIM_upperbound} as

\begin{align}
    \sigma_\mathrm{EIM}(\Vec{h})  \leq & \frac{\sigma_\mathrm{ROB}(\Vec{h})}{\lambda_\mathrm{min}(\hat{A} \hat{A}^\dagger)} \, .
    \label{eq:sigmaEIM_upperbound_rewritten}
\end{align}

Therefore, given a maximum error of the ROB, the error of the EIM model is bounded from above by Eq.~\eqref{eq:sigmaEIM_upperbound_rewritten}. To make this bound as stringent as possible, we could maximize the smallest eigenvalue of $\hat{A} \hat{A}^\dagger$. Using the definition of $\hat{A}$ from Eq.~\eqref{eq:Amatrix_def} we can write

\begin{equation}
    (\hat{A} \hat{A}^\dagger)_{i j} = \sum_{k=1}^n e_k(X_i) e^{*}_k(X_j) = \langle \Vec{v}_j, \Vec{v}_i \rangle \, ,
    \label{eq:AA_dagger}
\end{equation}

\noindent where we have defined the vectors $\{(\Vec{v}_i)_k = e_k(X_i)|k=1,\dots,n\}_{i=1}^M$ as the rows of $\hat{V}$ corresponding to the interpolation nodes $X_i$. We then observe that $\hat{A} \hat{A}^\dagger$ is the same as the scalar product between the corresponding selected rows of $\hat{V}$.

If the vectors $\Vec{v}_i$ were orthonormal, we would obtain that $(\hat{A} \hat{A}^\dagger)_{i j} = \delta_{i j}$, and therefore $\lambda_\mathrm{min}(\hat{A} \hat{A}^\dagger) = 1$ and the EIM would not introduce additional error over the ROB. Selecting $n$ orthonormal rows of $\hat{V}$ is in general not possible, however, we can try to minimize the EIM error by picking rows which are as close to orthogonal as possible using algorithm~\ref{alg:create_EIM_orto}.

% New algorithm 
\begin{algorithm}[H]
\caption{Selection of interpolation nodes}
\label{alg:create_EIM_orto}
\begin{algorithmic}[1]
\State {\bf Input:} Evaluated basis $\{ \vec{e}_i \}_{i=1}^n$
\vskip 10pt
\State Define row vectors: $\{ \vec{v}_\alpha = \{ e_i(x_\alpha)\}_{i=1}^n \}_{\alpha=1}^M$ 
\State Initialize ortonormal base of columns: $\mathrm{OB} = \{\Vec{w}_i\}_{i=1}^0$
\State Initialize the norm of the orthogonal part of $\vec{v}_\alpha$ to OB: $\{ N_\alpha = |\vec{v}_\alpha|^2 \}_{\alpha=1}^M$ 
\vskip 10pt
\For{$j = 1 \to n$}
\State Choose vector with largest $N_\alpha$: $\beta_j = \mathrm{argmax}(N_\alpha)$
\State Append $\vec{v}_{\beta_j}$ to OB using Gram-Schmidt
\State Update $N_\alpha$: $\{N_\alpha \gets N_\alpha - |\langle \Vec{w}_j, \vec{v}_\alpha\rangle| \}_{\alpha=1}^M$
\EndFor
\vskip 10pt
\State {\bf Output:} EIM interpolation nodes $\{ \beta_i \}_{i=1}^n$
\end{algorithmic}
\end{algorithm}

We observe that Algorithm~\ref{alg:create_EIM_orto} is equivalent to picking the EIM nodes that maximize the determinant of $\hat{A} \hat{A}^\dagger$, since

\begin{equation}
    \det(\hat{A} \hat{A}^\dagger) = \det(\hat{A}) \det(\hat{A}^\dagger) = |\det(\hat{A})|^2 = \prod_{j=1}^{n} |\langle\vec{w}_j, \vec{v}_{\beta_j}\rangle|^2 \, .
    \label{eq:det_AAH_orto}
\end{equation}

\begin{figure}
    \centering
    \includegraphics[width=0.5\textwidth]{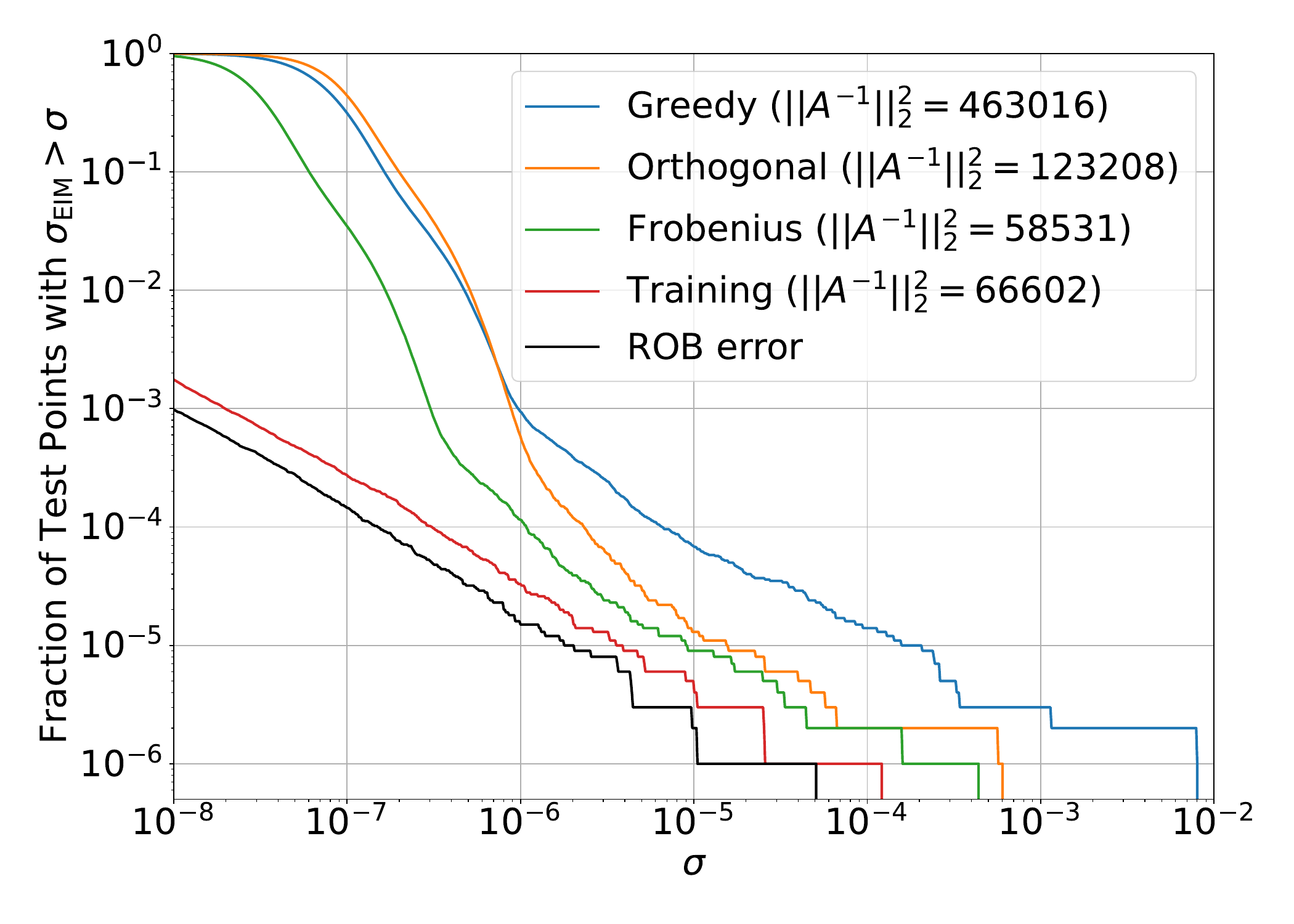}
    \includegraphics[width=0.5\textwidth]{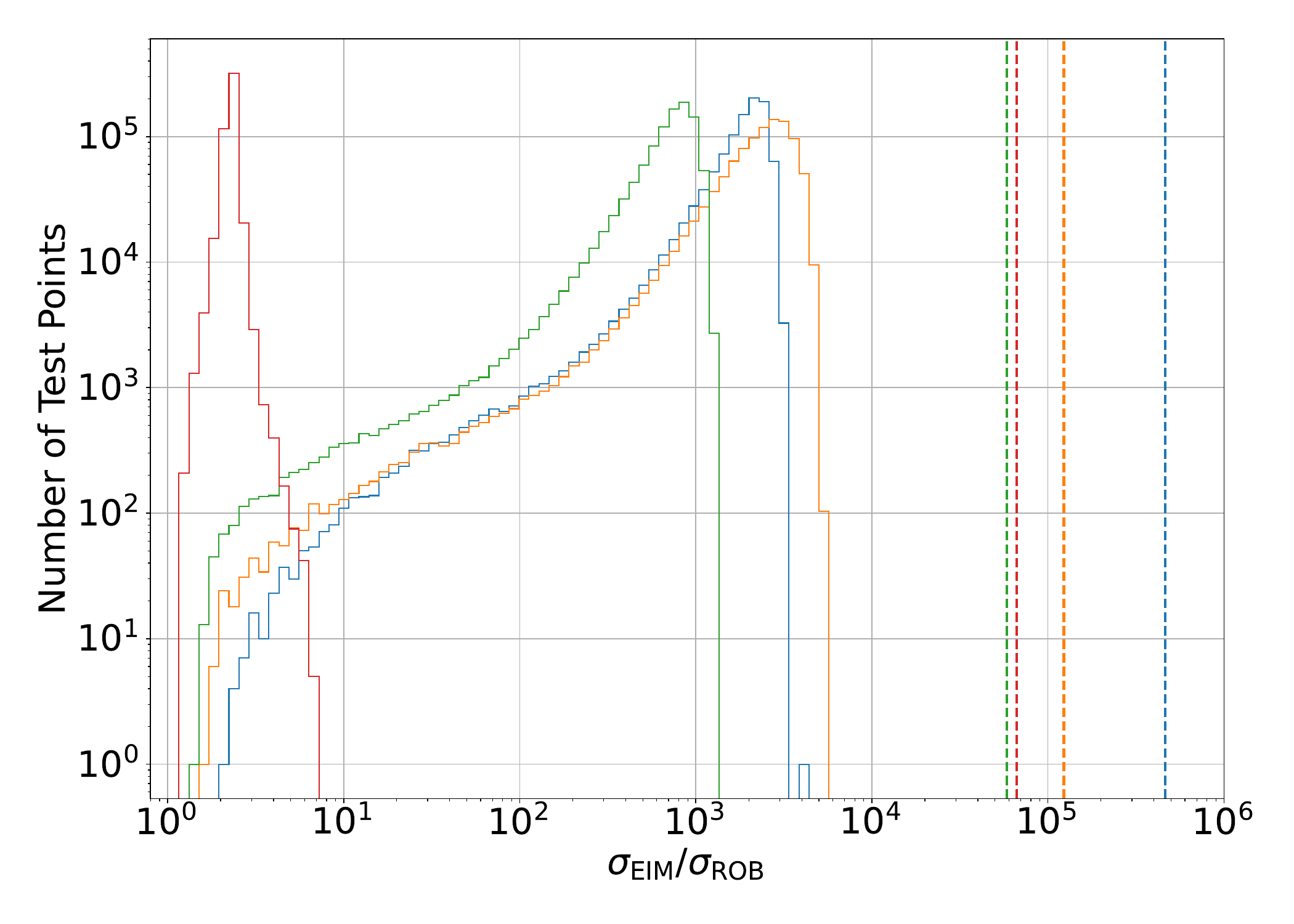}
    \caption{Comparison of methods to compute the EIM for the 256s \texttt{IMRPhenomPv2} ROB of Table~\ref{table:Pv2Basis}. We test the different EIMs on the same $10^6$ samples randomly drawn from the parameter space over which the ROB is generated (see Table~\ref{table:Pv2Basis}). The \textit{Greedy} method is the one outlined in~\cite{Canizares:2013ywa}, the \textit{Orthogonal} method stands for algorithm~\ref{alg:create_EIM_orto}, the \textit{Frobenius} method corresponds to using algorithm~\ref{alg:create_EIM_walk} to minimize $\Vert \hat{A}^{-1} \Vert_F$ and the \textit{Training} method is the one used to construct the EIM of Table~\ref{table:Pv2Basis} with algorithm~\ref{alg:create_EIM_walk_train}. Upper panel: Fraction of samples with an EIM error larger than a tolerance $\sigma$ as a function of $\sigma$. For comparison purposes, we also show the distribution of the ROB error. Lower panel: Histogram of the ratio between the EIM error and the ROB error for the same methods and test samples as in the upper panel. The vertical dashed lines represent an upper bound, defined by the value of $\Vert \hat{A}^{-1} \Vert_2^2$ for each method.}
    \label{fig:compare_EIM}
\end{figure}

Algorithm~\ref{alg:create_EIM_orto} does not directly maximize the minimum eigenvalue of $\hat{A} \hat{A}^\dagger$. However, based on the expression for the determinant of $\hat{A} \hat{A}^\dagger$

\begin{equation}
    \det(\hat{A} \hat{A}^\dagger) = \prod_{j=1}^n \lambda_i \, ,
    \label{eq:det_AAH_eig}
\end{equation}

\noindent to maximize it, the values of the individual eigenvalues have to be large, and thus, the output of the algorithm is near to the minimum of $\Vert\hat{A}^{-1}\Vert_2^2$. When compared to the greedy algorithm typically used in the literature (e.g. Refs.~\cite{Antil:2012wf,Canizares:2013ywa,Smith:2016qas,Qi:2020lfr}) to compute the interpolation nodes, we observe a superior performance of algorithm~\ref{alg:create_EIM_orto}, as we will later discuss in relation to figure~\ref{fig:compare_EIM}.

If we wanted to create an EIM with a tolerance smaller than $\sigma$, from Eq.~\eqref{eq:sigmaEIM_upperbound_rewritten} we could in principle just construct a ROB with a tolerance better than $\lambda_\mathrm{min}(\hat{A} \hat{A}^\dagger) \sigma$. However, in real settings, we observe that Eq.~\eqref{eq:sigmaEIM_upperbound_rewritten} is a loose upper bound on the EIM error, and we can obtain an EIM with a tolerance better than $\sigma$ using fewer basis elements.

Instead of bounding $\sigma_\mathrm{EIM}(\Vec{h})$ using the inequality of Eq.~\eqref{eq:sigmaEIM}, we can refine this expression by doing:
\begin{align}
    \sigma_\mathrm{EIM}(\Vec{h}) & = \left\Vert \left[\mathbb{1} - \hat{V} (\hat{P}^\dagger \hat{V})^{-1} \hat{P}^\dagger\right](\Vec{h} -\vec{h}^\mathrm{ROB})  \right\Vert^2 \nonumber \\
    & = \left\Vert \Vec{h} -\vec{h}^\mathrm{ROB} \right\Vert^2 + \left\Vert \hat{V} (\hat{P}^\dagger \hat{V})^{-1} \hat{P}^\dagger (\Vec{h} -\vec{h}^\mathrm{ROB}) \right\Vert^2 \nonumber \\
    & = \sigma_\mathrm{ROB}(\Vec{h}) + \left\Vert (\hat{P}^\dagger \hat{V})^{-1} \hat{P}^\dagger (\Vec{h} -\vec{h}^\mathrm{ROB}) \right\Vert^2 \, ,
    \label{eq:sigmaEIM_v2}
\end{align}

\noindent where we have used that $\hat{V}^\dagger \hat{V} = \mathbb{1}$ and that the EIM projects the waveform onto the ROB, and therefore  $\langle \hat{V} (\hat{P}^\dagger \hat{V})^{-1} \hat{P}^\dagger (\Vec{h} -\vec{h}^\mathrm{ROB}), \, \Vec{h} -\vec{h}^\mathrm{ROB} \rangle = 0$.  From Eq.~\eqref{eq:sigmaEIM_v2} we have that the EIM error is always larger than or equal to the ROB error. We also observe that for the bound of Eq.~\eqref{eq:sigmaEIM_upperbound_rewritten} to be saturated we need $\hat{P}^\dagger ( \Vec{h} -\vec{h}^\mathrm{ROB})$ to be the eigenvector of $\hat{A}^\dagger \hat{A}$ with the maximum eigenvalue, which is extremely unlikely in general. To explore this we assume that $\Vec{h} -\vec{h}^\mathrm{ROB} \equiv \delta \Vec{h}$ is a random variable, such that:

\begin{equation}
    \mathbb{E}\left[\delta h^{*}_\alpha \delta h_\beta \right] = c_\alpha \delta_{\alpha \beta} \, ,
    \label{eq:ROB_err_dist_assumption}
\end{equation}

\noindent where $\mathbb{E}\left[ \, \cdot \, \right]$ denotes the expected value (i.e. the average over random waveform realizations). Using Eq.~\eqref{eq:ROB_err_dist_assumption}, we compute the expected value of $\sigma_\mathrm{EIM}$ as

\begin{align}
    \mathbb{E}\left[\sigma_\mathrm{EIM}\right] & = \sum_{\alpha=1}^M \mathbb{E}\left[ \delta h^{*}_\alpha \delta h_\alpha \right] \nonumber \\ 
    & \quad + \sum_{k=1}^n \sum_{q=1}^n \sum_{l=1}^n (A^{-1})^{*}_{l k} (A^{-1})_{l q} \mathbb{E}\left[ \delta h^{*}_{\beta_k} \delta h_{\beta_q} \right] \nonumber \\
    & = \sum_{\alpha=1}^M c_\alpha + \sum_{k=1}^n \sum_{l=1}^n c_{\beta_k} |(A^{-1})_{l k}|^2 \nonumber \\
    & = \sum_{\alpha=1}^M c_\alpha  + \left\Vert \Hat{\tilde{A}}^{-1} \right\Vert^2_F\, ,
    \label{eq:E_sigma_EIM}
\end{align}

\noindent where $\hat{\tilde{A}}$ is the matrix

\begin{equation}
    \tilde{A}_{k l} = \frac{1}{\sqrt{c_{\beta_k}}} A_{k l} = \frac{1}{\sqrt{c_{\beta_k}}} e_l (x_{\beta_k}) \, .
    \label{eq:A_tilde}
\end{equation}

Such that $(\tilde{A}^{-1})_{l k} = \sqrt{c_{\beta_k}} (A^{-1})_{l k}$ and $\Vert \cdot \Vert_F$ is the Frobenius norm, defined as:

\begin{equation}
    ||\hat{M}||_F = \sqrt{\sum_{k=1}^n \sum_{l=1}^n |M_{k l}|^2} = \sqrt{\mathrm{Tr}\left\{ \hat{M}^\dagger \hat{M} \right\}} \, .
    \label{eq:F_norm}
\end{equation}

Therefore, to optimize the EIM such that the expected value of $\sigma_\mathrm{EIM}$ is minimum, we want to minimize the value of the Frobenius norm of $\hat{\Tilde{A}}^{-1}$. Using the properties of the trace we can rewrite it as:

\begin{equation}
    \left\Vert \hat{\Tilde{A}}^{-1} \right\Vert_F = \sqrt{\sum_{k=1}^n \frac{1}{\lambda_k (\hat{\Tilde{A}}^\dagger \hat{\Tilde{A}})}}
    \label{eq:F_norm_invA}
\end{equation}

To minimize the Frobenius norm of $\hat{\Tilde{A}})$ we can start from the EIM given by Algorithm.~\ref{alg:create_EIM_orto} and allow the interpolation nodes to ``walk'' in the direction of diminishing $\left\Vert \hat{\Tilde{A}}^{-1} \right\Vert_F$, as outlined in Algorithm~\ref{alg:create_EIM_walk}. 

\begin{algorithm}[H]
\caption{Selection of interpolation nodes to minimize target function of the EIM $F(\cdot)$}
\label{alg:create_EIM_walk}
\begin{algorithmic}[1]
\State {\bf Input:} Maximum number of rounds $N_\mathrm{rounds}$, initial interpolation nodes $\Vec{\beta}$, function to be minimized $F(\Vec{\beta})$. 
\vskip 10pt
\For{$j = 1 \to N_\mathrm{rounds}$}
\For{$k = 1 \to n$}
\For{$\delta\beta$ {\bf in} $[-1, 1]$}
\State Copy interpolation nodes: $\Vec{\beta}' = \Vec{\beta}$
\Repeat
\State Test new EIM: $\beta'_k \gets \beta'_k + \delta\beta$
\If{$F(\Vec{\beta}') \leq F(\Vec{\beta})$}
\State Update reference EIM: $\Vec{\beta} \gets \Vec{\beta}'$
\EndIf
\Until{$F(\Vec{\beta}') > F(\Vec{\beta})$}
\EndFor
\EndFor
\If{$\left\{ \beta_i\right\}_{i=1}^n$ didn't change this iteration}
\State {\bf break for loop}
\EndIf
\EndFor
\vskip 10pt
\State {\bf Output:} EIM interpolation nodes $\{ \beta_i \}_{i=1}^n$
\end{algorithmic}
\end{algorithm}

The time complexity of Algorithm.~\ref{alg:create_EIM_walk} is $O(\!N_\mathrm{rounds} n N_F\!)$, where $N_F$ denotes the number of operations required to compute $F(\vec{\beta})$. Given that our target function is $F(\Vec{\beta}) = \left\Vert \hat{\Tilde{A}}^{-1} \right\Vert_F$, one could naively expect that, based on the size $n \times n$ of the matrix $\hat{A}$, directly inverting it would take $O(n^3)$ operations, and therefore the time complexity of Algorithm~\ref{alg:create_EIM_walk} would be $O(N_\mathrm{rounds} n^4)$. This can be computationally very expensive even if $n \ll M$. However, updating the value of $\Vert \hat{A}^{-1}\Vert_F$ when  only one row of the matrix changes, can be done in $O(n^2)$ by following the procedure of Appendix~\ref{sec:anex:fast_FrobinvA}, and we can implement Algorithm~\ref{alg:create_EIM_walk} with target function $F(\Vec{\beta}) = \left\Vert \hat{\Tilde{A}}^{-1} \right\Vert_F$ in a way that takes $O(N_\mathrm{rounds} n^3)$ operations.

Even though Algorithm~\ref{alg:create_EIM_walk} is considerably better than the greedy algorithms used in the literature, as we will later discuss in relation to figure~\ref{fig:compare_EIM}, it can still be improved by training the EIM directly on the waveform data. For this purpose, we assume that we have an initial ROB $\{\vec{e}_i\}_{i=1}^n$ with a corresponding EIM that can be computed with e.g. algorithm~\ref{alg:create_EIM_walk}. We want to update this EIM to better fit a training set of waveforms $\{h_A\}_{A=1}^N$. We first generate a ROB for the part of the training set orthogonal to the initial ROB ($\vec{h}-\vec{h}^\mathrm{ROB}$), which can be done by diagonalizing the matrix of Eq.~\eqref{eq:MAB^ROB}.  Analogously to Eq.~\eqref{eq:hA_ROB} we can write:

\begin{equation}
    \vec{h}_A - \vec{h}_A^\mathrm{ROB} = \sum_{B=1}^N \sqrt{\lambda_B} E^{*}_{A B} \vec{u}_B \, , 
    \label{eq:h-hROB}
\end{equation}

\noindent where $\lambda_b$ and $E_{A B}$ are the eigenvalues and eigenvectors of the matrix $M_{AB}^\mathrm{ROB}$ defined in Eq.~\eqref{eq:MAB^ROB} and $\vec{u}_B$ represent the eigenvectors in the waveform domain. Substituting Eq.~\eqref{eq:h-hROB} in the expression for $\sigma_\mathrm{EIM}$ derived in Eq.~\eqref{eq:sigmaEIM_v2}, we obtain 

\begin{align}
    & \sigma_{\mathrm{EIM},A}  = \left\Vert \Vec{h}_A - I_n[\vec{h}_A] \right\Vert^2 \nonumber \\
    & = \left\Vert \sum_{B=1}^N \! \sqrt{\lambda_B} E^{*}_{A B} \vec{u}_B \right\Vert^2 \!\! + \left\Vert \sum_{B=1}^N \!\sqrt{\lambda_B} E^{*}_{A B} (\hat{P}^\dagger \hat{V})^{-1} \hat{P}^\dagger \vec{u}_B \right\Vert^2  \nonumber \\
    & = \sum_{B=1}^N \!\lambda_B |E_{AB}|^2 + \sum_{B=1}^N \sum_{C=1}^N \!\sqrt{\lambda_B \lambda_C} E^{*}_{A B} E_{A C} \langle  \vec{w}_C, \vec{w}_B \rangle \, , %\left \langle (\hat{P}^\dagger \hat{V})^{-1} \hat{P}^\dagger \vec{u}_C , (\hat{P}^\dagger \hat{V})^{-1} \hat{P}^\dagger \vec{u}_B\right \rangle
    \label{eq:sigmaEIM_ROB}
\end{align}

\noindent where we have defined

\begin{equation}
    \vec{w}_B = (\hat{P}^\dagger \hat{V})^{-1} \hat{P}^\dagger \vec{u}_B
    \label{eq:wB_def}
\end{equation}

From Eq.~\eqref{eq:sigmaEIM_ROB},  we can compute the sum of all the EIM errors of the waveforms in the training set. That is:

\begin{align}
    \sigma_\mathrm{EIM}^\mathrm{tot} & = \sum_{A=1}^N \sigma_{\mathrm{EIM},A} = \sum_{B=1}^N \lambda_B \left( 1 + \left\langle \vec{w}_B, \vec{w}_B \right\rangle \right) \nonumber \\
    & = \sum_{B=1}^N \lambda_B \left( 1 + \left\Vert  (\hat{P}^\dagger \hat{V})^{-1} \hat{P}^\dagger \vec{u}_B \right\Vert^2 \right) \nonumber \\
    & \approx \sum_{B=1}^{n_\lambda} \lambda_B \left( 1 + \left\Vert  (\hat{P}^\dagger \hat{V})^{-1} \hat{P}^\dagger \vec{u}_B \right\Vert^2 \right) 
    \label{eq:sigmaEIM_ROB_sum}
\end{align}

Where we have used that $E_{A B}$ is unitary and that the matrix $M_{A B}^\mathrm{ROB}$ will usually have a small number of large eigenvalues, with the rest of the eigenvalues close to 0. Therefore, we can truncate the sum to be made only over the largest $n_\lambda$ eigenvalues and obtain a very good approximation of $\sigma_\mathrm{EIM}^\mathrm{tot}$.

\begin{algorithm}[H]
\caption{Selection of interpolation nodes trained on a set of waveforms $\{h_A\}_{A=1}^N$}
\label{alg:create_EIM_walk_train}
\begin{algorithmic}[1]
\State {\bf Input:} Evaluated basis $\{ \vec{e}_i \}_{i=1}^n$, maximum number of rounds $N_\mathrm{rounds}$, $n_\lambda$ eigenvalues $\lambda_B$ and eigenvectors in waveform domain $\vec{u}_{B}$ of the matrix $M_{AB}^\mathrm{ROB} = \langle \vec{h}_A - \vec{h}_A^\mathrm{ROB} , \vec{h}_B - \vec{h}_B^\mathrm{ROB} \rangle$.
\vskip 10pt
\State Compute weights: $c_\alpha = \sum_{B=1}^{n_\lambda} \lambda_B |u_{B, \alpha}|^2$
\State Compute weighted basis: \newline $\left\{ \left\{w_i (x_\alpha) \right\}_{\alpha=1}^M \right\}_{i=1}^n = \left\{ \left\{ \frac{1}{\sqrt{c_{\alpha}}} e_i (x_\alpha) \right \}_{\alpha=1}^M \right\}_{i=1}^n$
\State Get initial EIM $\vec{\beta}$ inputting $\{ \vec{w}_i \}_{i=1}^n$ in Algorithm~\ref{alg:create_EIM_orto}
\State Update $\Vec{\beta}$ using Algorithm.~\ref{alg:create_EIM_walk} with maximum rounds $N_\mathrm{rounds}$ and target function $F(\Vec{\beta}) = \Vert \hat{\Tilde{A}}^{-1} \Vert_F$, where $\Tilde{A}_{i j} = w_j(x_{\beta_i})$
\State Update $\Vec{\beta}$ again with Algorithm.~\ref{alg:create_EIM_walk} with maximum rounds $N_\mathrm{rounds}$ and target function $F(\Vec{\beta}) = \sigma_\mathrm{EIM}^\mathrm{tot}$, where \newline $\sigma_\mathrm{EIM}^\mathrm{tot} = \sum_{B=1}^{n_\lambda} \lambda_B \left( 1 + \sum_{i=1}^n \left| \sum_{j=1}^n  (\hat{A}^{-1})_{i j} u_{B, \beta_j} \right|^2 \right) $ \newline and $A_{i j} = e_j(x_{\beta_i})$
\vskip 10pt
\State {\bf Output:} EIM interpolation nodes $\{ \beta_i \}_{i=1}^n$
\end{algorithmic}
\end{algorithm}

To minimize the value of $\sigma_\mathrm{EIM}^\mathrm{tot}$, we follow Algorithm~\ref{alg:create_EIM_walk_train}, in which we start with an EIM and perform walks around the initial solution in the direction of diminishing $\sigma_\mathrm{EIM}^\mathrm{tot}$. For the initial solution, we will use the EIM generated by Algorithm~\ref{alg:create_EIM_walk} with target function $F(\Vec{\beta}) = \left\Vert \hat{\Tilde{A}}^{-1} \right\Vert_F$. Since we want to fit $\{ h_A \}_{A=1}^N$, following Eq.~\eqref{eq:ROB_err_dist_assumption}, the weights $c_\alpha$ of Eq.~\eqref{eq:A_tilde} are

\begin{align}
    c_\alpha & = \mathbb{E}\left[\delta h^{*}_\alpha \delta h_\alpha \right] = \frac{1}{N} \sum_{A=1}^N |h_{A, \alpha} - h^\mathrm{ROB}_{A, \alpha}|^2 \nonumber \\
    & = \frac{1}{N} \sum_{A=1}^N \sum_{B=1}^N \sum_{C=1}^N \sqrt{\lambda_B \lambda_C} E_{A B}^{*} E_{A C} u^{*}_{C, \alpha} u_{B, \alpha} \nonumber \\
    & = \frac{1}{N} \sum_{B=1}^N \lambda_B |u_{B, \alpha}|^2 \approx \frac{1}{N} \sum_{B=1}^{n_\lambda} \lambda_B |u_{B, \alpha}|^2 \, ,
    \label{eq:weighted_evaluated_bases}
\end{align}

\noindent where we have once again used that $E_{A B}$ is unitary and that the sum can be approximated by taking only the largest $n_\lambda$ eigenvalues. In algorithm~\ref{alg:create_EIM_walk}, using $ \sigma_\mathrm{EIM}^\mathrm{tot}$ as target function, the value of $\sigma_\mathrm{EIM}^\mathrm{tot}$ can be efficiently updated with $O(n n_\lambda)$ operations, as described in appendix~\ref{sec:anex:fast_FrobinvA}. Therefore, the algorithm~\ref{alg:create_EIM_walk} to walk around an initial solution minimizing $\sigma_\mathrm{EIM}^\mathrm{tot}$ will require $O(N_\mathrm{rounds} n^2 n_\lambda)$ operations.

In Figure~\ref{fig:compare_EIM} we show for the 256s \texttt{IMRPhenomPv2} ROB listed in Table~\ref{table:Pv2Basis} a comparison between algorithms~\ref{alg:create_EIM_orto}~\ref{alg:create_EIM_walk}~\ref{alg:create_EIM_walk_train} proposed in this paper, the usual greedy algorithm used in the literature and the lower bound imposed by the ROB error. We show only the analysis for the 256s \texttt{IMRPhenomPv2} basis of Table~\ref{table:Pv2Basis}, but we find similar results for all the other cases in Tables~\ref{table:Pv2Basis}~\ref{table:XPHMBasis}. In the upper panel of Figure~\ref{fig:compare_EIM} we show the fraction of points with an EIM error larger than a tolerance $\sigma$ as a function of $\sigma$.  Comparing the methods we observe that the \textit{Training} one (algorithm~\ref{alg:create_EIM_walk_train}) outperforms the others, which is expected since it has been trained on the waveform data to reduce the EIM error. The worst performer is the \textit{Greedy} method since it induces the largest EIM error in all cases tested. We also observe that the \textit{Frobenius} method, which uses algorithm~\ref{alg:create_EIM_walk} to minimize $\Vert \hat{A}^{-1} \Vert_F$ induces the smallest EIM error among the algorithms that do not train on waveforms, which could make it more robust against overfitting.

In the lower panel of Figure~\ref{fig:compare_EIM} we show the ratio between the EIM and the ROB error for the same methods and test samples as in the upper panel. We observe that this ratio is in the range $1 \leq \sigma_\mathrm{EIM}/\sigma_\mathrm{ROB} \leq \Vert \hat{A}^{-1} \Vert_2^2$, as was derived in Eqs.~(\ref{eq:sigmaEIM_upperbound},\ref{eq:sigmaEIM_v2}). In general, we observe that the EIM errors obtained with the different methods are always considerably below the upper limit imposed by Eq.~\eqref{eq:sigmaEIM_upperbound} ($\sigma_\mathrm{EIM}/\sigma_\mathrm{ROB} \ll \Vert \hat{A}^{-1} \Vert_2^2$). This is expected since to saturate this upper bound we need $\hat{P}^\dagger ( \Vec{h} -\vec{h}^\mathrm{ROB})$ to be the eigenvector of $\hat{A}^\dagger \hat{A}$ with the maximum eigenvalue, which is hard to get in practice. We also observe that the \textit{Training} method is almost optimal since most samples are close to the lower bound of $\sigma_\mathrm{EIM}/\sigma_\mathrm{ROB} \geq 1$. In contrast, most of the samples for the methods that do not involve training on waveform data, concentrate at values of $\sigma_\mathrm{EIM}/\sigma_\mathrm{ROB}\gtrapprox 10^3$. This is probably because when we train on the waveform data, we are selecting an EIM that avoids coincidences between $\hat{P}^\dagger ( \Vec{h} -\vec{h}^\mathrm{ROB})$ and eigenvectors of $\hat{A}^\dagger \hat{A}$ with large eigenvalues.

\subsection{Construction of the ROQ}
\label{sec:Algorithm:EigROQ}

In this section, we describe how we use the methods of sections \ref{sec:Algorithm:ROB} and \ref{sec:Algorithm:EIM} to create, in an efficient way, an EIM that fits a waveform model over a parameter space with a tolerance better than $\sigma$. 

We obtain an initial ROB $\{\vec{e}_i\}_{i=1}^n$ using Algorithm~\ref{alg:create_ROB} and construct its corresponding EIM with Algorithm~\ref{alg:create_EIM_walk_train}, where the set of training waveforms is the $\{h_A\}_{A=1}^N$, selected in the last step of Algorithm~\ref{alg:create_ROB}. We add elements to this initial ROB following a similar philosophy to that of Algorithm~\ref{alg:create_ROB}, in which we generate $N_\mathrm{lim}$ random waveforms, compute their EIM error $\sigma_\mathrm{EIM}$, and select the $N$ waveforms with largest EIM error for further study. Again, we want $N$ to be as large as allowed by the memory (see Eq.~\eqref{eq:Memory_MAB}). We then compute the matrix $M_{A B}^\mathrm{ROB}$ for the $N$ selected waveforms, find its Eigenvalues $\lambda_B$ and compute the $n_\lambda < N$ most relevant eigenvectors in the waveform domain $\{\vec{u}_B\}_{B=1}^{n_\lambda}$, where the value of $n_\lambda$ is again limited by the memory of the system. We iteratively select the eigenvector with the largest contribution to the EIM error, add it to the ROB and construct a new EIM with Algorithm~\ref{alg:create_EIM_walk_train} until all $N$ waveforms are fitted with a tolerance better than the required one. The process is summarized in Algorithm.~\ref{alg:eig_EIM}.

\begin{algorithm}[H]
\caption{Enrich ROB to construct an EIM under tolerance}
\label{alg:eig_EIM}
\begin{algorithmic}[1]
\State {\bf Input:} Initial ROB $\{\vec{e}_i\}_{i=1}^n$ and EIM $\{\beta_i\}_{i=1}^n$, maximum number of waveforms selected $N$, tolerance $\sigma$, maximum number of waveforms computed $N_\mathrm{lim}$, maximum number of eigenvectors used $n_\lambda$
\vskip 10pt
\Repeat
   \State Generate $N_{\mathrm{lim},j}$ waveforms $\{h_A \}_{A=1}^{N_{\mathrm{lim},j}}$ and compute their EIM error $\sigma_{\mathrm{EIM},A}$
   \State Select the N waveforms $\{h_A\}_{A=1}^N$ with largest $\sigma_\mathrm{EIM}$
   \State Save the minimum value of $\sigma_\mathrm{EIM}$ for the selected waveforms: $\sigma_{\mathrm{EIM},\mathrm{min}}$
   \State $M^\mathrm{ROB}_{A B} = \langle h_A, h_B \rangle - \sum_{c=1}^{n_{j-1}} \langle h_A, e_c \rangle \langle e_c, h_B \rangle$
   \State Diagonalize $M^\mathrm{ROB}_{A B}$ and obtain eigenvalues $\lambda_A$ and eigenvectors $E_{A B}$
   \State Compute the $n_\lambda$ normalized eigenvectors in waveform domain with largest $\delta \sigma_{A,\mathrm{max}} = \lambda_A \underset{B}{\max} |E_{A B}|^2$:  $\{\vec{u}_{A}\}_{A=1}^{n_\lambda}$
   \Repeat
   \State Compute the maximum contribution of each eigenvector to $\sigma_{\mathrm{EIM},A}$:\newline $\{\delta\sigma^\mathrm{EIM}_{A,\mathrm{max}} = \lambda_A (1 + \Vert  (\hat{P}^\dagger \hat{V})^{-1} \hat{P}^\dagger \vec{u}_B \Vert^2) \underset{B}{\mathrm{max}} |E_{B A}|^2\}_{A=1}^{n_\lambda}$
   \State Find largest $\delta\sigma^\mathrm{EIM}_{A,\mathrm{max}}$: $A_\mathrm{sel} = \underset{A}{\mathrm{argmax}} (\delta\sigma^\mathrm{EIM}_{A,\mathrm{max}})$
   \State Add the corresponding eigenvector to the ROB: $\{\vec{e}_i\}_{i=1}^n \gets \{\vec{e}_i\}_{i=1}^n \cup \{\vec{u}_{A_\mathrm{sel}} \}$ 
   \State Remove the selected eigenvector from the eigenvector list: $\{\vec{u}_{A}\}_{A=1}^{n_\lambda} \gets \{\vec{u}_{A}\}_{A=1}^{n_\lambda} \setminus \{\vec{u}_{A_\mathrm{sel}}\}$
   \State Input $\{\vec{e}_i\}_{i=1}^n$, $N_\mathrm{rounds}$, $\{\vec{u}_{A}\}_{A=1}^{n_\lambda}$ and their corresponding eigenvalues $\{\lambda_A\}_{A=1}^{n_\lambda}$ into Algorithm.~\ref{alg:create_EIM_walk_train} to obtain a new EIM $\{\beta_i\}_{i=1}^n$.
   \State Find new error of selected waveforms $\{\sigma^\mathrm{new}_{\mathrm{EIM},A}\}_{A=1}^N$
   \Until{$\underset{A}{\max} \; \sigma^\mathrm{new}_{\mathrm{EIM},A} \leq \sigma$}
\Until{$\sigma_{\mathrm{EIM},\mathrm{min}}<\sigma$}
\vskip 10pt
\State {\bf Output:} ROB $\{ e_i \}_{i=1}^n$
\end{algorithmic}
\end{algorithm}

\section{Code Validation}
\label{sec:CodeValidation}

\begin{figure}
    \centering
    \includegraphics[width=0.5\textwidth]{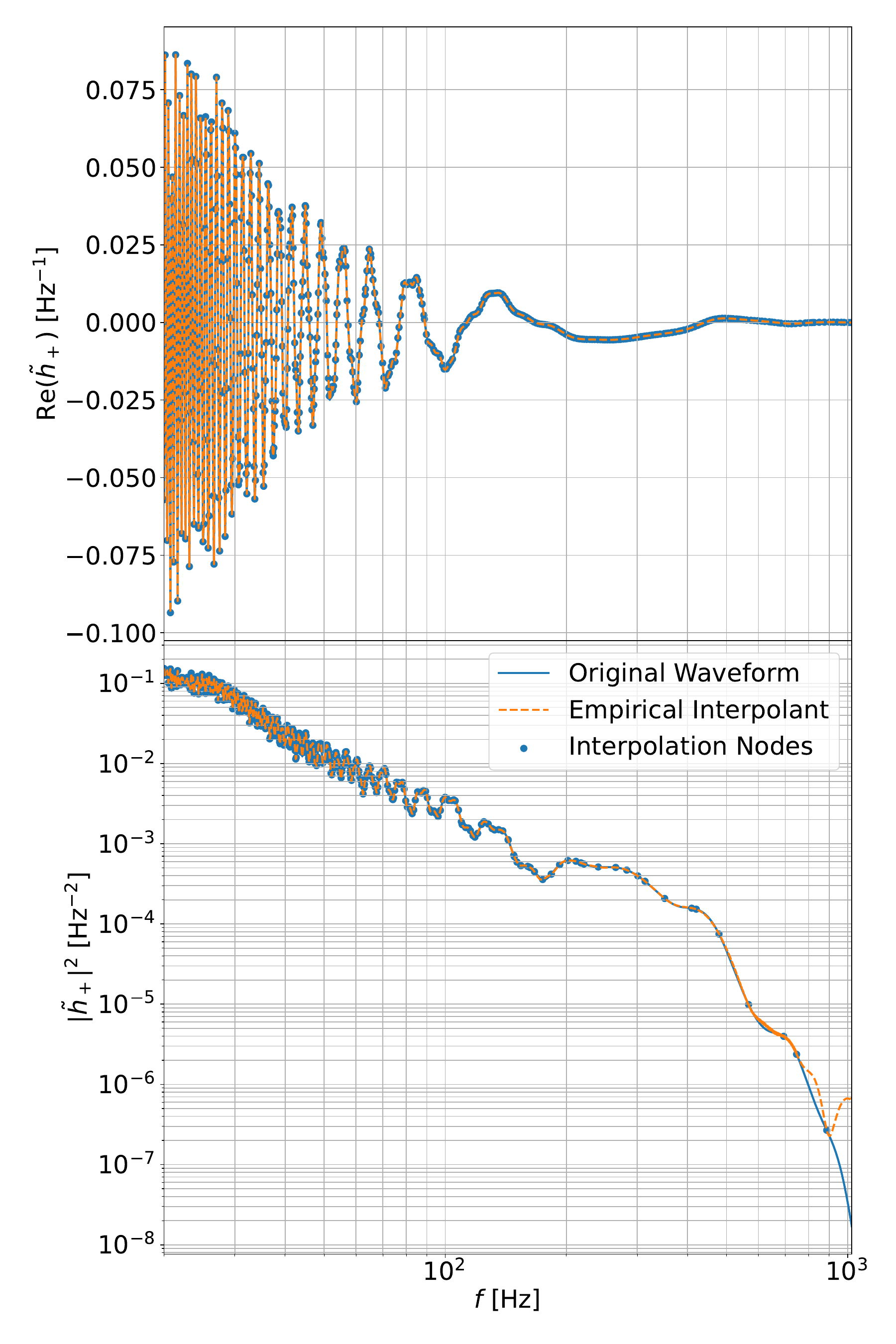}
    \caption{Example of an \texttt{IMRPhenomXPHM} template and its empirical interpolant. In the upper panel, we show the real part of the plus polarization of the template $\mathrm{Re}(\tilde{h}_+)$ as a function of the frequency and in the lower panel we depict its square $|\tilde{h}_+|^2$. We superimpose in each panel the corresponding interpolation nodes and empirical interpolants as defined in~\eqref{eq:Interpolant_def}. The template shown has $\mathcal{M} = 13.6 M_\odot$, $q=2.61$, $\chi_\mathrm{eff} = -0.011$~\cite{PhysRevLett.106.241101,PhysRevD.82.064016}, $\chi_p = 0.208$~\cite{PhysRevD.91.024043} and inclination angle $\iota = 61.6^\circ$. Using the ROQ basis of Table~\ref{table:XPHMBasis} covering $\mathcal{M} \in [10, 15] M_\odot$, we have linear and quadratic EIM errors of $\sigma^{EIM}_\mathrm{linear}= 1.26\cdot10^{-9}$ and $\sigma^{EIM}_\mathrm{linear}= 6.06\cdot10^{-8}$ respectively.}
    \label{fig:waveform_interp_panel}
\end{figure}

In this section, we aim to quantify and assess the validity of the ROQ basis obtained using the algorithm described in~\ref{sec:Algorithm}. For that matter, we would like to evaluate the accuracy of the different basis in reconstructing the original waveform as well as the speed up gained. First, in Sec.~\ref{sec:CodeValidation:BasisGeneration} we describe the bases to be tested and compare them with examples found in the literature, in Sec.~\ref{sec:CodeValidation:StatisticalTests} we show the results of two statistical tests for the various bases, in Sec.~\ref{sec:CodeValidation:Speeduptests} we comment on the theoretical and empirical speedups using the ROQ, and finally in Sec.~\ref{sec:CodeValidation:AplicationtoGWevents} we compare the results of doing a parameter estimation analysis with the standard and the ROQ likelihoods.

\subsection{Basis Generation and comparison with other ROQ methods}
\label{sec:CodeValidation:BasisGeneration}

In this section, we describe how we generate the bases that will be used for testing and parameter estimation. We construct bases for both \texttt{IMRPhenomPv2}~\cite{PhysRevLett.113.151101} and \texttt{IMRPhenomXPHM}~\cite{Pratten_2021}. Both waveform models take into account the effects of spin precession and \texttt{IMRPhenomXPHM} also includes higher order mode GW emission.

For \texttt{IMRPhenomPv2} we generate the bases listed in Table~\ref{table:Pv2Basis}, covering a chirp mass ($\mathcal{M}$) range between $0.95M_\odot$ and $45M_\odot$. Given that integration is performed from a low-frequency cutoff of 20Hz, we find bases duration ranging from 256s to 4s. For \texttt{IMRPhenomXPHM} we generate the bases listed in Table~\ref{table:Pv2Basis}, with chirp masses ranging between $2.18M_\odot$ and $110M_\odot$ and corresponding durations between 64s and 4s from 20Hz. We show in Fig~\ref{fig:waveform_interp_panel} an example of an \texttt{IMRPhenomXPHM} waveform and its corresponding empirical interpolant. More specifically, the upper panel shows the real part of the plus polarization $\mathrm{Re}(\tilde{h}_+)$ and the lower panel, its square $|\tilde{h}_+|^2$ in the frequency domain. The corresponding interpolation nodes and empirical interpolant are shown to visually confirm the goodness of the fit to the original waveform. The parameters of the template are shown in the caption of Figure~\ref{fig:waveform_interp_panel} and are selected so that the quadratic EIM error is equal to the median quadratic EIM error over the testing set of waveforms of the basis of Table~\ref{table:XPHMBasis} covering $\mathcal{M} \in [10, 15] M_\odot$. We can observe how both the linear and quadratic parts have a complicated dependence on frequency, coming from the interference of the higher order modes with the main (2,2) mode. This is the principal reason for the larger number of linear and specially quadratic elements when comparing the basis of \texttt{IMRPhenomPv2} and \texttt{IMRPhenomXPHM}.

The 4s and 8s basis of \texttt{IMRPhenomPv2} and \texttt{IMRPhenomXPHM} are directly comparable with those published in Ref~\cite{Qi:2020lfr} computed using \texttt{PyROQ}, since they cover the exact same parameter space and frequency range. We observe that the number of basis elements in \texttt{PyROQ} and \texttt{EigROQ} is generally similar and we expect it to be smaller than that of a comparable basis constructed with \texttt{GreedyCPP}. However, the number of test points over the set tolerance is about an order of magnitude smaller in our bases than in \texttt{PyROQ}'s ones.\footnote{Note that while in Ref~\cite{Qi:2020lfr} the bases use $10^6$ points for testing, we use $10^7$ points.} We attribute this improvement to the way we approach the minimization in the error of the Empirical Interpolant. In the \texttt{PyROQ} algorithm, it is implicitly assumed that once a template is below the tolerance it will remain like this throughout the computation, which would be true if the EIM error were monotonically decreasing. This, however, is not true in general as adding new templates to the base, can deteriorate the fit, and in particular it can bring some of the waveforms which were under the tolerance, back over tolerance. The fact that this is happening can be explicitly seen in Ref~\cite{Qi:2020lfr} because the maximum EIM error in the training set is over the tolerance. To alleviate this problem, we simultaneously use the $N$ waveforms with initially more EIM error, even if some of them are already below tolerance.

We have also extended the parameter space of the ROQ bases with respect to those computed by \texttt{PyROQ} in Ref~\cite{Qi:2020lfr}, with durations up to 256s for \texttt{IMRPhenomPv2} and 64s for \texttt{IMRPhenomXPHM}. Doing this in \texttt{PyROQ} is computationally challenging since finding the template with the largest associated EIM error requires the recomputation of the waveforms in the training set many times. With our methods, this is no longer the case as we only need to compute any given waveform once. This allows more complex case studies to be feasible.

\begin{table*}
    \centering
    \input{Pv2BasisTable}
        \caption{Summary of the reduced bases constructed with \texttt{EigROQ} for the \texttt{IMRPhenomPv2} waveform model. We limit the mass ratio $1\leq q \leq 8$, the magnitudes of the two spins $-0.8\leq \chi_i\leq0.8$ for $i \in [1,2]$, and the full range for the spin angles $(0,0) \leq (\theta_J , \alpha_0) \leq (\pi, 2\pi)$. For the first base ($\Delta f = 0.25 \mathrm{Hz}$) we extend the coverage in spins to $-0.88\leq \chi_i\leq0.88$. For the creation of all the basis, we run \texttt{EigROQ} with the same configuration. In algorithm~\ref{alg:create_ROB} we set the maximum number of waveform selected $N=20000$ , tolerances of each step $\sigma_i = [10^{-1}, 10^{-3}, 10^{-5}]$ and maximum number of waveforms computed per step $N_{\mathrm{lim}, i} = [10^{6}, 3.16 \cdot 10^6]$, and in algorithm~\ref{alg:eig_EIM} we set $N=10^7$, $\sigma=10^{-5}$, $N_\mathrm{lim}=10^7$ and the maximum number of eigenvectors used $n_\lambda=5000$, except for the 256s basis where we set  $n_\lambda=4000$ due to memory limitations. The basis are tested on $10^7$ randomly generated waveforms in the same parameter space as the training was done on. The ``Theoretical'' speedup has been computed with Eq.~\eqref{eq:TheoreticalSpeedup} while the ``Empirical'' speedup is the median and 90$\%$ credible interval of the corresponding points in the upper panel of Figure~\ref{fig:Speedup}.}
    \label{table:Pv2Basis}
\end{table*}

\begin{table*}
    \centering
    \input{XPHMBasisTable}

        \caption{Summary of the reduced bases constructed with \texttt{EigROQ} for the \texttt{IMRPhenomXPHM} waveform model. We limit the mass ratio $1\leq q \leq 4$, the magnitudes of the two spins $-0.8\leq \chi_i\leq0.8$ for $i \in [1,2]$, and the full range for the spin angles $(0,0) \leq (\theta_J , \alpha_0) \leq (\pi, 2\pi)$.  For the creation of all the basis, we run \texttt{EigROQ} with the same configuration. In algorithm~\ref{alg:create_ROB} we set the maximum number of waveform selected $N=20000$ , tolerances of each step $\sigma_i = [10^{-2}, 10^{-3}, 10^{-4}]$ and maximum number of waveforms computed per step $N_{\mathrm{lim}, i} = [10^{6}, 3.16 \cdot 10^6]$, and in algorithm~\ref{alg:eig_EIM} we set $N=10^7$, $\sigma=10^{-4}$, $N_\mathrm{lim}=10^7$ and the maximum number of eigenvectors used $n_\lambda=5000$. The basis are tested on $10^7$ randomly generated waveforms in the same parameter space as the training was done on. The ``Theoretical'' speedup has been computed with Eq.~\eqref{eq:TheoreticalSpeedup} while the ``Empirical'' speedup is the median and 90$\%$ credible interval of the corresponding points in the lower panel of Figure~\ref{fig:Speedup}. For the empirical speedups, we show the values both without (Emp.) and with (MB) the \texttt{IMRPhenomXPHM} multibanding option enabled~\cite{Garcia-Quiros:2020qlt}.}
    \label{table:XPHMBasis}
\end{table*}

\subsection{Statistical tests}
\label{sec:CodeValidation:StatisticalTests}

In this section, we perform 2 different statistical tests to check the faithfulness of the ROQ basis in gravitational waves inference, a likelihood test and a P-P test.

The likelihood test consists of a comparison of the log-likelihood ratios evaluated using the standard waveform with those obtained using the ROQ approximation. The log-likelihood ratio is defined as the ratio between the likelihood of Eq.~\eqref{eq:Likelihood_def} and the likelihood of the noise hypothesis ($h=0$), that is
\begin{equation}
    \log\mathcal{L}_\mathrm{ratio}(d|\vec{\theta}) = \log\frac{\mathcal{L}(d|h(\vec{\theta}))}{\mathcal{L}(d|0)} = (d, h(\vec{\theta})) - \frac{1}{2} (h(\vec{\theta}), h(\vec{\theta})) \, .
    \label{eq:Likelihood_ratio_def}
\end{equation}

\begin{figure}
    \centering
    \includegraphics[width=0.5\textwidth]{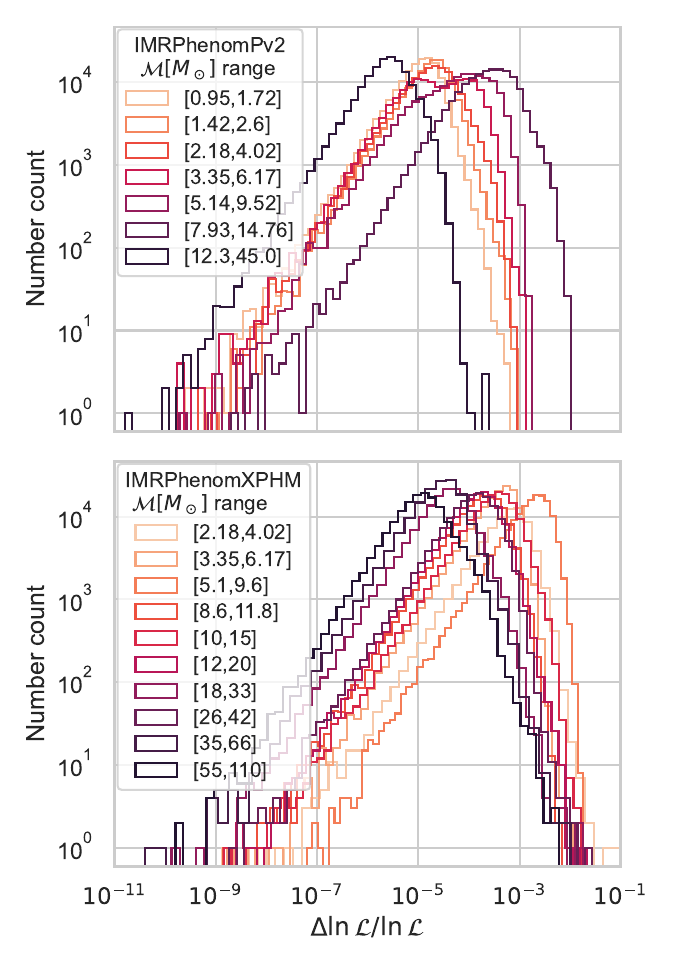}
    \caption{Likelihood error tests for various $\mathcal{M}$ ranges. Specifically, we plot $\Delta \ln \mathcal{L} / \ln \mathcal{L}$, that is, the fractional error of the $\ln\mathcal{L}$ when calculated with and without the ROQ. Upper pannel: \texttt{IMRPhenomPv2}. Lower pannel: \texttt{IMRPhenomXPHM}. }
    \label{fig:Likelihoodtests}
\end{figure}

This quantity, which is just the likelihood of Eq.~\eqref{eq:Likelihood_def} removing the constant part that only depends on the data, is what we will be referring to as the log-likelihood throughout the rest of the text. The log-likelihood is the crucial quantity used in estimating the parameters of a given GW event, which is the ultimate end for which the ROQ is created. We perform likelihood tests on the \texttt{IMRPhenomPv2} and \texttt{IMRPhenomXPHM} bases described in table~\ref{table:Pv2Basis} and table~\ref{table:XPHMBasis} respectively, and show the results on Figure~\ref{fig:Likelihoodtests}. To obtain the difference in the log-likelihood, we create a random realization of Gaussian noise and inject a waveform calculated using the corresponding approximant. The injected waveforms' parameters are randomly sampled from uniform distributions whose boundaries are the respective ROQs' ranges of validity. We use a fixed distance of 100Mpc and randomly sample the incoming direction of the GW from a uniform distribution in the sky. We then compute the standard log-likelihood and the ROQ log-likelihood using the same injected waveform and compare them. What we plot is the relative difference between both logarithms for a total of $1.5\cdot10^5$ realizations. We see the maximum discrepancy lies below 0.1 for every case considered here, and the bulk of the samples lie below $10^{-3}$. 

Given that the likelihood is the only signal-dependent quantity that enters the computation of the posterior (Eq.~\eqref{eq:BayesTheorem}), as long as the ROQ and standard likelihoods agree reasonably well, we can expect the PE posteriors with and without the ROQ to be virtually the same. According to Wilks theorem~\cite{WilksTheorem} in the frequentist and large sample size limits, the quantity $-2\log\{\mathcal{L}/\mathcal{L}_\mathrm{max}\}$ is distributed as a $\chi^2$ with a number of degrees of freedom equal to the number of parameters being fitted by the PE. In the case of a CBC, we need 15 parameters to fully characterize the binary, although, since the azimuthal spin angles and phase of coalescence are usually so poorly constrained, in most cases the effective number of parameters is reduced to 12. We then expect $\log\{\mathcal{L}/\mathcal{L}_\mathrm{max}\} = -5.7^{+3.1}_{-4.8}$, which is in accordance with most of the GW observations, specially those with high signal to noise ratio (SNR). Under the same model, the standard deviation of $\log\mathcal{L}$ is $\sigma_\mathcal{L} \sim \sqrt{N_\mathrm{eff}/2} \sim \sqrt{6} \sim 2.4$, where $N_\mathrm{eff}$ is the effective number of parameters. Therefore, as long as the difference between the logarithm of the standard and the ROQ likelihoods is much smaller than $\sigma_\mathcal{L} \sim 2.4$, we expect the posteriors to be similar.

From Eq.~\eqref{eq:Likelihood_ratio_def}, we observe that the likelihood ratio of a GW signal will approximately be given by $\log\mathcal{L}\sim \rho^2/2$, where $\rho$ is the SNR. Therefore, the condition that $\Delta\log\mathcal{L} \ll 2.3$ can be translated into a condition on the SNR:

\begin{equation}
    \rho \ll 2 \sqrt{\frac{\log\mathcal{L}}{\Delta\log\mathcal{L}}}\, ,
    \label{eq:max_snr_validity}
\end{equation}

\noindent which can be used to interpret figure~\ref{fig:Likelihoodtests} in terms of up to which SNR we can trust the posteriors obtained when using the corresponding ROQ. If we want the ROQ to be valid for the analysis of larger SNRs, we can always decrease the tolerance $\sigma$ with which we generate it, at the expense of having more basis elements.

\begin{figure}
    \centering
    \includegraphics[width=0.5\textwidth]{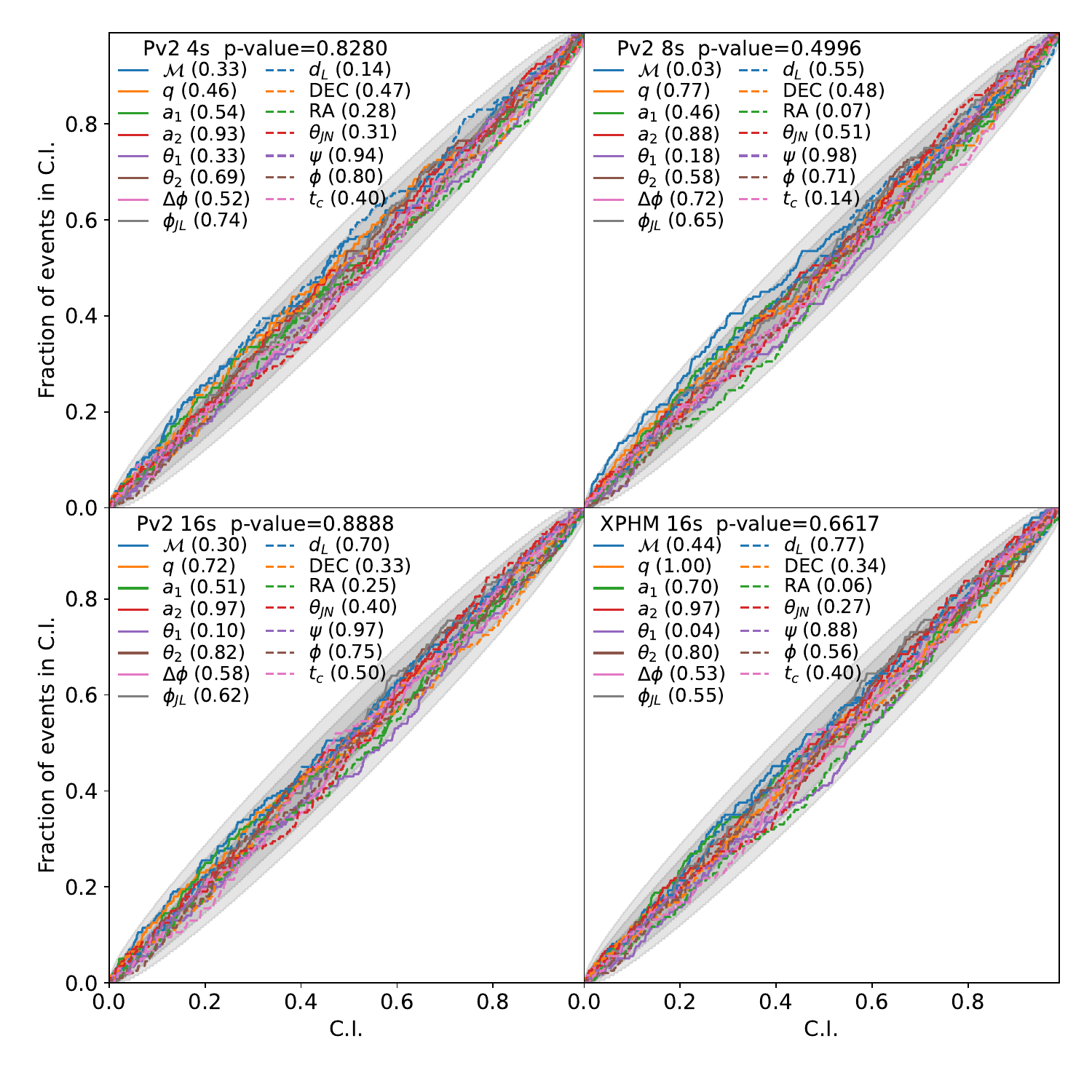}
    \caption{P-P plots performed with different ROQ basis as stated in each subplot's legend. We show here the result of 200 injections being drawn from the corresponding ROQ-compatible prior, as stated in tables~[\ref{table:Pv2Basis},\ref{table:XPHMBasis}]. The contours in grey delimit the 1$\sigma$, 2$\sigma$ and 3$\sigma$ regions. We plot a line for every parameter that uniquely characterizes a given CBC with consistent colors and styles across subplots. The lines represent the cumulative fraction of events. }
    \label{fig:PP_Plot}
\end{figure}

The second of the tests is the percent-percent (P-P) plot~\cite{Gelman_pp_plots,Talts_pp_plots}. P-P plots have been widely used in the literature~\cite{Romero_shaw_bilby_validation} to validate codes that perform Bayesian parameter estimation (PE). Therefore, we use the P-P plots to directly test the ROQ's faithfulness in its intended use. In this specific case, to make the P-P plots shown in Fig.~\ref{fig:PP_Plot}, we use the posteriors pdfs resulting from performing PE on 200 injections. The PE analyses are done using the ROQ likelihood and the \texttt{dynesty}~\cite{speagle2020dynesty} sampler within the \texttt{Bilby}~\cite{Ashton:2018jfp} framework and the injections use the same waveform model for which the corresponding ROQ was constructed. The priors of the PE and the distribution from which the injections are drawn are the same and coincide with the parameter space in which each ROQ basis has been constructed. For the extrinsic parameters we put priors which are uniform in the sky and in comoving volume, going to a maximum distance tailored for each chirp mass range to have detectable signals.

In the P-P plots of Fig.~\ref{fig:PP_Plot} we show the fraction of posterior pdfs for which the injected value of the parameter is found in a given confidence interval as a function of that same confidence interval. We expect the fraction of injected parameter values that fall into a particular confidence interval of the posterior pdfs to be drawn from a uniform distribution. We can thus assign a p-value quantifying such claim~\cite{Talts_pp_plots}, individually for each binary parameter and jointly for all the parameters. The p-values are shown in the legends of figure~\ref{fig:PP_Plot}. For all the PP-plots shown, the cdfs of the majority of the parameters fall within the $3-\sigma$ regions, leading to p-values that are consistent with a uniform distribution. The combined p-values lie between 0.49 and 0.89, indicating that the posterior pdfs produced using these ROQs are well-calibrated. 

\subsection{Speedup analysis}\label{sec:CodeValidation:Speeduptests}

\begin{figure}
\centering
\includegraphics[width=0.5\textwidth]{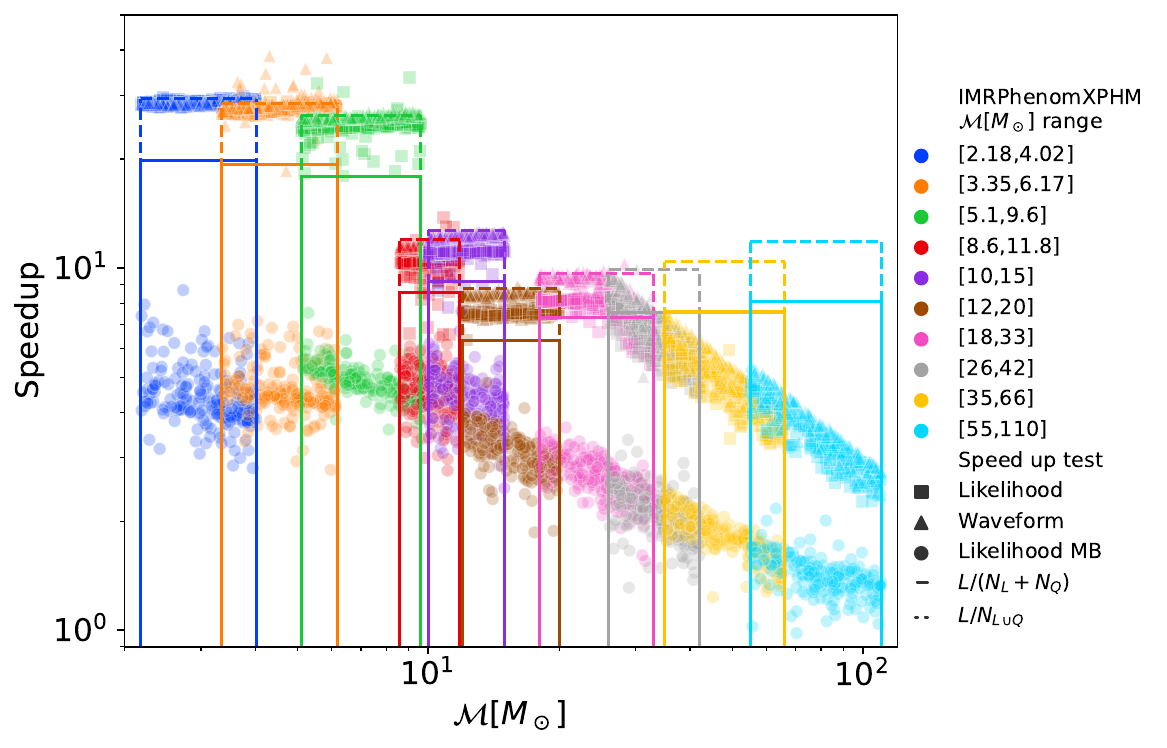}\hspace{0em}%
\includegraphics[width=0.5\textwidth]{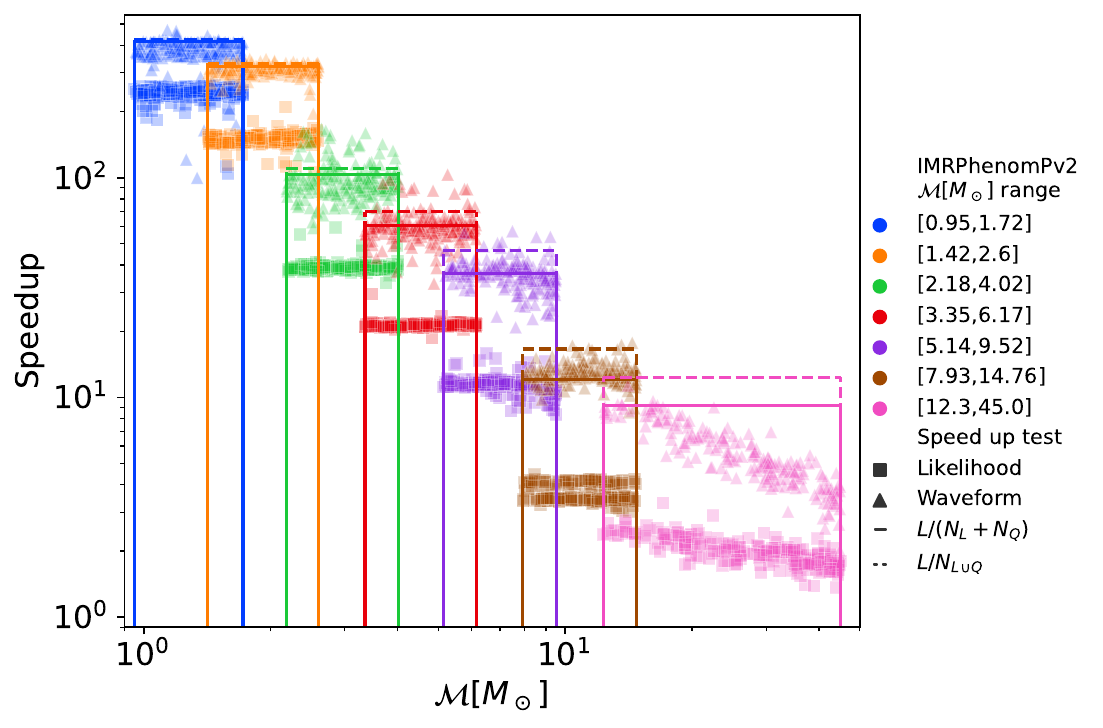}\hspace{0em}
\caption{Speed up factor for the \texttt{IMRPhenomXPHM} (upper panel) and  \texttt{IMRPhenomPv2} (lower panel) waveforms in the different regions in chirp mass where the ROQ has been computed. We can differentiate theoretical and empirical speedups. The empirical speedups are calculated as the ratio between the time spent in computing the waveform without the ROQ and with it and are plotted as triangles. Squares are obtained in the same way but employing the Likelihood. In the \texttt{IMRPhenomXPHM} case, we include the likelihood speedups with the multibanding option enabled (circles) and disabled(squares). The theoretical speedups are drawn as bars. The dashed bars represent the speedup when array frequency duplications are accounted for while solid bars don't.}
\label{fig:Speedup}
\end{figure}

The main purpose of the ROQ is to accelerate the computation of the GW likelihood. To test how good it is in this regard we perform a series of speed-up trials shown in figure~\ref{fig:Speedup}. There are two quantities which we evaluate for the benchmarking test, the waveform and the Gaussian log-likelihood described in Eq~\eqref{eq:Likelihood_def}. The tests consist in timing several calculations of both quantities for the standard case and the ROQ case. The sets of parameters used as inputs are drawn from uniform distributions with boundaries based on the range of validity of the corresponding ROQ basis. The ratio between the time for the standard method and the ROQ is what we call the empirical speedup, where we use the term empirical because we perform the actual likelihood and waveform computations using \texttt{python}~\cite{Python_citation} and the \texttt{Bilby}~\cite{Ashton:2018jfp} framework. For \texttt{IMRPhenomXPHM} waveform speedups, we disable the default multibanding~\cite{Garcia-Quiros:2020qlt}, which is used to speed up the full waveform computation by reducing the number of frequencies the model is evaluated at, and then interpolates between them. Therefore, we disable this to test if the model is linear with the number of frequencies at which it is evaluated. However, for the likelihood test, we compute the speedups both without and with multibanding enabled, to explore real-world speedup gains.

In figure~\ref{fig:Speedup} we differentiate the speedups using triangles for the waveform, squares for the log-likelihood and in the \texttt{IMRPhenomXPHM} case, circles for the log-likelihood with multibanding enabled. We can also compare with the theoretical speed-ups that are plotted as histogram-like bars in the figure. There are two kinds of bars, the solid ones represent the quantity 
\begin{equation}
    \mathrm{Theoretical~Speedup} = \frac{L}{N_L+N_Q}
    \label{eq:TheoreticalSpeedup}
\end{equation}
where $L$ is the number of frequencies for the waveform evaluation in the standard computation and $N_L$ and $N_Q$ are the frequency nodes for the linear and quadratic ROQ bases without factoring out repeated frequencies. This is the theoretical speedup that is usually attributed to the ROQ in the literature~\cite{Smith:2016qas}. The dashed bars are the same quantity as in Eq.~\eqref{eq:TheoreticalSpeedup} when the frequencies belonging to both the linear and the quadratic interpolation list of frequency nodes are just considered once, thus the notation $L/N_{L\cup Q}$. In the ROQ likelihood we need to call the waveform model only once at the frequencies defined by $\{f_i \}_{i=1}^{N_{L\cup Q}} = \{F_j\}_{j=1}^{N_\mathrm{L}} \cup \{\mathcal{F}_k\}_{k=1}^{N_\mathrm{Q}}$, as is done in \texttt{Bilby}. Therefore, $L/N_{L\cup Q}$ will be the theoretical speedup of the waveform evaluation if we assume that its computation time is proportional to the number of sampling points. For the \texttt{IMRPhenomXPHM} case, the difference between $N_L+N_Q$ and $N_{L\cup Q}$ can be significant since there are many repeated interpolation nodes at low frequencies. The reason is that in the low-frequency region, the amplitude is larger and the waveform oscillates more rapidly than in the high-frequency part. Consequently, the interpolation nodes tend to concentrate at low frequencies leading some of them to coincide in the linear and quadratic ROQ. This behavior can be seen in figure~\ref{fig:waveform_interp_panel}. 

For $\mathcal{M}$ smaller than $\sim20 M_\odot$, we see that the waveform speedups are constant in the entire $\mathcal{M}$ range of a given basis and are always close to the theoretical value of $L/N_{L\cup Q}$. This is in agreement with our expectations, since the \texttt{IMRPhenom} models describe the inspiral in a way that the computation time is linear with the number of sampling points, and their implementation in \texttt{LALsimulation}~\cite{lalsuite} being tested is written efficiently in \texttt{C}~\cite{Performance}, with minimal overheads. In the case of large $\mathcal{M}$, above $\sim20 M_\odot$, the waveforms start being dominated by the merger and ringdown, the last two phases of a CBC, which are harder to model, and the speed-up of the \texttt{IMRPhenom} models is smaller than the theoretical expectation. This can be due to the waveform generation stopping above the ringdown frequency, meaning that the model is evaluated at fewer frequency points for high mass signals. Furthermore, when the waveform uses sufficiently few frequency points, fixed-costs associated with calculating post-Newtonian and phenomenological parameters of the model become important. Therefore, as $\mathcal{M}$ increases, the trend of the waveform speedup is to decrease until a value of $\mathcal{O}(1)$ is reached and we have no speed up at all.

For the \texttt{IMRPhenomXPHM} likelihood speedups, we show both the results with and without disabling the default multibanding~\cite{Garcia-Quiros:2020qlt}, which is used in the standard likelihood to speed up the full waveform computation. We observe that without multibanding \texttt{IMRPhenomXPHM} has a likelihood speedup very close to the theoretical value. This is due to the fact that the computation time of the likelihood is dominated by the waveform evaluation, and the \texttt{Bilby} implementation of the ROQ likelihood only generates the waveform once at the frequencies $\{f_i \}_{i=1}^{N_{L\cup Q}} = \{F_j\}_{j=1}^{N_\mathrm{L}} \cup \{\mathcal{F}_k\}_{k=1}^{N_\mathrm{Q}}$. However, when one includes the multibanding option, the \texttt{IMRPhenomXPHM} is already internally being evaluated in fewer frequency points, and therefore the speedup can be significantly lower than the expected one, although it still reaches median values that can be as large as 5, and which will be noticeable in PE applications. Looking at the targeted bases that are introduced in Table~\ref{table:TargetedBasis}, we observe that in this case, the speedup over the standard multibanded case can be even larger, reaching a value of $29.2^{+1.4}_{-4.6}$ for the base targeted at GW170817~\cite{GW170817}.

In the \texttt{IMRPhenomPv2} case, we observe that the likelihood speedup is significantly below the waveform speedup and therefore, also below the theoretical speedup. To understand this discrepancy, we note that for the standard likelihood case, the computation time is dominated by evaluating the waveform in all the required frequencies and computing the overlap integrals of Eq.~\eqref{eq:Likelihood_def}, both of which will be proportional to $L$. However, for the ROQ likelihood, the time to compute the waveform and overlap integrals is significantly reduced since they are proportional to $N_\mathrm{L} + N_\mathrm{Q} \ll L$. Given the fact that \texttt{IMRPhenomPv2} is much faster to generate than \texttt{IMRPhenomXPHM}, the computation time starts to be dominated by fixed-cost operations, which for example include computing the parameters of the waveform models, finding the detector responses as well as possible overheads.

\begin{figure}
    \centering
    \includegraphics[width=0.5\textwidth]{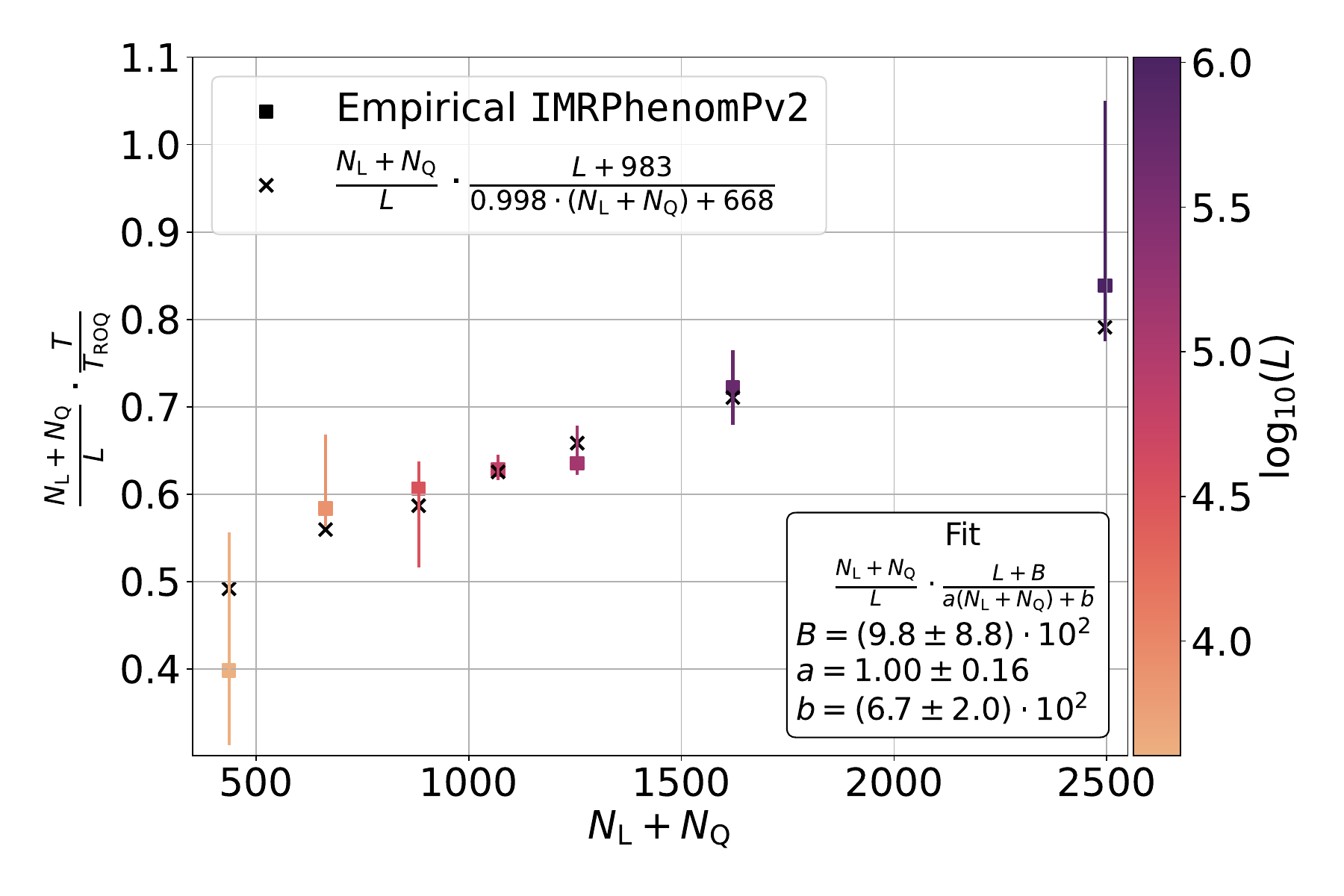}
    \caption{Ratio between empirical speedup and the theoretical speedup of Eq.~\eqref{eq:TheoreticalSpeedup}, plotted as a function of the total elements of the ROQ basis ($N_L+N_Q$) for \texttt{IMRPhenomPv2}. The colour of the error bars encodes the logarithm of the number of frequencies where the waveform is evaluated in the standard computation $\log\mathcal{L}$. In the bottom right box, we show the functional form we fit, which comes from Eq.~\eqref{eq:Speedup_model}, as well as the $1\sigma$ uncertainty for the three fitted parameters.  We also plot as black crosses the results obtained evaluating the best fit in the data points.}
    \label{fig:Speedup_regression}
\end{figure}

To further explore this hypothesis, we model the computation time of the likelihood as a coefficient multiplying the number of frequencies being evaluated plus a constant term which represents the fixed-cost operations. Since for \texttt{IMRPhenomPv2}, $N_{L\cup Q} \sim N_L+N_Q$, we have,
\begin{align}
    T&= A\cdot L + B \label{eq:Tnormal}\\
    T_{ROQ} &= a \cdot (N_\mathrm{L} + N_\mathrm{Q}) + b \label{eq:TROQ} \, .   
\end{align}
To compute the speedup, we divide Eq.~\eqref{eq:Tnormal} by Eq.~\eqref{eq:TROQ}, obtaining
\begin{equation}
    f(L,N_L,N_Q;B,a,b)=\frac{L+B}{a(N_\mathrm{L} + N_\mathrm{Q})+b} \, ,
    \label{eq:Speedup_model}
\end{equation}
\noindent where we have divided all the coefficients by $A$, which is not expected to be 0. In figure~\ref{fig:Speedup_regression} we show the ratio between the empirical and theoretical likelihood speedups, together with the best fit of our model in Eq.~\eqref{eq:Speedup_model}. We observe very good agreement between the model and the data. From the fitted values of $B$, $a$ and $b$, also displayed in the plot, we can substantiate our hypothesis that the fixed-cost operations in the ROQ likelihood is making the empirical speedup of the \texttt{IMRPhenomPv2} smaller than the theoretical value. We find a value of $a = 1.00 \pm 0.16$, and therefore, from Eq.~\eqref{eq:Speedup_model}, we observe that if the coefficients $B$ and $b$ describing the fixed-costs were 0, we would recover the theoretical speedup result. However, since we find a value of $b = (6.7 \pm 2.0)\cdot 10^{2}$, the \texttt{IMRPhenomPv2} speedup is reduced with respect to the theoretical unless we have a very large number of basis elements such that $a\cdot(N_\mathrm{L} + N_\mathrm{Q}) \gg b$. 

\begin{figure}
    \centering
    \includegraphics[width=0.5\textwidth]{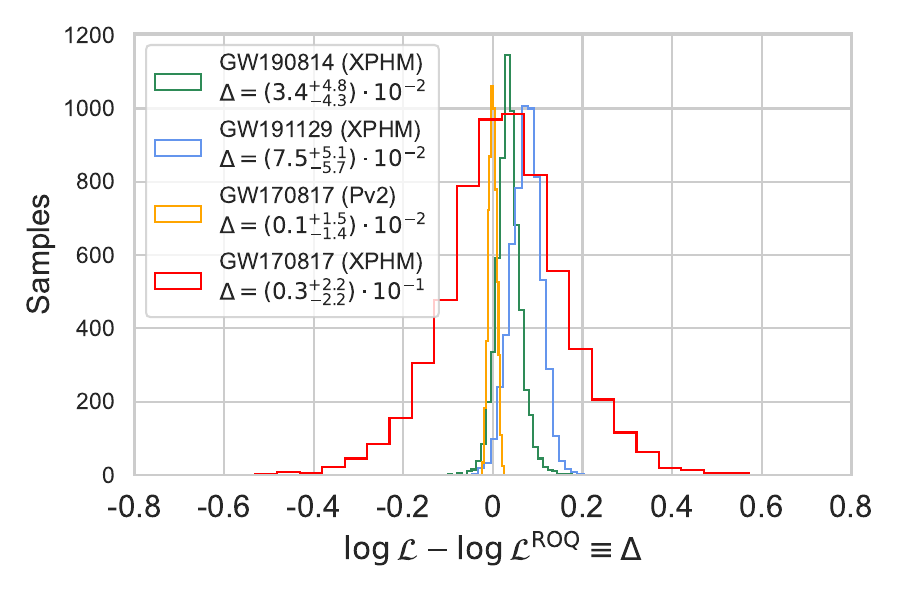}
    \caption{Difference between the logarithm of the standard Likelihood and the logarithm of the ROQ Likelihood for the three events analysed. }
    \label{fig:PE_Likelihood_Comparison}
\end{figure}

\subsection{Aplication to GW events}
\label{sec:CodeValidation:AplicationtoGWevents}

We now perform four PE analyses~\cite{Thrane_2019} on three confirmed GW events using the ROQ approximation. More specifically, we use the \texttt{IMRPhenomXPHM} 16s basis described in table~\ref{table:XPHMBasis} for the GW191129\_134029~\cite{GWTC-3} event and the \texttt{IMRPhenomPv2} 256s basis of table~\ref{table:Pv2Basis} for the GW170817~\cite{GW170817} event. For the other two PE analyses of GW190814~\cite{GW190814} and GW170817 with \texttt{IMRPhenomXPHM}, in a similar spirit to Refs~\cite{Morisaki:2020oqk,Morisaki:2023kuq}, we construct targeted ROQ bases with narrow $\mathcal{M}$ ranges, listed in Table~\ref{table:TargetedBasis}. These bases are centered on the search $\mathcal{M}$ value and have a narrow width tuned to be larger than the expected chirp mass resolution. Note that the bases have been generated using a factor of 10 times fewer waveforms than that of Tables~\ref{table:Pv2Basis}~\ref{table:XPHMBasis}, since the parameter space they cover is smaller.

The analyses use the ROQ likelihood and the \texttt{dynesty}~\cite{speagle2020dynesty} sampler within version 2.1.0 of \texttt{Bilby}~\cite{Ashton:2018jfp} and the version 5.1.0. of \texttt{LALSimulation}. The PSDs employed were estimated using \texttt{BayesWave}~\cite{Cornish_2015,PhysRevD.91.084034} and are those used by the LVK collaboration for the public analysis of the events. We also include the effects of calibration uncertainties~\cite{Cahillane_2017,Acernese_2022,Sun_2020} in the phase and the amplitude.

\begin{table*}
    \centering
    \input{TargetedBasisTable}
        \caption{Focused \texttt{IMRPhenomXPHM} bases for GW190814 ($\Delta f = 1/16 \mathrm{Hz}$) and GW170817 ($\Delta f = 1/256 \mathrm{Hz}$). We limit the magnitudes of the two spins $-0.8\leq \chi_i\leq0.8$ for $i \in [1,2]$, and the full range for the spin angles $(0,0) \leq (\theta_J , \alpha_0) \leq (\pi, 2\pi)$. For the GW190814 we limit the mass ratio $q \leq 16$ while for GW170817 we limit it $q \leq 4$. For the creation of the two basis, we run \texttt{EigROQ} with the same configuration. In algorithm~\ref{alg:create_ROB} we set the maximum number of waveform selected $N=20000$ , tolerances of each step $\sigma_i = [10^{-2}, 10^{-3}, 10^{-4}]$ and maximum number of waveforms computed per step $N_{\mathrm{lim}, i} = [10^{5}, 3.16 \cdot 10^5]$, and in algorithm~\ref{alg:eig_EIM} we set $N=10^6$, $\sigma=10^{-4}$, $N_\mathrm{lim}=10^7$ and the maximum number of eigenvectors used $n_\lambda=5000$. The ``Theoretical'' speedup has been computed with Eq.~\eqref{eq:TheoreticalSpeedup} while the ``Empirical'' speedup is the median and 90$\%$ credible interval of the corresponding points in the lower panel of Figure~\ref{fig:Speedup}. For the empirical speedups, we show the values both without (Emp.) and with (MB) the \texttt{IMRPhenomXPHM} multibanding option enabled~\cite{Garcia-Quiros:2020qlt}.}
    \label{table:TargetedBasis}
\end{table*}

The first event we discuss is GW191129\_134029~\cite{GWTC-3,LIGOScientific:2023vdi}. This is an event with $\mathcal{M}^{\rm detector}=8.48_{-0.05}^{+0.06}M_\odot$ so we can use the 16 seconds \texttt{IMRPhenomXPHM} ROQ basis. It has a relatively big median network SNR of $13.1$, allowing us to put tight constraints on the parameters and better see if any differences arise between the ROQ and the standard posterior. We perform two \texttt{Bilby} runs with the exact same configuration, one using the standard GW likelihood and the other using the ROQ likelihood.

\begin{figure}
    \centering
    \includegraphics[width=0.5\textwidth]{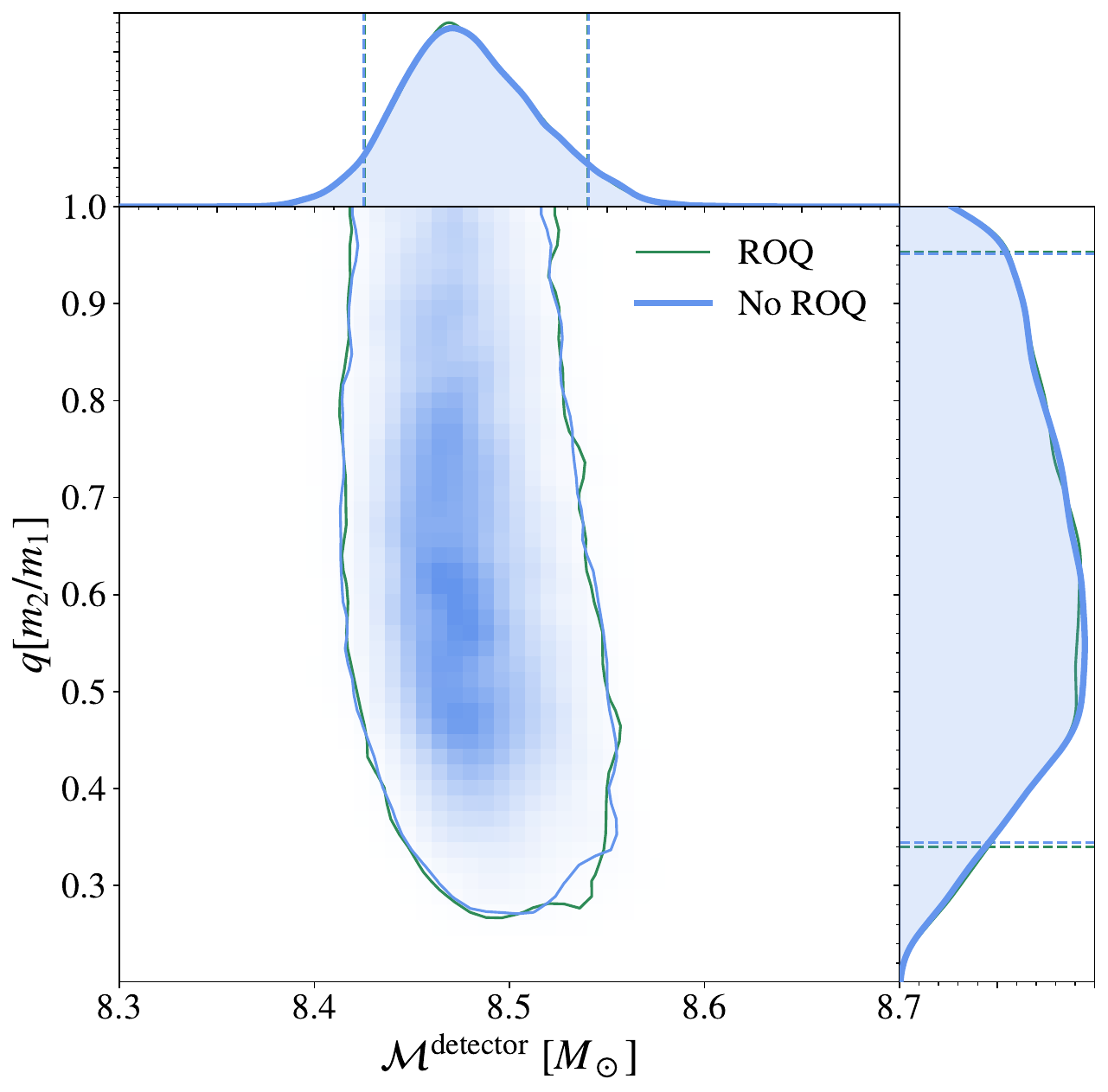}
    \caption{Posterior distributions for the mass ratio and $\mathcal{M}$ in the detector frame for the ROQ and non-ROQ analysis of GW191129\_134029. The 90$\%$ credible regions are indicated by the solid contour in the joint distribution, and by the dashed vertical and horizontal lines in the marginalized distributions.}
    \label{fig:MC_Q_Cornerplot}
\end{figure}

In figure~\ref{fig:PE_Likelihood_Comparison} we show the difference between the logarithm of the standard and the ROQ likelihoods, for the posterior samples of the PE with the ROQ likelihood. This difference corresponds to the ROQ error in the log-likelihood. We find a 90\% c.l. error of $\Delta\log\mathcal{L} = 0.075_{-0.057}^{+0.051}$. Since $\Delta\log\mathcal{L} \ll 1$, we expect the posteriors with and without the ROQ to be almost the same. Using that the log likelihood of this event is $\log\mathcal{L}=84.2^{+2.9}_{-4.1}$, the fractional error in the ROQ log-likelihood computation is $\delta_\mathcal{L} =(9.1^{+6.3}_{-6.8}) \cdot 10^{-4}$.\footnote{We define the  fractional error in the ROQ log-likelihood computation as $\delta_\mathcal{L} = \Delta\log\mathcal{L}/\log\mathcal{L}$} The distribution of errors is centered at a positive value, as one would expect if the waveform model were a good representation of reality since any error in the ROQ modelization of the waveform would push it away from the true GW and thus, to a lower likelihood value. 

In figure~\ref{fig:MC_Q_Cornerplot} we corroborate that indeed the posteriors with and without the ROQ are similar by showing the corresponding distributions for the detector frame chirp mass $\mathcal{M}$ and the mass-ratio $q$. We find a Jensen-Shannon Divergence (JSD)~\cite{JSDivergence} of $1.3\cdot10^{-4}$ and $1.9\cdot10^{-4}$ respectively, robustly assessing the similarity between the distributions with and without the ROQ approximation.

The second event we analyze is GW190814~\cite{GW190814,LIGOScientific:2023vdi}. This event was measured to have a chirp mass of $\mathcal{M}=6.42^{-0.02}_{0.02}M_\odot$ and a very unequal mass ratio of $0.11^{-0.01}_{0.01}$, which is below the mass ratios of $q>0.25$ explored in the bases of Table~\ref{table:XPHMBasis}. Therefore we create a targeted ROQ base with 16 seconds of duration, and chirp mass range from $6.2M_\odot$ to $6.6M_\odot$ for the \texttt{IMRPhenomXPHM} waveform. In figure~\ref{fig:PE_Likelihood_Comparison} we show the ROQ log-likelihood errors of the posterior samples of the PE performed using this targeted basis. We have that $\Delta\log\mathcal{L} = 0.034_{-0.043}^{+0.048}$ which is similar in magnitude to that of GW191129\_134029.  Again, since $\Delta\log\mathcal{L} \ll 1$, we expect the posteriors with and without the ROQ to be almost the same. However, for this event, the log-likelihood is larger, at $\log\mathcal{L}=310.3e+02^{+3.1}_{-5.0}$, and therefore the relative error in the ROQ log-likelihood computation is smaller, at $\delta_{\mathcal{L}}=(1.1^{+1.5}_{-1.4}) \cdot 10^{-4}$.

\begin{figure}
    \centering
    \includegraphics[width=0.5\textwidth]{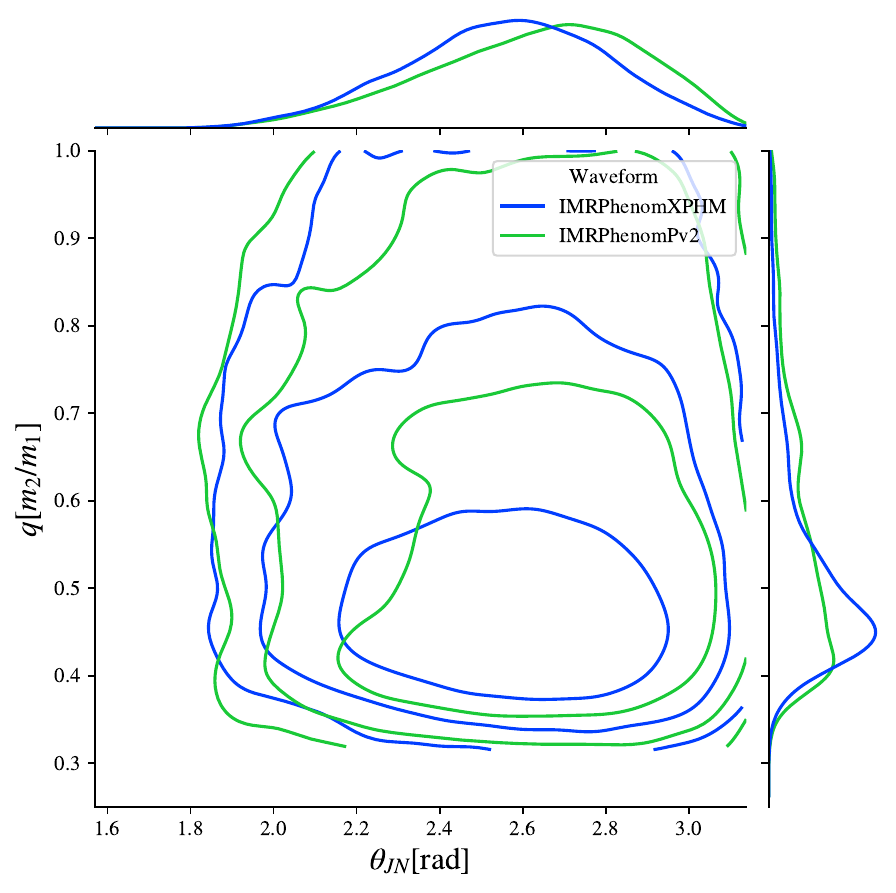}
    \caption{Posterior distributions for the mass ratio $q$ and the inclination angle $\theta_{JN}$ for the ROQ analysis of GW170817. In blue we plot the \texttt{IMRPhenomXPHM} run and in green \texttt{IMRPhenomPv2}. Three contours per run delimit the 1$\sigma$ (68.3\% C.L.), 2$\sigma$ (95.4\% C.L.) and 3$\sigma$ (99.7\% C.L.) credible regions in the joint $q-\theta_{JN}$ distribution. Note that the non-continuous behaviour of the contours near the border is an artefact of the Gaussian kernel employed in the drawing. This is expected whenever the parameter is bounded and presents many samples close to the border.}
    \label{fig:q_iota_corner}
\end{figure}

The last GW event we analyze is GW170817~\cite{GW170817}, the event with the largest Network SNR ($\sim 33$) ever detected. It was identified as a binary neutron star with $\mathcal{M}=1.1976^{+0.0004}_{-0.0002}$~\cite{LIGOScientific:2018hze} and we use it to probe the longest of our \texttt{IMRPhenomPv2} bases with 256s in duration as well as a targeted ROQ using \texttt{IMRPhenomXPHM} for such long signals. For our analysis, we make use of the public strain data after noise subtraction~\cite{GW170817_data}. In figure~\ref{fig:PE_Likelihood_Comparison} we show the ROQ log-likelihood errors of the posterior samples of both PEs. For both cases, we do not expect the ROQ error to significantly impact the posterior, since $\Delta\log\mathcal{L} \ll 2.3$. The \texttt{IMRPhenomPv2} PE has an order of magnitude smaller ROQ error than the \texttt{IMRPhenomXPHM} case. This is most likely the result of the \texttt{IMRPhenomPv2} basis being constructed with a tolerance $\sigma=10^{-5}$, which is an order of magnitude smaller than the tolerance $\sigma=10^{-4}$ used in the \texttt{IMRPhenomXPHM} case. In the \texttt{IMRPhenomPv2} case, the log-likelihood is $536.1^{+3.2}_{-4.3}$ and the corresponding fractional error is $\delta_{\mathcal{L}}=(-0.1^{+2.8}_{-2.6}) \cdot 10^{-5}$. In the \texttt{IMRPhenomXPHM} case, we find a larger likelihood of $538.1^{+4.3}_{-5.1}$, which is expected since the higher order modes give more freedom to the waveform to fit the data. The corresponding fractional error is $\delta_{\mathcal{L}}=(0.5^{+4.1}_{-4.0}) \cdot 10^{-4}$. Comparing the Bayes Factors of both PE runs, adjusted to have the same priors, we find $\log\mathcal{B} = 1.1 \pm 0.3$ in favour of \texttt{IMRPhenomXPHM}, which can be taken as evidence for Higher Order Modes in the signal. This highlights the importance of considering all physical effects of the waveform. To further make this point, we show in figure~\ref{fig:q_iota_corner} how the addition of the Higher Order Modes improves the determination of the mass ratio and the inclination angle $\theta_{JN}$, even for this low mass CBC for which the Higher Order Modes are harder to measure in LIGO-Virgo~\cite{Mills:2020thr}. 

\section{Conclusions}
\label{sec:Conclusions}

In this paper, we have explored in-depth Reduced Order Quadrature (ROQ) methods applied to GW data analysis and have presented novel algorithms to improve different aspects of the ROQ bases construction. ROQ methods offer a significant advantage by reducing the computational burden associated with likelihood evaluations, especially for long-duration waveforms, and therefore can greatly speed up parameter estimation analyses. Existing procedures for constructing ROQ bases encounter challenges in approximating waveforms that include complicated features such as precession or Higher Order Modes. We present algorithms to address these limitations by making use of SVD methods to characterize the waveform space and choose a reduced order basis close to optimal. We also propose improved methods to select the empirical interpolation nodes, greatly reducing the error induced by the empirical interpolation model.

We have demonstrated the effectiveness of our algorithm by constructing multiple ROQ bases for the \texttt{IMRPhenomPv2} and \texttt{IMRPhenomXPHM} waveforms, ranging in duration from 4s to 256s. These bases have been subjected to various tests, including likelihood error tests and P-P tests, to validate their accuracy and trustworthiness for data analysis applications. The speedup of these bases has also been empirically explored, confirming that ROQ methods provide close to the expected reduction in computational time compared to traditional likelihood calculations.

Furthermore, we have performed end-to-end parameter estimation analyses on several confirmed GW events. The results provide compelling evidence of the algorithm's ability to generate ROQ bases that accurately represent complex waveform models over both broad and targeted parameter spaces. By directly comparing the posterior distributions using the ROQ and standard methods and understanding the log-likelihood error distributions, we validate that our bases can straightforwardly be incorporated into current pipelines to produce precise and unbiased Parameter Estimations in real gravitational wave detector data.

In conclusion, the algorithms introduced in this paper represent a step forward in the quest to efficiently exploit the capabilities of advanced gravitational wave detectors. We improve upon previous ROQ construction algorithms allowing for more efficient bases in regions of parameter space that were previously inaccessible.  As gravitational wave astronomy continues to evolve, and the number of events detected per year continues to grow, having fast and accurate techniques to perform Parameter Estimation will undoubtedly play a vital role in maximizing the scientific potential of future observatories and advancing our knowledge of the Universe.

\section*{Acknowledgements}

We would like to thank every member of the LVK Collaboration who has taken the time to analyze and discuss the results presented here, as it has helped us understand them better. In particular, we would like to extend special thanks to Colm Talbot, Geraint Pratten, Cecilio García-Quiros and Soichiro Morisaki for their exceptional comments and feedback. Their input has played a role in shaping and improving the ideas and content of this paper. We would also like to thank Colm Talbot and Jose Antonio Font for their work reviewing this paper within the LIGO and Virgo Collaborations respectively.
The authors acknowledge the use of the publicly available codes: \texttt{lalsuite} \cite{lalsuite_code} and \texttt{Bilby} \cite{Ashton:2018jfp}. 
They acknowledge support from the research project  PID2021-123012NB-C43 and the Spanish Research Agency (Agencia Estatal de Investigaci\'on) through the Grant IFT Centro de Excelencia Severo Ochoa No CEX2020-001007-S, funded by MCIN/AEI/10.13039/501100011033. 
GM acknowledges support from the Ministerio de Universidades through Grant No. FPU20/02857 
and JFNS acknowledges support from MCIN through Grant No. PRE2020-092571.
The authors acknowledge use of the Hydra cluster at the IFT, on which some of the numerical computations for this paper took place.
We acknowledge the support of the Supercomputing Wales project, which is part-funded by the European Regional Development Fund (ERDF) via Welsh Government.
This research has made use of data or software obtained from the Gravitational Wave Open Science Center \cite{LIGOScientific:2019lzm, LIGOScientific:2023vdi} (gw-openscience.org), a service of LIGO Laboratory, the LIGO Scientific Collaboration, the Virgo Collaboration, and KAGRA. LIGO Laboratory and Advanced LIGO are funded by the United States National Science Foundation (NSF) as well as the Science and Technology Facilities Council (STFC) of the United Kingdom, the Max-Planck-Society (MPS), and the State of Niedersachsen/Germany for support of the construction of Advanced LIGO and construction and operation of the GEO600 detector. Additional support for Advanced LIGO was provided by the Australian Research Council. Virgo is funded, through the European Gravitational Observatory (EGO), by the French Centre National de Recherche Scientifique (CNRS), the Italian Istituto Nazionale di Fisica Nucleare (INFN) and the Dutch Nikhef, with contributions by institutions from Belgium, Germany, Greece, Hungary, Ireland, Japan, Monaco, Poland, Portugal, Spain. The construction and operation of KAGRA are funded by Ministry of Education, Culture, Sports, Science and Technology (MEXT), and Japan Society for the Promotion of Science (JSPS), National Research Foundation (NRF) and Ministry of Science and ICT (MSIT) in Korea, Academia Sinica (AS) and the Ministry of Science and Technology (MoST) in Taiwan.

\appendix

\section{Fast way to update $\Vert \hat{A}^{-1}\Vert_F$ and $\sigma_\mathrm{EIM}^\mathrm{tot}$}
\label{sec:anex:fast_FrobinvA}

In this section we assume that we have the inverse of the matrix $\Bar{A}_{i j} = e_j (X_i)$ and its Frobenius norm $\Vert \hat{\Bar{A}}^{-1} \Vert_F$, defined in Eq.~\eqref{eq:F_norm}, and we want to compute the inverse and Frobenius norm of the inverse of the matrix $A_{i j}$, defined as:

\begin{equation}
    A_{i j} = 
     \begin{cases}
       e_j (x_{\beta_i}) &\; i \neq k\\
       e_j (x_{\beta'_k}) &\; i = k \\
     \end{cases}
     \label{eq:Aij_row_change}
\end{equation}  

\noindent which is just the result of changing the row $k$ of $\Bar{A}_{i j}$. We then use the fact that, from the properties of the inverse $\Bar{A}_{i j}$, we have:

\begin{equation}
    (\hat{A} \hat{\Bar{A}}^{-1})_{i j} = 
     \begin{cases}
      \delta_{i j} &\; i \neq k\\
      \sum_{l=1}^n e_l(x_{\beta'_k}) (\hat{\Bar{A}}^{-1})_{l j} \equiv c_j  &\; i = k \\
     \end{cases}
     \label{eq:AinvA}
\end{equation}

Since the matrix of Eq.~\eqref{eq:AinvA} has such a simple structure, it can be analytically inverted as:

\begin{equation}
    ((\hat{A} \hat{\Bar{A}}^{-1})^{-1})_{i j} = 
     \begin{cases}
       \delta_{i j} &\; i \neq k\\
       - \frac{c_j}{c_k} &\; i = k, \, j \neq k  \\
       \frac{1}{c_k} &\; i = j = k  \\
     \end{cases}
    \label{eq:inv_AinvA}
\end{equation}

And we can use that $\hat{A}^{-1} = \hat{\Bar{A}}^{-1} (\hat{A} \hat{\Bar{A}}^{-1})^{-1}$ to show that:

\begin{equation}
    (\hat{A}^{-1})_{i j} = 
     \begin{cases}
       (\hat{\Bar{A}}^{-1})_{i j} - \frac{c_j}{c_k} (\hat{\Bar{A}}^{-1})_{i k}  &\; j \neq k\\
       \frac{1}{c_k} (\hat{\Bar{A}}^{-1})_{i k}  &\; j = k \\
     \end{cases}
    \label{eq:invA_fast}
\end{equation}

\noindent We observe that this way of computing the inverse will require $O(n^2)$ for computing $c_j$ with Eq.~\eqref{eq:AinvA} and also $O(n^2)$ operations to update each element of $\hat{A}^{-1}$ using Eq.~\eqref{eq:invA_fast}. So the total number of operations will be $O(n^2)$, much smaller than the $O(n^3)$ required to directly invert the matrix.

Using this expression for the updated inverse we can find a way to update also the Frobenius norm of the inverse, which is given by:

\begin{align}
    \Vert \hat{A}^{-1} \Vert_F^2 & = \sum_{i,j =1}^{n} | (\hat{A}^{-1})_{i j}|^2 \nonumber \\ 
    & = \sum_{i,j =1}^{n} \left| (\hat{\Bar{A}}^{-1})_{i j} - \frac{c_j}{c_k} (\hat{\Bar{A}}^{-1})_{i k} \right|^2 + \sum_{i=1}^n \left| \frac{1}{c_k} (\hat{\Bar{A}}^{-1})_{i k} \right|^2 \nonumber \\
    & = \Vert \hat{\Bar{A}}^{-1} \Vert_F^2 + \frac{1}{|c_k|^2} \left( 1 + \sum_{j=1}^n |c_j|^2 \right) \left[ \sum_{i=1}^n |(\hat{\Bar{A}}^{-1})_{i k}|^2 \right] \nonumber \\
    & \quad - 2 \mathrm{Re}\left( \frac{1}{c_k} \sum_{j=1}^{n}\left[\sum_{i=1}^{n}  (\hat{\Bar{A}}^{-1})_{i k} (\hat{\Bar{A}}^{-1})^{*}_{i j}\right] c_j \right) \, ,
    \label{eq:FrobinvA_fast}
\end{align}

\noindent where we can precompute whith $O(n^2)$ operations the factors in square brackets that only depend on $\hat{\Bar{A}}$ for each row $k$ which we will change, and afterwards, updating the Frobenius norm will only need $O(n)$ operations on top of the $O(n^2)$ operations needed to compute $c_j$ for each new row $q$ we want to test. Since with Eq.~\eqref{eq:FrobinvA_fast} we do not need to update the inverse each time that we want to update its Frobenius norm, we can avoid the $O(n^2)$ memory allocations that are needed in Eq.~\eqref{eq:invA_fast}.

We will now also look for a method to rapidly compute $\sigma_\mathrm{EIM}^\mathrm{tot}$. We assume that we have the value computed for an EIM whose variables we denote with a bar over them:

\begin{equation}
    \Bar{\sigma}_\mathrm{EIM}^\mathrm{tot} = \sum_{B=1}^{n_\lambda} \left(  \lambda_B + \left\Vert  \hat{\Bar{A}}^{-1} \vec{\Bar{v}}_B \right\Vert^2 \right)\, ,
    \label{eq:sigmaEIM_ROB_sum_old}
\end{equation}

\noindent where we have defined

\begin{equation}
    \Bar{v}_{B, i} = \sqrt{\lambda_B} u_{B, \beta_i} \, .
    \label{eq:v_sigmaEIM_ROB_sum_anex_def}
\end{equation}

When we change the $k$'th interpolation node of the EIM from $\beta_k$ to $\beta'_k$, this becomes:

\begin{equation}
    v_{B, i} =
     \begin{cases}
       \sqrt{\lambda_B} u_{B, \beta_i} &\; i \neq k\\
       \sqrt{\lambda_B} u_{B, \beta'_k} &\; i = k \\
     \end{cases}
    \label{eq:v_sigmaEIM_ROB_sum_anex_def_change}
\end{equation}

And the value of multiplying $\hat{A}$ by $\vec{v}_B$ will change to:

\begin{align}
     (\hat{A}^{-1} \vec{v}_B)_i & = \sum_{j=1}^n (\hat{A}^{-1})_{i j} v_{B, j}  \nonumber \\
     & = \sum_{j=1}^n \left(\hat{\Bar{A}}^{-1}_{i j}  - \frac{c_j}{c_k} \hat{\Bar{A}}^{-1}_{i k} \right) v_{B, j} + \frac{1}{c_k} \hat{\Bar{A}}^{-1}_{i k} v_{B, k} \nonumber \\
     & =\underbrace{\sum_{j=1}^n \hat{\Bar{A}}^{-1}_{i j} \Bar{v}_{B, j}}_{\Bar{\Omega}_{B, i}}  + \underbrace{\left[\frac{1}{c_k} \left(v_{B, k} - \sum_{i=1}^n c_j \Bar{v}_{B, j}  \right) \right]}_{\Theta_B} \underbrace{\hat{\Bar{A}}^{-1}_{i k}}_{\Bar{\Gamma}_i} \, ,
    \label{eq:invA_vB} 
\end{align}

\noindent where we have used the updated value of $\hat{A}^{-1}$ computed in Eq.~\eqref{eq:invA_fast} and we put bar over the variables that do not depend on the value of the new interpolation node. Using Eq.~\eqref{eq:invA_vB}, $\Bar{\sigma}_\mathrm{EIM}^\mathrm{tot}$ becomes:

\begin{align}
    \sigma_\mathrm{EIM}^\mathrm{tot} & = \sum_{B=1}^{n_\lambda} \left( \lambda_B + \sum_{i=1}^n \left| \Bar{\Omega}_{B, i} + \Theta_B  \Bar{\Gamma}_i \right|^2 \right) \nonumber \\
    & = \Bar{\sigma}_\mathrm{EIM}^\mathrm{tot} +  \left(\sum_{B=1}^{n_\lambda} |\Theta_B|^2\right)   \left[\sum_{i=1}^n |\Bar{\Gamma}_i|^2 \right] \nonumber \\
    & \quad + 2 \mathrm{Re}\left\{ \sum_{B=1}^{n_\lambda}\left( \Theta_B\left[\sum_{i=1}^n \Bar{\Omega}^{*}_{B,i} \Bar{\Gamma}_i \right]\right) \right\} \, .
    \label{eq:sigmaEIM_ROB_sum_new} 
\end{align}

In general we will assume that $n_\lambda \gg n$. For each row $k$ that we change, we can precompute with $O(n n_\lambda)$operations all the factors in square brackets that will stay constant. Afterwards, the computational complexity of updating the value of $\sigma_\mathrm{EIM}^\mathrm{tot}$ will require $O(n n_\lambda)$ operations for computing $\Theta_B$ and only $O(n_\lambda)$ additional operations to evaluate Eq.~\eqref{eq:sigmaEIM_ROB_sum_new}.

\bibliography{Refs}

\end{document}

%% file: Pv2BasisTable.tex
\begin{tabular}{ c | c | c | c | c | c | c} 
\hline
\hline
\rule{0pt}{22pt} \shortstack{Freq. range (Hz) \\ Min\ \  Max}  
&  $\Delta{f} (\mathrm{Hz})$ &  \shortstack{$M_c (M_\odot)$ \\ Min\ \ Max}  & \shortstack{Basis size \\ Linear\ \ Quadratic} &  \shortstack{Test set $\sigma_{\text{EI,max}}$\\Linear Quadratic} &  \shortstack{Test set $\sigma_{\text{EI}}>10^{-5}$\\Linear\ \ Quadratic} & \shortstack{Likelihood Speedup \\ Theoretical \  Empirical}\\
\hline
\rule{0pt}{12pt}$20\ \ \ \ 1024$ & $1/4$ & $12.3\ \ 45$ & 242\ \ \ \ 194 &\sqz$1.00\times 10^{-3}$\ \ \ $1.09\times 10^{-4}$ &$31$\ \ \ \ $19$ & 9.2 \hspace{20pt} $3.7^{+1.0}_{-0.6}$ \\
\hline
\rule{0pt}{12pt}$20\ \ \ \ 1024$ & $1/8$ & $7.93 \ \ 14.76$ & 369\ \ \ \ 294 &\sqz$4.91\times 10^{-4}$\ \ \ $1.46\times 10^{-4}$ &$55$\ \ \ \ $31$ & 12.1 \hspace{20pt} $7.1^{+0.7}_{-0.1}$ \\
\hline
\rule{0pt}{12pt}$\ 20 \ \ \ \ 2048$ & $1/16$ & $5.14\ \ 9.52$ & 493\ \ \ \ 389 &\sqz$6.85\times 10^{-4}$\ \ \ $5.72\times 10^{-4}$ &$110$\ \ \ \ $59$ & 36.8 \hspace{20pt} $22.3^{+0.6}_{-1.4}$ \\
\hline
\rule{0pt}{12pt}$\ 20 \ \ \ \ 2048$ & $1/32$ & $3.35\ \ 6.17$ & 631\ \ \ \ 438 &\sqz$6.88\times 10^{-4}$\ \ \ $5.83\times 10^{-4}$ &$98$\ \ \ \ $75$ & 60.7 \hspace{15pt} $38.1^{+0.5}_{-0.4}$ \\
\hline
\rule{0pt}{12pt}$\ 20 \ \ \ \ 2048$ & $1/64$ & $2.18\ \ 4.02$ & 848\ \ \ \ 407 &\sqz$1.51\times 10^{-3}$\ \ \ $5.71\times 10^{-4}$ &$103$\ \ \ \ $71$ & 103.4 \hspace{15pt} $65.7^{+1.6}_{-0.9}$ \\
\hline
\rule{0pt}{12pt}$\ 20 \ \ \ \ 4096$ & $1/128$ & $1.42\ \ 2.60$ & 1315\ \ \ \ 306 &\sqz$6.4\times 10^{-4}$\ \ \ $2.46\times 10^{-3}$ &$83$\ \ \ \ $50$ & 321.9 \hspace{8pt} $232.3^{+8.0}_{-7.0}$ \\
\hline
\rule{0pt}{12pt}$\ 20 \ \ \ \ 4096$ & $1/256$ & $0.95\ \ 1.72$ & 2196\ \ \ \ 300 &\sqz$1.43\times 10^{-4}$\ \ \ $6.32\times 10^{-5}$ &$69$\ \ \ \ $28$ & 418.1 \hspace{8pt} $350.7^{+49.8}_{-17.8}$ \\
\hline
\hline
\end{tabular}

%% file: XPHMBasisTable.tex
\begin{tabular}{ c | c | c | c | c | c| c} 
\hline
\hline
\rule{0pt}{22pt} \shortstack{Frequency \\ range (Hz) \\ Min\ \  Max}  
&  $\Delta{f} (\mathrm{Hz})$ &  \shortstack{$M_c (M_\odot)$ \\ Min\ \ Max}  & \shortstack{Basis size \\ Linear\ \ Quadratic} &  \shortstack{Test set $\sigma_{\text{EI,max}}$ \\Linear Quadratic} &  \shortstack{Test set $\sigma_{\text{EI}}>10^{-4}$ \\Linear\ \ Quadratic} & \shortstack{Likelihood Speedup \\ Th.  \hspace{14pt}  Emp.  \hspace{12pt} MB}\\
\hline
\rule{0pt}{12pt}$20\ \ \ \ 1024$ & $1/4$ & $55\ \ 110$ & 303\ \ \ \ 195 &\sqz$3.67\times 10^{-2}$\ \ \ $2.47\times 10^{-2}$ & $119$\ \ \ \ $86$ & 8.1 \hspace{10pt} $3.2^{+1.2}_{-0.6}$   \hspace{10pt} $1.4^{+0.3}_{-0.3}$ \\
\hline
\rule{0pt}{12pt}$20\ \ \ \ 1024$ & $1/4$ & $35\ \ 66$ & 339\ \ \ \ 192 &\sqz$6.95\times 10^{-2}$\ \ \ $2.47\times 10^{-2}$ & $115$\ \ \ \ $64$ & 7.6 \hspace{10pt} $4.5^{+1.7}_{-1.0}$ \hspace{10pt} $1.7^{+0.5}_{-0.3}$ \\
\hline
\rule{0pt}{12pt}$20\ \ \ \ 1024$ & $1/4$ & $26\ \ 42$ & 328\ \ \ \ 204 &\sqz$9.57\times 10^{-3}$\ \ \ $1.04\times 10^{-2}$ & $84$\ \ \ \ $21$ & 7.6 \hspace{10pt} $6.1^{+1.8}_{-1.1}$ \hspace{10pt} $2.2^{+0.6}_{-0.5}$\\
\hline
\rule{0pt}{12pt}$20\ \ \ \ 1024$ & $1/4$ & $18\ \ 33$ & 348\ \ \ \ 201 &\sqz$1.80\times 10^{-2}$\ \ \ $1.32\times 10^{-3}$ & $70$\ \ \ \ $19$ & 7.3 \hspace{10pt} $7.8^{+0.4}_{-1.7}$ \hspace{10pt} $2.6^{+0.7}_{-0.6}$\\
\hline
\rule{0pt}{12pt}$20\ \ \ \ 1024$ & $1/4$ & $12\ \ 20$ & 371\ \ \ \ 264 &\sqz$1.18\times 10^{-2}$\ \ \ $1.03\times 10^{-3}$ & $67$\ \ \ \ $16$ & 6.3 \hspace{10pt} $7.5^{+0.3}_{-1.6}$ \hspace{10pt} $3.1^{+0.7}_{-0.6}$\\
\hline
\rule{0pt}{12pt}$20\ \ \ \ 1024$ & $1/8$ & $10\ \ 15$ & 491\ \ \ \ 386 &\sqz$4.32\times 10^{-3}$\ \ \ $4.39\times 10^{-4}$ & $50$\ \ \ \ $6$ & 9.2 \hspace{10pt} $11.1^{+0.3}_{-0.4}$ \hspace{10pt} $4.3^{+1.2}_{-1.0}$\\
\hline
\rule{0pt}{12pt}$20\ \ \ \ 1024$ & $1/8$ & $8.6\ \ 11.8$ & 505\ \ \ \ 435 &\sqz$9.33\times 10^{-3}$\ \ \ $1.96\times 10^{-4}$ & $48$\ \ \ \ $3$ & 8.5 \hspace{10pt} $10.5^{+0.3}_{-0.9}$ \hspace{10pt} $4.8^{+0.8}_{-1.0}$\\
\hline
\rule{0pt}{12pt}$\ 20 \ \ \ \ 2048$ & $1/16$ & $5.1\ \ 9.6$ & 868\ \ \ \ 942 &\sqz$2.95\times 10^{-3}$\ \ \ $2.38\times 10^{-3}$ & $56$\ \ \ \ $11$ & 17.9 \hspace{10pt} $24.6^{+2.3}_{-4.6}$ \hspace{10pt} $4.8^{+1.2}_{-0.8}$\\
\hline
\rule{0pt}{12pt}$\ 20 \ \ \ \ 2048$ & $1/32$ & $3.35\ \ 6.17$ & 1539\ \ \ \ 1826 &\sqz$9.62\times 10^{-4}$\ \ \ $2.53\times 10^{-4}$ & $46$\ \ \ \ $1$ & 19.3 \hspace{10pt}  $27.6^{+1.1}_{-0.8}$ \hspace{10pt} $4.6^{+1.8}_{-0.9}$\\
\hline
\rule{0pt}{12pt}$\ 20 \ \ \ \ 2048$ & $1/64$ & $2.18\ \ 4.02$ & 2924\ \ \ \ 3636 &\sqz$6.37\times 10^{-4}$\ \ \ $2.68\times 10^{-4}$ & $19$\ \ \ \ $7$ & 19.8 \hspace{10pt} $28.6^{+0.7}_{-0.5}$ \hspace{10pt} $4.2^{+1.7}_{-0.7}$\\
%\hline
%\rule{0pt}{12pt}$\ 20 \ \ \ \ 4096$ & $1/128$ & $1.42\ \ 2.60$ & xxx\ \ \ \ xxx &\sqz$x.x\times 10^{-x}$\ \ \ $x.x\times 10^{-x}$ &$xx$\ \ \ \ $xx$ & xx \hspace{10pt} xx \\
\hline
\hline
\end{tabular}

%% file: TargetedBasisTable.tex
\begin{tabular}{ c | c | c | c | c } 
\hline
\hline
\rule{0pt}{22pt} \shortstack{Freq. range (Hz) \\ Min\ \  Max}  
&  $\Delta{f} (\mathrm{Hz})$ &  \shortstack{$M_c (M_\odot)$ \\ Min\ \ Max}  & \shortstack{Basis size \\ Linear\ \ Quadratic} & \shortstack{Likelihood Speedup \\ Th.  \hspace{14pt}  Emp.  \hspace{12pt} MB}\\
\hline
\rule{0pt}{12pt}$20\ \ \ \ 2048$ & $1/16$ & $6.2\ \ 6.6$ & 1090\ \ \ \ 816 &  17.0 \hspace{8pt} $21.6^{+3.8}_{-3.6}$ \hspace{5pt} $4.8^{+1.1}_{-0.6}$ \\
\hline
\rule{0pt}{12pt}$20\ \ \ \ 2048$ & $1/256$ & $1.195\ \ 1.200$ & 1392\ \ \ \ 2007 &  152.7 \hspace{8pt} $151.8^{+4.5}_{-4.1}$ \hspace{5pt} $29.2^{+1.4}_{-4.6}$ \\
\hline
\hline
\end{tabular}

%% file: main.bbl
\begin{thebibliography}{62}%
\makeatletter
\providecommand \@ifxundefined [1]{%
 \@ifx{#1\undefined}
}%
\providecommand \@ifnum [1]{%
 \ifnum #1\expandafter \@firstoftwo
 \else \expandafter \@secondoftwo
 \fi
}%
\providecommand \@ifx [1]{%
 \ifx #1\expandafter \@firstoftwo
 \else \expandafter \@secondoftwo
 \fi
}%
\providecommand \natexlab [1]{#1}%
\providecommand \enquote  [1]{``#1''}%
\providecommand \bibnamefont  [1]{#1}%
\providecommand \bibfnamefont [1]{#1}%
\providecommand \citenamefont [1]{#1}%
\providecommand \href@noop [0]{\@secondoftwo}%
\providecommand \href [0]{\begingroup \@sanitize@url \@href}%
\providecommand \@href[1]{\@@startlink{#1}\@@href}%
\providecommand \@@href[1]{\endgroup#1\@@endlink}%
\providecommand \@sanitize@url [0]{\catcode `\\12\catcode `\$12\catcode
  `\&12\catcode `\#12\catcode `\^12\catcode `\_12\catcode `\%12\relax}%
\providecommand \@@startlink[1]{}%
\providecommand \@@endlink[0]{}%
\providecommand \url  [0]{\begingroup\@sanitize@url \@url }%
\providecommand \@url [1]{\endgroup\@href {#1}{\urlprefix }}%
\providecommand \urlprefix  [0]{URL }%
\providecommand \Eprint [0]{\href }%
\providecommand \doibase [0]{http://dx.doi.org/}%
\providecommand \selectlanguage [0]{\@gobble}%
\providecommand \bibinfo  [0]{\@secondoftwo}%
\providecommand \bibfield  [0]{\@secondoftwo}%
\providecommand \translation [1]{[#1]}%
\providecommand \BibitemOpen [0]{}%
\providecommand \bibitemStop [0]{}%
\providecommand \bibitemNoStop [0]{.\EOS\space}%
\providecommand \EOS [0]{\spacefactor3000\relax}%
\providecommand \BibitemShut  [1]{\csname bibitem#1\endcsname}%
\let\auto@bib@innerbib\@empty
%</preamble>
\bibitem [{\citenamefont {Collaboration}\ \emph {et~al.}(2015)\citenamefont
  {Collaboration}, \citenamefont {Aasi}, \citenamefont {Abbott}, \citenamefont
  {Abbott} \emph {et~al.}}]{AdvancedLigo}%
  \BibitemOpen
  \bibfield  {author} {\bibinfo {author} {\bibfnamefont {T.~L.~S.}\
  \bibnamefont {Collaboration}}, \bibinfo {author} {\bibfnamefont
  {J.}~\bibnamefont {Aasi}}, \bibinfo {author} {\bibfnamefont {B.~P.}\
  \bibnamefont {Abbott}}, \bibinfo {author} {\bibfnamefont {R.}~\bibnamefont
  {Abbott}},  \emph {et~al.},\ }\href {\doibase 10.1088/0264-9381/32/7/074001}
  {\bibfield  {journal} {\bibinfo  {journal} {Classical and Quantum Gravity}\
  }\textbf {\bibinfo {volume} {32}},\ \bibinfo {pages} {074001} (\bibinfo
  {year} {2015})}\BibitemShut {NoStop}%
\bibitem [{\citenamefont {Acernese}\ \emph {et~al.}(2014)\citenamefont
  {Acernese}, \citenamefont {Agathos} \emph {et~al.}}]{AdvancedVirgo}%
  \BibitemOpen
  \bibfield  {author} {\bibinfo {author} {\bibfnamefont {F.}~\bibnamefont
  {Acernese}}, \bibinfo {author} {\bibfnamefont {M.}~\bibnamefont {Agathos}},
  \emph {et~al.},\ }\href {\doibase 10.1088/0264-9381/32/2/024001} {\bibfield
  {journal} {\bibinfo  {journal} {Classical and Quantum Gravity}\ }\textbf
  {\bibinfo {volume} {32}},\ \bibinfo {pages} {024001} (\bibinfo {year}
  {2014})}\BibitemShut {NoStop}%
\bibitem [{\citenamefont {Akutsu}\ \emph {et~al.}(2019)\citenamefont {Akutsu}
  \emph {et~al.}}]{KAGRA:2018plz}%
  \BibitemOpen
  \bibfield  {author} {\bibinfo {author} {\bibfnamefont {T.}~\bibnamefont
  {Akutsu}} \emph {et~al.} (\bibinfo {collaboration} {KAGRA}),\ }\href
  {\doibase 10.1038/s41550-018-0658-y} {\bibfield  {journal} {\bibinfo
  {journal} {Nature Astron.}\ }\textbf {\bibinfo {volume} {3}},\ \bibinfo
  {pages} {35} (\bibinfo {year} {2019})},\ \Eprint
  {http://arxiv.org/abs/1811.08079} {arXiv:1811.08079 [gr-qc]} \BibitemShut
  {NoStop}%
\bibitem [{\citenamefont {others}(2020)}]{Abbott_2020}%
  \BibitemOpen
  \bibfield  {author} {\bibinfo {author} {\bibfnamefont {B.~P.~A.}\
  \bibnamefont {others}},\ }\href {\doibase 10.1007/s41114-020-00026-9}
  {\bibfield  {journal} {\bibinfo  {journal} {Living Reviews in Relativity}\
  }\textbf {\bibinfo {volume} {23}} (\bibinfo {year} {2020}),\
  10.1007/s41114-020-00026-9}\BibitemShut {NoStop}%
\bibitem [{\citenamefont {Maggiore}\ \emph {et~al.}(2020)\citenamefont
  {Maggiore}, \citenamefont {Broeck}, \citenamefont {Bartolo}, \citenamefont
  {Belgacem}, \citenamefont {Bertacca}, \citenamefont {Bizouard}, \citenamefont
  {Branchesi}, \citenamefont {Clesse}, \citenamefont {Foffa}, \citenamefont
  {García-Bellido}, \citenamefont {Grimm}, \citenamefont {Harms},
  \citenamefont {Hinderer}, \citenamefont {Matarrese}, \citenamefont {Palomba},
  \citenamefont {Peloso}, \citenamefont {Ricciardone},\ and\ \citenamefont
  {Sakellariadou}}]{Maggiore_2020}%
  \BibitemOpen
  \bibfield  {author} {\bibinfo {author} {\bibfnamefont {M.}~\bibnamefont
  {Maggiore}}, \bibinfo {author} {\bibfnamefont {C.~V.~D.}\ \bibnamefont
  {Broeck}}, \bibinfo {author} {\bibfnamefont {N.}~\bibnamefont {Bartolo}},
  \bibinfo {author} {\bibfnamefont {E.}~\bibnamefont {Belgacem}}, \bibinfo
  {author} {\bibfnamefont {D.}~\bibnamefont {Bertacca}}, \bibinfo {author}
  {\bibfnamefont {M.~A.}\ \bibnamefont {Bizouard}}, \bibinfo {author}
  {\bibfnamefont {M.}~\bibnamefont {Branchesi}}, \bibinfo {author}
  {\bibfnamefont {S.}~\bibnamefont {Clesse}}, \bibinfo {author} {\bibfnamefont
  {S.}~\bibnamefont {Foffa}}, \bibinfo {author} {\bibfnamefont
  {J.}~\bibnamefont {García-Bellido}}, \bibinfo {author} {\bibfnamefont
  {S.}~\bibnamefont {Grimm}}, \bibinfo {author} {\bibfnamefont
  {J.}~\bibnamefont {Harms}}, \bibinfo {author} {\bibfnamefont
  {T.}~\bibnamefont {Hinderer}}, \bibinfo {author} {\bibfnamefont
  {S.}~\bibnamefont {Matarrese}}, \bibinfo {author} {\bibfnamefont
  {C.}~\bibnamefont {Palomba}}, \bibinfo {author} {\bibfnamefont
  {M.}~\bibnamefont {Peloso}}, \bibinfo {author} {\bibfnamefont
  {A.}~\bibnamefont {Ricciardone}}, \ and\ \bibinfo {author} {\bibfnamefont
  {M.}~\bibnamefont {Sakellariadou}},\ }\href {\doibase
  10.1088/1475-7516/2020/03/050} {\bibfield  {journal} {\bibinfo  {journal}
  {Journal of Cosmology and Astroparticle Physics}\ }\textbf {\bibinfo {volume}
  {2020}},\ \bibinfo {pages} {050} (\bibinfo {year} {2020})}\BibitemShut
  {NoStop}%
\bibitem [{\citenamefont {Evans}\ \emph {et~al.}(2021)\citenamefont {Evans},
  \citenamefont {Adhikari}, \citenamefont {Afle}, \citenamefont {Ballmer},
  \citenamefont {Biscoveanu}, \citenamefont {Borhanian}, \citenamefont {Brown},
  \citenamefont {Chen}, \citenamefont {Eisenstein}, \citenamefont {Gruson},
  \citenamefont {Gupta}, \citenamefont {Hall}, \citenamefont {Huxford},
  \citenamefont {Kamai}, \citenamefont {Kashyap}, \citenamefont {Kissel},
  \citenamefont {Kuns}, \citenamefont {Landry}, \citenamefont {Lenon},
  \citenamefont {Lovelace}, \citenamefont {McCuller}, \citenamefont {Ng},
  \citenamefont {Nitz}, \citenamefont {Read}, \citenamefont {Sathyaprakash},
  \citenamefont {Shoemaker}, \citenamefont {Slagmolen}, \citenamefont {Smith},
  \citenamefont {Srivastava}, \citenamefont {Sun}, \citenamefont {Vitale},\
  and\ \citenamefont {Weiss}}]{evans2021horizon}%
  \BibitemOpen
  \bibfield  {author} {\bibinfo {author} {\bibfnamefont {M.}~\bibnamefont
  {Evans}}, \bibinfo {author} {\bibfnamefont {R.~X.}\ \bibnamefont {Adhikari}},
  \bibinfo {author} {\bibfnamefont {C.}~\bibnamefont {Afle}}, \bibinfo {author}
  {\bibfnamefont {S.~W.}\ \bibnamefont {Ballmer}}, \bibinfo {author}
  {\bibfnamefont {S.}~\bibnamefont {Biscoveanu}}, \bibinfo {author}
  {\bibfnamefont {S.}~\bibnamefont {Borhanian}}, \bibinfo {author}
  {\bibfnamefont {D.~A.}\ \bibnamefont {Brown}}, \bibinfo {author}
  {\bibfnamefont {Y.}~\bibnamefont {Chen}}, \bibinfo {author} {\bibfnamefont
  {R.}~\bibnamefont {Eisenstein}}, \bibinfo {author} {\bibfnamefont
  {A.}~\bibnamefont {Gruson}}, \bibinfo {author} {\bibfnamefont
  {A.}~\bibnamefont {Gupta}}, \bibinfo {author} {\bibfnamefont {E.~D.}\
  \bibnamefont {Hall}}, \bibinfo {author} {\bibfnamefont {R.}~\bibnamefont
  {Huxford}}, \bibinfo {author} {\bibfnamefont {B.}~\bibnamefont {Kamai}},
  \bibinfo {author} {\bibfnamefont {R.}~\bibnamefont {Kashyap}}, \bibinfo
  {author} {\bibfnamefont {J.~S.}\ \bibnamefont {Kissel}}, \bibinfo {author}
  {\bibfnamefont {K.}~\bibnamefont {Kuns}}, \bibinfo {author} {\bibfnamefont
  {P.}~\bibnamefont {Landry}}, \bibinfo {author} {\bibfnamefont
  {A.}~\bibnamefont {Lenon}}, \bibinfo {author} {\bibfnamefont
  {G.}~\bibnamefont {Lovelace}}, \bibinfo {author} {\bibfnamefont
  {L.}~\bibnamefont {McCuller}}, \bibinfo {author} {\bibfnamefont {K.~K.~Y.}\
  \bibnamefont {Ng}}, \bibinfo {author} {\bibfnamefont {A.~H.}\ \bibnamefont
  {Nitz}}, \bibinfo {author} {\bibfnamefont {J.}~\bibnamefont {Read}}, \bibinfo
  {author} {\bibfnamefont {B.~S.}\ \bibnamefont {Sathyaprakash}}, \bibinfo
  {author} {\bibfnamefont {D.~H.}\ \bibnamefont {Shoemaker}}, \bibinfo {author}
  {\bibfnamefont {B.~J.~J.}\ \bibnamefont {Slagmolen}}, \bibinfo {author}
  {\bibfnamefont {J.~R.}\ \bibnamefont {Smith}}, \bibinfo {author}
  {\bibfnamefont {V.}~\bibnamefont {Srivastava}}, \bibinfo {author}
  {\bibfnamefont {L.}~\bibnamefont {Sun}}, \bibinfo {author} {\bibfnamefont
  {S.}~\bibnamefont {Vitale}}, \ and\ \bibinfo {author} {\bibfnamefont
  {R.}~\bibnamefont {Weiss}},\ }\href@noop {} {\enquote {\bibinfo {title} {A
  horizon study for cosmic explorer: Science, observatories, and community},}\
  } (\bibinfo {year} {2021}),\ \Eprint {http://arxiv.org/abs/2109.09882}
  {arXiv:2109.09882 [astro-ph.IM]} \BibitemShut {NoStop}%
\bibitem [{\citenamefont {Klein}\ \emph {et~al.}(2015)\citenamefont {Klein},
  \citenamefont {Barausse}, \citenamefont {Sesana}, \citenamefont {Petiteau},
  \citenamefont {Berti}, \citenamefont {Babak}, \citenamefont {Gair},
  \citenamefont {Aoudia}, \citenamefont {Hinder}, \citenamefont {Ohme},\ and\
  \citenamefont {Wardell}}]{LISA_I}%
  \BibitemOpen
  \bibfield  {author} {\bibinfo {author} {\bibfnamefont {A.}~\bibnamefont
  {Klein}}, \bibinfo {author} {\bibfnamefont {E.}~\bibnamefont {Barausse}},
  \bibinfo {author} {\bibfnamefont {A.}~\bibnamefont {Sesana}}, \bibinfo
  {author} {\bibfnamefont {A.}~\bibnamefont {Petiteau}}, \bibinfo {author}
  {\bibfnamefont {E.}~\bibnamefont {Berti}}, \bibinfo {author} {\bibfnamefont
  {S.}~\bibnamefont {Babak}}, \bibinfo {author} {\bibfnamefont
  {J.}~\bibnamefont {Gair}}, \bibinfo {author} {\bibfnamefont {S.}~\bibnamefont
  {Aoudia}}, \bibinfo {author} {\bibfnamefont {I.}~\bibnamefont {Hinder}},
  \bibinfo {author} {\bibfnamefont {F.}~\bibnamefont {Ohme}}, \ and\ \bibinfo
  {author} {\bibfnamefont {B.}~\bibnamefont {Wardell}},\ }\href {\doibase
  10.1103/PhysRevD.93.024003} {\bibfield  {journal} {\bibinfo  {journal}
  {Physical Review D}\ }\textbf {\bibinfo {volume} {93}} (\bibinfo {year}
  {2015}),\ 10.1103/PhysRevD.93.024003}\BibitemShut {NoStop}%
\bibitem [{\citenamefont {Tamanini}\ \emph {et~al.}(2016)\citenamefont
  {Tamanini}, \citenamefont {Caprini}, \citenamefont {Barausse}, \citenamefont
  {Sesana}, \citenamefont {Klein},\ and\ \citenamefont {Petiteau}}]{LISA_III}%
  \BibitemOpen
  \bibfield  {author} {\bibinfo {author} {\bibfnamefont {N.}~\bibnamefont
  {Tamanini}}, \bibinfo {author} {\bibfnamefont {C.}~\bibnamefont {Caprini}},
  \bibinfo {author} {\bibfnamefont {E.}~\bibnamefont {Barausse}}, \bibinfo
  {author} {\bibfnamefont {A.}~\bibnamefont {Sesana}}, \bibinfo {author}
  {\bibfnamefont {A.}~\bibnamefont {Klein}}, \ and\ \bibinfo {author}
  {\bibfnamefont {A.}~\bibnamefont {Petiteau}},\ }\href {\doibase
  10.1088/1475-7516/2016/04/002} {\bibfield  {journal} {\bibinfo  {journal}
  {Journal of Cosmology and Astroparticle Physics}\ }\textbf {\bibinfo {volume}
  {2016}},\ \bibinfo {pages} {002} (\bibinfo {year} {2016})}\BibitemShut
  {NoStop}%
\bibitem [{\citenamefont {Babak}\ \emph {et~al.}(2017)\citenamefont {Babak},
  \citenamefont {Gair}, \citenamefont {Sesana}, \citenamefont {Barausse},
  \citenamefont {Sopuerta}, \citenamefont {Berry}, \citenamefont {Berti},
  \citenamefont {Amaro-Seoane}, \citenamefont {Petiteau},\ and\ \citenamefont
  {Klein}}]{LISA_V}%
  \BibitemOpen
  \bibfield  {author} {\bibinfo {author} {\bibfnamefont {S.}~\bibnamefont
  {Babak}}, \bibinfo {author} {\bibfnamefont {J.}~\bibnamefont {Gair}},
  \bibinfo {author} {\bibfnamefont {A.}~\bibnamefont {Sesana}}, \bibinfo
  {author} {\bibfnamefont {E.}~\bibnamefont {Barausse}}, \bibinfo {author}
  {\bibfnamefont {C.~F.}\ \bibnamefont {Sopuerta}}, \bibinfo {author}
  {\bibfnamefont {C.~P.}\ \bibnamefont {Berry}}, \bibinfo {author}
  {\bibfnamefont {E.}~\bibnamefont {Berti}}, \bibinfo {author} {\bibfnamefont
  {P.}~\bibnamefont {Amaro-Seoane}}, \bibinfo {author} {\bibfnamefont
  {A.}~\bibnamefont {Petiteau}}, \ and\ \bibinfo {author} {\bibfnamefont
  {A.}~\bibnamefont {Klein}},\ }\href {\doibase 10.1103/physrevd.95.103012}
  {\bibfield  {journal} {\bibinfo  {journal} {Physical Review D}\ }\textbf
  {\bibinfo {volume} {95}} (\bibinfo {year} {2017}),\
  10.1103/physrevd.95.103012}\BibitemShut {NoStop}%
\bibitem [{\citenamefont {Thrane}\ and\ \citenamefont
  {Talbot}(2019)}]{Thrane_2019}%
  \BibitemOpen
  \bibfield  {author} {\bibinfo {author} {\bibfnamefont {E.}~\bibnamefont
  {Thrane}}\ and\ \bibinfo {author} {\bibfnamefont {C.}~\bibnamefont
  {Talbot}},\ }\href {\doibase 10.1017/pasa.2019.2} {\bibfield  {journal}
  {\bibinfo  {journal} {Publications of the Astronomical Society of Australia}\
  }\textbf {\bibinfo {volume} {36}} (\bibinfo {year} {2019}),\
  10.1017/pasa.2019.2}\BibitemShut {NoStop}%
\bibitem [{\citenamefont {Maggiore}(2007)}]{Maggiore_Vol1}%
  \BibitemOpen
  \bibfield  {author} {\bibinfo {author} {\bibfnamefont {M.}~\bibnamefont
  {Maggiore}},\ }\href@noop {} {\emph {\bibinfo {title} {{Gravitational Waves.
  Vol. 1: Theory and Experiments}}}},\ Oxford Master Series in Physics\
  (\bibinfo  {publisher} {Oxford University Press},\ \bibinfo {year} {2007})\
  p.\ \bibinfo {pages} {572}\BibitemShut {NoStop}%
\bibitem [{\citenamefont {Mills}\ and\ \citenamefont
  {Fairhurst}(2021)}]{Mills:2020thr}%
  \BibitemOpen
  \bibfield  {author} {\bibinfo {author} {\bibfnamefont {C.}~\bibnamefont
  {Mills}}\ and\ \bibinfo {author} {\bibfnamefont {S.}~\bibnamefont
  {Fairhurst}},\ }\href {\doibase 10.1103/PhysRevD.103.024042} {\bibfield
  {journal} {\bibinfo  {journal} {Phys. Rev. D}\ }\textbf {\bibinfo {volume}
  {103}},\ \bibinfo {pages} {024042} (\bibinfo {year} {2021})},\ \Eprint
  {http://arxiv.org/abs/2007.04313} {arXiv:2007.04313 [gr-qc]} \BibitemShut
  {NoStop}%
\bibitem [{\citenamefont {Morisaki}(2021)}]{PhysRevD.104.044062}%
  \BibitemOpen
  \bibfield  {author} {\bibinfo {author} {\bibfnamefont {S.}~\bibnamefont
  {Morisaki}},\ }\href {\doibase 10.1103/PhysRevD.104.044062} {\bibfield
  {journal} {\bibinfo  {journal} {Phys. Rev. D}\ }\textbf {\bibinfo {volume}
  {104}},\ \bibinfo {pages} {044062} (\bibinfo {year} {2021})}\BibitemShut
  {NoStop}%
\bibitem [{\citenamefont {Zackay}\ \emph {et~al.}(2018)\citenamefont {Zackay},
  \citenamefont {Dai},\ and\ \citenamefont {Venumadhav}}]{Zackay:2018qdy}%
  \BibitemOpen
  \bibfield  {author} {\bibinfo {author} {\bibfnamefont {B.}~\bibnamefont
  {Zackay}}, \bibinfo {author} {\bibfnamefont {L.}~\bibnamefont {Dai}}, \ and\
  \bibinfo {author} {\bibfnamefont {T.}~\bibnamefont {Venumadhav}},\
  }\href@noop {} {\  (\bibinfo {year} {2018})},\ \Eprint
  {http://arxiv.org/abs/1806.08792} {arXiv:1806.08792 [astro-ph.IM]}
  \BibitemShut {NoStop}%
\bibitem [{\citenamefont {Cornish}(2021)}]{Cornish:2021lje}%
  \BibitemOpen
  \bibfield  {author} {\bibinfo {author} {\bibfnamefont {N.~J.}\ \bibnamefont
  {Cornish}},\ }\href {\doibase 10.1103/PhysRevD.104.104054} {\bibfield
  {journal} {\bibinfo  {journal} {Phys. Rev. D}\ }\textbf {\bibinfo {volume}
  {104}},\ \bibinfo {pages} {104054} (\bibinfo {year} {2021})},\ \Eprint
  {http://arxiv.org/abs/2109.02728} {arXiv:2109.02728 [gr-qc]} \BibitemShut
  {NoStop}%
\bibitem [{\citenamefont {Chua}\ and\ \citenamefont
  {Vallisneri}(2020)}]{Chua:2019wwt}%
  \BibitemOpen
  \bibfield  {author} {\bibinfo {author} {\bibfnamefont {A.~J.~K.}\
  \bibnamefont {Chua}}\ and\ \bibinfo {author} {\bibfnamefont {M.}~\bibnamefont
  {Vallisneri}},\ }\href {\doibase 10.1103/PhysRevLett.124.041102} {\bibfield
  {journal} {\bibinfo  {journal} {Phys. Rev. Lett.}\ }\textbf {\bibinfo
  {volume} {124}},\ \bibinfo {pages} {041102} (\bibinfo {year} {2020})},\
  \Eprint {http://arxiv.org/abs/1909.05966} {arXiv:1909.05966 [gr-qc]}
  \BibitemShut {NoStop}%
\bibitem [{\citenamefont {Green}\ \emph {et~al.}(2020)\citenamefont {Green},
  \citenamefont {Simpson},\ and\ \citenamefont {Gair}}]{Green:2020hst}%
  \BibitemOpen
  \bibfield  {author} {\bibinfo {author} {\bibfnamefont {S.~R.}\ \bibnamefont
  {Green}}, \bibinfo {author} {\bibfnamefont {C.}~\bibnamefont {Simpson}}, \
  and\ \bibinfo {author} {\bibfnamefont {J.}~\bibnamefont {Gair}},\ }\href
  {\doibase 10.1103/PhysRevD.102.104057} {\bibfield  {journal} {\bibinfo
  {journal} {Phys. Rev. D}\ }\textbf {\bibinfo {volume} {102}},\ \bibinfo
  {pages} {104057} (\bibinfo {year} {2020})},\ \Eprint
  {http://arxiv.org/abs/2002.07656} {arXiv:2002.07656 [astro-ph.IM]}
  \BibitemShut {NoStop}%
\bibitem [{\citenamefont {Antil}\ \emph {et~al.}(2013)\citenamefont {Antil},
  \citenamefont {Field}, \citenamefont {Herrmann}, \citenamefont {Nochetto},\
  and\ \citenamefont {Tiglio}}]{Antil:2012wf}%
  \BibitemOpen
  \bibfield  {author} {\bibinfo {author} {\bibfnamefont {H.}~\bibnamefont
  {Antil}}, \bibinfo {author} {\bibfnamefont {S.~E.}\ \bibnamefont {Field}},
  \bibinfo {author} {\bibfnamefont {F.}~\bibnamefont {Herrmann}}, \bibinfo
  {author} {\bibfnamefont {R.~H.}\ \bibnamefont {Nochetto}}, \ and\ \bibinfo
  {author} {\bibfnamefont {M.}~\bibnamefont {Tiglio}},\ }\href {\doibase
  10.1007/s10915-013-9722-z} {\bibfield  {journal} {\bibinfo  {journal} {J.
  Sci. Comput.}\ }\textbf {\bibinfo {volume} {57}},\ \bibinfo {pages} {604}
  (\bibinfo {year} {2013})},\ \Eprint {http://arxiv.org/abs/1210.0577}
  {arXiv:1210.0577 [cs.NA]} \BibitemShut {NoStop}%
\bibitem [{\citenamefont {Canizares}\ \emph {et~al.}(2013)\citenamefont
  {Canizares}, \citenamefont {Field}, \citenamefont {Gair},\ and\ \citenamefont
  {Tiglio}}]{Canizares:2013ywa}%
  \BibitemOpen
  \bibfield  {author} {\bibinfo {author} {\bibfnamefont {P.}~\bibnamefont
  {Canizares}}, \bibinfo {author} {\bibfnamefont {S.~E.}\ \bibnamefont
  {Field}}, \bibinfo {author} {\bibfnamefont {J.~R.}\ \bibnamefont {Gair}}, \
  and\ \bibinfo {author} {\bibfnamefont {M.}~\bibnamefont {Tiglio}},\ }\href
  {\doibase 10.1103/PhysRevD.87.124005} {\bibfield  {journal} {\bibinfo
  {journal} {Phys. Rev. D}\ }\textbf {\bibinfo {volume} {87}},\ \bibinfo
  {pages} {124005} (\bibinfo {year} {2013})},\ \Eprint
  {http://arxiv.org/abs/1304.0462} {arXiv:1304.0462 [gr-qc]} \BibitemShut
  {NoStop}%
\bibitem [{\citenamefont {Smith}\ \emph {et~al.}(2016)\citenamefont {Smith},
  \citenamefont {Field}, \citenamefont {Blackburn}, \citenamefont {Haster},
  \citenamefont {P\"urrer}, \citenamefont {Raymond},\ and\ \citenamefont
  {Schmidt}}]{Smith:2016qas}%
  \BibitemOpen
  \bibfield  {author} {\bibinfo {author} {\bibfnamefont {R.}~\bibnamefont
  {Smith}}, \bibinfo {author} {\bibfnamefont {S.~E.}\ \bibnamefont {Field}},
  \bibinfo {author} {\bibfnamefont {K.}~\bibnamefont {Blackburn}}, \bibinfo
  {author} {\bibfnamefont {C.-J.}\ \bibnamefont {Haster}}, \bibinfo {author}
  {\bibfnamefont {M.}~\bibnamefont {P\"urrer}}, \bibinfo {author}
  {\bibfnamefont {V.}~\bibnamefont {Raymond}}, \ and\ \bibinfo {author}
  {\bibfnamefont {P.}~\bibnamefont {Schmidt}},\ }\href {\doibase
  10.1103/PhysRevD.94.044031} {\bibfield  {journal} {\bibinfo  {journal} {Phys.
  Rev. D}\ }\textbf {\bibinfo {volume} {94}},\ \bibinfo {pages} {044031}
  (\bibinfo {year} {2016})},\ \Eprint {http://arxiv.org/abs/1604.08253}
  {arXiv:1604.08253 [gr-qc]} \BibitemShut {NoStop}%
\bibitem [{\citenamefont {Qi}\ and\ \citenamefont
  {Raymond}(2021)}]{Qi:2020lfr}%
  \BibitemOpen
  \bibfield  {author} {\bibinfo {author} {\bibfnamefont {H.}~\bibnamefont
  {Qi}}\ and\ \bibinfo {author} {\bibfnamefont {V.}~\bibnamefont {Raymond}},\
  }\href {\doibase 10.1103/PhysRevD.104.063031} {\bibfield  {journal} {\bibinfo
   {journal} {Phys. Rev. D}\ }\textbf {\bibinfo {volume} {104}},\ \bibinfo
  {pages} {063031} (\bibinfo {year} {2021})},\ \Eprint
  {http://arxiv.org/abs/2009.13812} {arXiv:2009.13812 [gr-qc]} \BibitemShut
  {NoStop}%
\bibitem [{\citenamefont {Morisaki}\ and\ \citenamefont
  {Raymond}(2020)}]{Morisaki:2020oqk}%
  \BibitemOpen
  \bibfield  {author} {\bibinfo {author} {\bibfnamefont {S.}~\bibnamefont
  {Morisaki}}\ and\ \bibinfo {author} {\bibfnamefont {V.}~\bibnamefont
  {Raymond}},\ }\href {\doibase 10.1103/PhysRevD.102.104020} {\bibfield
  {journal} {\bibinfo  {journal} {Phys. Rev. D}\ }\textbf {\bibinfo {volume}
  {102}},\ \bibinfo {pages} {104020} (\bibinfo {year} {2020})},\ \Eprint
  {http://arxiv.org/abs/2007.09108} {arXiv:2007.09108 [gr-qc]} \BibitemShut
  {NoStop}%
\bibitem [{\citenamefont {Morisaki}\ \emph {et~al.}(2023)\citenamefont
  {Morisaki}, \citenamefont {Smith}, \citenamefont {Tsukada}, \citenamefont
  {Sachdev}, \citenamefont {Stevenson}, \citenamefont {Talbot},\ and\
  \citenamefont {Zimmerman}}]{Morisaki:2023kuq}%
  \BibitemOpen
  \bibfield  {author} {\bibinfo {author} {\bibfnamefont {S.}~\bibnamefont
  {Morisaki}}, \bibinfo {author} {\bibfnamefont {R.}~\bibnamefont {Smith}},
  \bibinfo {author} {\bibfnamefont {L.}~\bibnamefont {Tsukada}}, \bibinfo
  {author} {\bibfnamefont {S.}~\bibnamefont {Sachdev}}, \bibinfo {author}
  {\bibfnamefont {S.}~\bibnamefont {Stevenson}}, \bibinfo {author}
  {\bibfnamefont {C.}~\bibnamefont {Talbot}}, \ and\ \bibinfo {author}
  {\bibfnamefont {A.}~\bibnamefont {Zimmerman}},\ }\href@noop {} {\  (\bibinfo
  {year} {2023})},\ \Eprint {http://arxiv.org/abs/2307.13380} {arXiv:2307.13380
  [gr-qc]} \BibitemShut {NoStop}%
\bibitem [{\citenamefont {Pankow}\ \emph {et~al.}(2015)\citenamefont {Pankow},
  \citenamefont {Brady}, \citenamefont {Ochsner},\ and\ \citenamefont
  {O'Shaughnessy}}]{Pankow_2015}%
  \BibitemOpen
  \bibfield  {author} {\bibinfo {author} {\bibfnamefont {C.}~\bibnamefont
  {Pankow}}, \bibinfo {author} {\bibfnamefont {P.}~\bibnamefont {Brady}},
  \bibinfo {author} {\bibfnamefont {E.}~\bibnamefont {Ochsner}}, \ and\
  \bibinfo {author} {\bibfnamefont {R.}~\bibnamefont {O'Shaughnessy}},\ }\href
  {\doibase 10.1103/physrevd.92.023002} {\bibfield  {journal} {\bibinfo
  {journal} {Physical Review D}\ }\textbf {\bibinfo {volume} {92}} (\bibinfo
  {year} {2015}),\ 10.1103/physrevd.92.023002}\BibitemShut {NoStop}%
\bibitem [{\citenamefont {Lange}\ \emph {et~al.}(2018)\citenamefont {Lange},
  \citenamefont {O'Shaughnessy},\ and\ \citenamefont {Rizzo}}]{Lange:2018pyp}%
  \BibitemOpen
  \bibfield  {author} {\bibinfo {author} {\bibfnamefont {J.}~\bibnamefont
  {Lange}}, \bibinfo {author} {\bibfnamefont {R.}~\bibnamefont
  {O'Shaughnessy}}, \ and\ \bibinfo {author} {\bibfnamefont {M.}~\bibnamefont
  {Rizzo}},\ }\href@noop {} {\  (\bibinfo {year} {2018})},\ \Eprint
  {http://arxiv.org/abs/1805.10457} {arXiv:1805.10457 [gr-qc]} \BibitemShut
  {NoStop}%
\bibitem [{\citenamefont {Pathak}\ \emph {et~al.}(2022)\citenamefont {Pathak},
  \citenamefont {Reza},\ and\ \citenamefont {Sengupta}}]{Pathak:2022ktt}%
  \BibitemOpen
  \bibfield  {author} {\bibinfo {author} {\bibfnamefont {L.}~\bibnamefont
  {Pathak}}, \bibinfo {author} {\bibfnamefont {A.}~\bibnamefont {Reza}}, \ and\
  \bibinfo {author} {\bibfnamefont {A.~S.}\ \bibnamefont {Sengupta}},\
  }\href@noop {} {\  (\bibinfo {year} {2022})},\ \Eprint
  {http://arxiv.org/abs/2210.02706} {arXiv:2210.02706 [gr-qc]} \BibitemShut
  {NoStop}%
\bibitem [{\citenamefont {Hannam}\ \emph {et~al.}(2014)\citenamefont {Hannam},
  \citenamefont {Schmidt}, \citenamefont {Boh\'e}, \citenamefont {Haegel},
  \citenamefont {Husa}, \citenamefont {Ohme}, \citenamefont {Pratten},\ and\
  \citenamefont {P\"urrer}}]{PhysRevLett.113.151101}%
  \BibitemOpen
  \bibfield  {author} {\bibinfo {author} {\bibfnamefont {M.}~\bibnamefont
  {Hannam}}, \bibinfo {author} {\bibfnamefont {P.}~\bibnamefont {Schmidt}},
  \bibinfo {author} {\bibfnamefont {A.}~\bibnamefont {Boh\'e}}, \bibinfo
  {author} {\bibfnamefont {L.}~\bibnamefont {Haegel}}, \bibinfo {author}
  {\bibfnamefont {S.}~\bibnamefont {Husa}}, \bibinfo {author} {\bibfnamefont
  {F.}~\bibnamefont {Ohme}}, \bibinfo {author} {\bibfnamefont {G.}~\bibnamefont
  {Pratten}}, \ and\ \bibinfo {author} {\bibfnamefont {M.}~\bibnamefont
  {P\"urrer}},\ }\href {\doibase 10.1103/PhysRevLett.113.151101} {\bibfield
  {journal} {\bibinfo  {journal} {Phys. Rev. Lett.}\ }\textbf {\bibinfo
  {volume} {113}},\ \bibinfo {pages} {151101} (\bibinfo {year}
  {2014})}\BibitemShut {NoStop}%
\bibitem [{\citenamefont {Pratten}\ \emph {et~al.}(2021)\citenamefont
  {Pratten}, \citenamefont {García-Quirós}, \citenamefont {Colleoni},
  \citenamefont {Ramos-Buades}, \citenamefont {Estellés}, \citenamefont
  {Mateu-Lucena}, \citenamefont {Jaume}, \citenamefont {Haney}, \citenamefont
  {Keitel}, \citenamefont {Thompson},\ and\ \citenamefont
  {et~al.}}]{Pratten_2021}%
  \BibitemOpen
  \bibfield  {author} {\bibinfo {author} {\bibfnamefont {G.}~\bibnamefont
  {Pratten}}, \bibinfo {author} {\bibfnamefont {C.}~\bibnamefont
  {García-Quirós}}, \bibinfo {author} {\bibfnamefont {M.}~\bibnamefont
  {Colleoni}}, \bibinfo {author} {\bibfnamefont {A.}~\bibnamefont
  {Ramos-Buades}}, \bibinfo {author} {\bibfnamefont {H.}~\bibnamefont
  {Estellés}}, \bibinfo {author} {\bibfnamefont {M.}~\bibnamefont
  {Mateu-Lucena}}, \bibinfo {author} {\bibfnamefont {R.}~\bibnamefont {Jaume}},
  \bibinfo {author} {\bibfnamefont {M.}~\bibnamefont {Haney}}, \bibinfo
  {author} {\bibfnamefont {D.}~\bibnamefont {Keitel}}, \bibinfo {author}
  {\bibfnamefont {J.~E.}\ \bibnamefont {Thompson}}, \ and\ \bibinfo {author}
  {\bibnamefont {et~al.}},\ }\href {\doibase 10.1103/physrevd.103.104056}
  {\bibfield  {journal} {\bibinfo  {journal} {Physical Review D}\ }\textbf
  {\bibinfo {volume} {103}} (\bibinfo {year} {2021}),\
  10.1103/physrevd.103.104056}\BibitemShut {NoStop}%
\bibitem [{\citenamefont {Finn}(1992)}]{GWLikelihood_Finn}%
  \BibitemOpen
  \bibfield  {author} {\bibinfo {author} {\bibfnamefont {L.~S.}\ \bibnamefont
  {Finn}},\ }\href {\doibase 10.1103/PhysRevD.46.5236} {\bibfield  {journal}
  {\bibinfo  {journal} {Phys. Rev. D}\ }\textbf {\bibinfo {volume} {46}},\
  \bibinfo {pages} {5236} (\bibinfo {year} {1992})},\ \Eprint
  {http://arxiv.org/abs/gr-qc/9209010} {arXiv:gr-qc/9209010} \BibitemShut
  {NoStop}%
\bibitem [{\citenamefont {Dietrich}\ \emph {et~al.}(2019)\citenamefont
  {Dietrich}, \citenamefont {Samajdar}, \citenamefont {Khan}, \citenamefont
  {Johnson-McDaniel}, \citenamefont {Dudi},\ and\ \citenamefont
  {Tichy}}]{Dietrich:2019kaq}%
  \BibitemOpen
  \bibfield  {author} {\bibinfo {author} {\bibfnamefont {T.}~\bibnamefont
  {Dietrich}}, \bibinfo {author} {\bibfnamefont {A.}~\bibnamefont {Samajdar}},
  \bibinfo {author} {\bibfnamefont {S.}~\bibnamefont {Khan}}, \bibinfo {author}
  {\bibfnamefont {N.~K.}\ \bibnamefont {Johnson-McDaniel}}, \bibinfo {author}
  {\bibfnamefont {R.}~\bibnamefont {Dudi}}, \ and\ \bibinfo {author}
  {\bibfnamefont {W.}~\bibnamefont {Tichy}},\ }\href {\doibase
  10.1103/PhysRevD.100.044003} {\bibfield  {journal} {\bibinfo  {journal}
  {Phys. Rev. D}\ }\textbf {\bibinfo {volume} {100}},\ \bibinfo {pages}
  {044003} (\bibinfo {year} {2019})},\ \Eprint
  {http://arxiv.org/abs/1905.06011} {arXiv:1905.06011 [gr-qc]} \BibitemShut
  {NoStop}%
\bibitem [{\citenamefont {Chiaramello}\ and\ \citenamefont
  {Nagar}(2020)}]{Chiaramello:2020ehz}%
  \BibitemOpen
  \bibfield  {author} {\bibinfo {author} {\bibfnamefont {D.}~\bibnamefont
  {Chiaramello}}\ and\ \bibinfo {author} {\bibfnamefont {A.}~\bibnamefont
  {Nagar}},\ }\href {\doibase 10.1103/PhysRevD.101.101501} {\bibfield
  {journal} {\bibinfo  {journal} {Phys. Rev. D}\ }\textbf {\bibinfo {volume}
  {101}},\ \bibinfo {pages} {101501} (\bibinfo {year} {2020})},\ \Eprint
  {http://arxiv.org/abs/2001.11736} {arXiv:2001.11736 [gr-qc]} \BibitemShut
  {NoStop}%
\bibitem [{\citenamefont {Jaranowski}\ \emph {et~al.}(1998)\citenamefont
  {Jaranowski}, \citenamefont {Krolak},\ and\ \citenamefont
  {Schutz}}]{Jaranowski:1998qm}%
  \BibitemOpen
  \bibfield  {author} {\bibinfo {author} {\bibfnamefont {P.}~\bibnamefont
  {Jaranowski}}, \bibinfo {author} {\bibfnamefont {A.}~\bibnamefont {Krolak}},
  \ and\ \bibinfo {author} {\bibfnamefont {B.~F.}\ \bibnamefont {Schutz}},\
  }\href {\doibase 10.1103/PhysRevD.58.063001} {\bibfield  {journal} {\bibinfo
  {journal} {Phys. Rev. D}\ }\textbf {\bibinfo {volume} {58}},\ \bibinfo
  {pages} {063001} (\bibinfo {year} {1998})},\ \Eprint
  {http://arxiv.org/abs/gr-qc/9804014} {arXiv:gr-qc/9804014} \BibitemShut
  {NoStop}%
\bibitem [{\citenamefont {P\"urrer}(2014)}]{Purrer:2014fza}%
  \BibitemOpen
  \bibfield  {author} {\bibinfo {author} {\bibfnamefont {M.}~\bibnamefont
  {P\"urrer}},\ }\href {\doibase 10.1088/0264-9381/31/19/195010} {\bibfield
  {journal} {\bibinfo  {journal} {Class. Quant. Grav.}\ }\textbf {\bibinfo
  {volume} {31}},\ \bibinfo {pages} {195010} (\bibinfo {year} {2014})},\
  \Eprint {http://arxiv.org/abs/1402.4146} {arXiv:1402.4146 [gr-qc]}
  \BibitemShut {NoStop}%
\bibitem [{\citenamefont {Cotesta}\ \emph {et~al.}(2020)\citenamefont
  {Cotesta}, \citenamefont {Marsat},\ and\ \citenamefont
  {P\"urrer}}]{Cotesta:2020qhw}%
  \BibitemOpen
  \bibfield  {author} {\bibinfo {author} {\bibfnamefont {R.}~\bibnamefont
  {Cotesta}}, \bibinfo {author} {\bibfnamefont {S.}~\bibnamefont {Marsat}}, \
  and\ \bibinfo {author} {\bibfnamefont {M.}~\bibnamefont {P\"urrer}},\ }\href
  {\doibase 10.1103/PhysRevD.101.124040} {\bibfield  {journal} {\bibinfo
  {journal} {Phys. Rev. D}\ }\textbf {\bibinfo {volume} {101}},\ \bibinfo
  {pages} {124040} (\bibinfo {year} {2020})},\ \Eprint
  {http://arxiv.org/abs/2003.12079} {arXiv:2003.12079 [gr-qc]} \BibitemShut
  {NoStop}%
\bibitem [{\citenamefont {Szyld}(2006)}]{Szyld2006}%
  \BibitemOpen
  \bibfield  {author} {\bibinfo {author} {\bibfnamefont {D.~B.}\ \bibnamefont
  {Szyld}},\ }\href {\doibase 10.1007/s11075-006-9046-2} {\bibfield  {journal}
  {\bibinfo  {journal} {Numerical Algorithms}\ }\textbf {\bibinfo {volume}
  {42}},\ \bibinfo {pages} {309} (\bibinfo {year} {2006})}\BibitemShut
  {NoStop}%
\bibitem [{\citenamefont {Ajith}\ \emph {et~al.}(2011)\citenamefont {Ajith},
  \citenamefont {Hannam}, \citenamefont {Husa}, \citenamefont {Chen},
  \citenamefont {Br\"ugmann}, \citenamefont {Dorband}, \citenamefont
  {M\"uller}, \citenamefont {Ohme}, \citenamefont {Pollney}, \citenamefont
  {Reisswig}, \citenamefont {Santamar\'{\i}a},\ and\ \citenamefont
  {Seiler}}]{PhysRevLett.106.241101}%
  \BibitemOpen
  \bibfield  {author} {\bibinfo {author} {\bibfnamefont {P.}~\bibnamefont
  {Ajith}}, \bibinfo {author} {\bibfnamefont {M.}~\bibnamefont {Hannam}},
  \bibinfo {author} {\bibfnamefont {S.}~\bibnamefont {Husa}}, \bibinfo {author}
  {\bibfnamefont {Y.}~\bibnamefont {Chen}}, \bibinfo {author} {\bibfnamefont
  {B.}~\bibnamefont {Br\"ugmann}}, \bibinfo {author} {\bibfnamefont
  {N.}~\bibnamefont {Dorband}}, \bibinfo {author} {\bibfnamefont
  {D.}~\bibnamefont {M\"uller}}, \bibinfo {author} {\bibfnamefont
  {F.}~\bibnamefont {Ohme}}, \bibinfo {author} {\bibfnamefont {D.}~\bibnamefont
  {Pollney}}, \bibinfo {author} {\bibfnamefont {C.}~\bibnamefont {Reisswig}},
  \bibinfo {author} {\bibfnamefont {L.}~\bibnamefont {Santamar\'{\i}a}}, \ and\
  \bibinfo {author} {\bibfnamefont {J.}~\bibnamefont {Seiler}},\ }\href
  {\doibase 10.1103/PhysRevLett.106.241101} {\bibfield  {journal} {\bibinfo
  {journal} {Phys. Rev. Lett.}\ }\textbf {\bibinfo {volume} {106}},\ \bibinfo
  {pages} {241101} (\bibinfo {year} {2011})}\BibitemShut {NoStop}%
\bibitem [{\citenamefont {Santamar\'{\i}a}\ \emph {et~al.}(2010)\citenamefont
  {Santamar\'{\i}a}, \citenamefont {Ohme}, \citenamefont {Ajith}, \citenamefont
  {Br\"ugmann}, \citenamefont {Dorband}, \citenamefont {Hannam}, \citenamefont
  {Husa}, \citenamefont {M\"osta}, \citenamefont {Pollney}, \citenamefont
  {Reisswig}, \citenamefont {Robinson}, \citenamefont {Seiler},\ and\
  \citenamefont {Krishnan}}]{PhysRevD.82.064016}%
  \BibitemOpen
  \bibfield  {author} {\bibinfo {author} {\bibfnamefont {L.}~\bibnamefont
  {Santamar\'{\i}a}}, \bibinfo {author} {\bibfnamefont {F.}~\bibnamefont
  {Ohme}}, \bibinfo {author} {\bibfnamefont {P.}~\bibnamefont {Ajith}},
  \bibinfo {author} {\bibfnamefont {B.}~\bibnamefont {Br\"ugmann}}, \bibinfo
  {author} {\bibfnamefont {N.}~\bibnamefont {Dorband}}, \bibinfo {author}
  {\bibfnamefont {M.}~\bibnamefont {Hannam}}, \bibinfo {author} {\bibfnamefont
  {S.}~\bibnamefont {Husa}}, \bibinfo {author} {\bibfnamefont {P.}~\bibnamefont
  {M\"osta}}, \bibinfo {author} {\bibfnamefont {D.}~\bibnamefont {Pollney}},
  \bibinfo {author} {\bibfnamefont {C.}~\bibnamefont {Reisswig}}, \bibinfo
  {author} {\bibfnamefont {E.~L.}\ \bibnamefont {Robinson}}, \bibinfo {author}
  {\bibfnamefont {J.}~\bibnamefont {Seiler}}, \ and\ \bibinfo {author}
  {\bibfnamefont {B.}~\bibnamefont {Krishnan}},\ }\href {\doibase
  10.1103/PhysRevD.82.064016} {\bibfield  {journal} {\bibinfo  {journal} {Phys.
  Rev. D}\ }\textbf {\bibinfo {volume} {82}},\ \bibinfo {pages} {064016}
  (\bibinfo {year} {2010})}\BibitemShut {NoStop}%
\bibitem [{\citenamefont {Schmidt}\ \emph {et~al.}(2015)\citenamefont
  {Schmidt}, \citenamefont {Ohme},\ and\ \citenamefont
  {Hannam}}]{PhysRevD.91.024043}%
  \BibitemOpen
  \bibfield  {author} {\bibinfo {author} {\bibfnamefont {P.}~\bibnamefont
  {Schmidt}}, \bibinfo {author} {\bibfnamefont {F.}~\bibnamefont {Ohme}}, \
  and\ \bibinfo {author} {\bibfnamefont {M.}~\bibnamefont {Hannam}},\ }\href
  {\doibase 10.1103/PhysRevD.91.024043} {\bibfield  {journal} {\bibinfo
  {journal} {Phys. Rev. D}\ }\textbf {\bibinfo {volume} {91}},\ \bibinfo
  {pages} {024043} (\bibinfo {year} {2015})}\BibitemShut {NoStop}%
\bibitem [{\citenamefont {Garc\'\i{}a-Quir\'os}\ \emph
  {et~al.}(2021)\citenamefont {Garc\'\i{}a-Quir\'os}, \citenamefont {Husa},
  \citenamefont {Mateu-Lucena},\ and\ \citenamefont
  {Borchers}}]{Garcia-Quiros:2020qlt}%
  \BibitemOpen
  \bibfield  {author} {\bibinfo {author} {\bibfnamefont {C.}~\bibnamefont
  {Garc\'\i{}a-Quir\'os}}, \bibinfo {author} {\bibfnamefont {S.}~\bibnamefont
  {Husa}}, \bibinfo {author} {\bibfnamefont {M.}~\bibnamefont {Mateu-Lucena}},
  \ and\ \bibinfo {author} {\bibfnamefont {A.}~\bibnamefont {Borchers}},\
  }\href {\doibase 10.1088/1361-6382/abc36e} {\bibfield  {journal} {\bibinfo
  {journal} {Class. Quant. Grav.}\ }\textbf {\bibinfo {volume} {38}},\ \bibinfo
  {pages} {015006} (\bibinfo {year} {2021})},\ \Eprint
  {http://arxiv.org/abs/2001.10897} {arXiv:2001.10897 [gr-qc]} \BibitemShut
  {NoStop}%
\bibitem [{\citenamefont {Wilks}(1938)}]{WilksTheorem}%
  \BibitemOpen
  \bibfield  {author} {\bibinfo {author} {\bibfnamefont {S.~S.}\ \bibnamefont
  {Wilks}},\ }\href {http://www.jstor.org/stable/2957648} {\bibfield  {journal}
  {\bibinfo  {journal} {The Annals of Mathematical Statistics}\ }\textbf
  {\bibinfo {volume} {9}},\ \bibinfo {pages} {60} (\bibinfo {year}
  {1938})}\BibitemShut {NoStop}%
\bibitem [{\citenamefont {Cook}\ \emph {et~al.}(2006)\citenamefont {Cook},
  \citenamefont {Gelman},\ and\ \citenamefont {Rubin}}]{Gelman_pp_plots}%
  \BibitemOpen
  \bibfield  {author} {\bibinfo {author} {\bibfnamefont {S.~R.}\ \bibnamefont
  {Cook}}, \bibinfo {author} {\bibfnamefont {A.}~\bibnamefont {Gelman}}, \ and\
  \bibinfo {author} {\bibfnamefont {D.~B.}\ \bibnamefont {Rubin}},\ }\href
  {\doibase 10.1198/106186006X136976} {\bibfield  {journal} {\bibinfo
  {journal} {Journal of Computational and Graphical Statistics}\ }\textbf
  {\bibinfo {volume} {15}},\ \bibinfo {pages} {675} (\bibinfo {year} {2006})},\
  \Eprint {http://arxiv.org/abs/https://doi.org/10.1198/106186006X136976}
  {https://doi.org/10.1198/106186006X136976} \BibitemShut {NoStop}%
\bibitem [{\citenamefont {{Talts}}\ \emph {et~al.}(2018)\citenamefont
  {{Talts}}, \citenamefont {{Betancourt}}, \citenamefont {{Simpson}},
  \citenamefont {{Vehtari}},\ and\ \citenamefont {{Gelman}}}]{Talts_pp_plots}%
  \BibitemOpen
  \bibfield  {author} {\bibinfo {author} {\bibfnamefont {S.}~\bibnamefont
  {{Talts}}}, \bibinfo {author} {\bibfnamefont {M.}~\bibnamefont
  {{Betancourt}}}, \bibinfo {author} {\bibfnamefont {D.}~\bibnamefont
  {{Simpson}}}, \bibinfo {author} {\bibfnamefont {A.}~\bibnamefont
  {{Vehtari}}}, \ and\ \bibinfo {author} {\bibfnamefont {A.}~\bibnamefont
  {{Gelman}}},\ }\href {\doibase 10.48550/arXiv.1804.06788} {\bibfield
  {journal} {\bibinfo  {journal} {arXiv e-prints}\ ,\ \bibinfo {eid}
  {arXiv:1804.06788}} (\bibinfo {year} {2018})},\ \Eprint
  {http://arxiv.org/abs/1804.06788} {arXiv:1804.06788 [stat.ME]} \BibitemShut
  {NoStop}%
\bibitem [{\citenamefont {Romero-Shaw}\ \emph {et~al.}(2020)\citenamefont
  {Romero-Shaw}, \citenamefont {Talbot}, \citenamefont {Biscoveanu},
  \citenamefont {d'Emilio}, \citenamefont {Ashton}, \citenamefont {Berry},
  \citenamefont {Coughlin}, \citenamefont {Galaudage}, \citenamefont {Hoy},
  \citenamefont {Huebner}, \citenamefont {Phukon}, \citenamefont {Pitkin},
  \citenamefont {Rizzo}, \citenamefont {Sarin}, \citenamefont {Smith},
  \citenamefont {Stevenson}, \citenamefont {Vajpeyi}, \citenamefont {Arene},
  \citenamefont {Athar},\ and\ \citenamefont
  {Xiao}}]{Romero_shaw_bilby_validation}%
  \BibitemOpen
  \bibfield  {author} {\bibinfo {author} {\bibfnamefont {I.}~\bibnamefont
  {Romero-Shaw}}, \bibinfo {author} {\bibfnamefont {C.}~\bibnamefont {Talbot}},
  \bibinfo {author} {\bibfnamefont {S.}~\bibnamefont {Biscoveanu}}, \bibinfo
  {author} {\bibfnamefont {V.}~\bibnamefont {d'Emilio}}, \bibinfo {author}
  {\bibfnamefont {G.}~\bibnamefont {Ashton}}, \bibinfo {author} {\bibfnamefont
  {C.}~\bibnamefont {Berry}}, \bibinfo {author} {\bibfnamefont
  {S.}~\bibnamefont {Coughlin}}, \bibinfo {author} {\bibfnamefont
  {S.}~\bibnamefont {Galaudage}}, \bibinfo {author} {\bibfnamefont
  {C.}~\bibnamefont {Hoy}}, \bibinfo {author} {\bibfnamefont {M.}~\bibnamefont
  {Huebner}}, \bibinfo {author} {\bibfnamefont {K.~S.}\ \bibnamefont {Phukon}},
  \bibinfo {author} {\bibfnamefont {M.}~\bibnamefont {Pitkin}}, \bibinfo
  {author} {\bibfnamefont {M.}~\bibnamefont {Rizzo}}, \bibinfo {author}
  {\bibfnamefont {N.}~\bibnamefont {Sarin}}, \bibinfo {author} {\bibfnamefont
  {R.}~\bibnamefont {Smith}}, \bibinfo {author} {\bibfnamefont
  {S.}~\bibnamefont {Stevenson}}, \bibinfo {author} {\bibfnamefont
  {A.}~\bibnamefont {Vajpeyi}}, \bibinfo {author} {\bibfnamefont
  {M.}~\bibnamefont {Arene}}, \bibinfo {author} {\bibfnamefont
  {K.}~\bibnamefont {Athar}}, \ and\ \bibinfo {author} {\bibfnamefont
  {L.}~\bibnamefont {Xiao}},\ }\href {\doibase 10.1093/mnras/staa2850}
  {\bibfield  {journal} {\bibinfo  {journal} {Monthly Notices of the Royal
  Astronomical Society}\ }\textbf {\bibinfo {volume} {499}} (\bibinfo {year}
  {2020}),\ 10.1093/mnras/staa2850}\BibitemShut {NoStop}%
\bibitem [{\citenamefont {Speagle}(2020)}]{speagle2020dynesty}%
  \BibitemOpen
  \bibfield  {author} {\bibinfo {author} {\bibfnamefont {J.~S.}\ \bibnamefont
  {Speagle}},\ }\href@noop {} {\bibfield  {journal} {\bibinfo  {journal}
  {Monthly Notices of the Royal Astronomical Society}\ }\textbf {\bibinfo
  {volume} {493}},\ \bibinfo {pages} {3132} (\bibinfo {year}
  {2020})}\BibitemShut {NoStop}%
\bibitem [{\citenamefont {Ashton}\ \emph {et~al.}(2019)\citenamefont {Ashton}
  \emph {et~al.}}]{Ashton:2018jfp}%
  \BibitemOpen
  \bibfield  {author} {\bibinfo {author} {\bibfnamefont {G.}~\bibnamefont
  {Ashton}} \emph {et~al.},\ }\href {\doibase 10.3847/1538-4365/ab06fc}
  {\bibfield  {journal} {\bibinfo  {journal} {Astrophys. J. Suppl.}\ }\textbf
  {\bibinfo {volume} {241}},\ \bibinfo {pages} {27} (\bibinfo {year} {2019})},\
  \Eprint {http://arxiv.org/abs/1811.02042} {arXiv:1811.02042 [astro-ph.IM]}
  \BibitemShut {NoStop}%
\bibitem [{\citenamefont {van Rossum}(2007)}]{Python_citation}%
  \BibitemOpen
  \bibfield  {author} {\bibinfo {author} {\bibfnamefont {G.}~\bibnamefont {van
  Rossum}},\ }in\ \href@noop {} {\emph {\bibinfo {booktitle} {Proceedings of
  the 2007 {USENIX} Annual Technical Conference, Santa Clara, CA, USA, June
  17-22, 2007}}},\ \bibinfo {editor} {edited by\ \bibinfo {editor}
  {\bibfnamefont {J.}~\bibnamefont {Chase}}\ and\ \bibinfo {editor}
  {\bibfnamefont {S.}~\bibnamefont {Seshan}}}\ (\bibinfo  {publisher}
  {{USENIX}},\ \bibinfo {year} {2007})\BibitemShut {NoStop}%
\bibitem [{\citenamefont {{LIGO Scientific
  Collaboration}}(2018{\natexlab{a}})}]{lalsuite}%
  \BibitemOpen
  \bibfield  {author} {\bibinfo {author} {\bibnamefont {{LIGO Scientific
  Collaboration}}},\ }\href {\doibase 10.7935/GT1W-FZ16} {\enquote {\bibinfo
  {title} {{LIGO} {A}lgorithm {L}ibrary - {LALS}uite},}\ }\bibinfo
  {howpublished} {free software (GPL)} (\bibinfo {year}
  {2018}{\natexlab{a}})\BibitemShut {NoStop}%
\bibitem [{\citenamefont {Prechelt}(2000)}]{Performance}%
  \BibitemOpen
  \bibfield  {author} {\bibinfo {author} {\bibfnamefont {L.}~\bibnamefont
  {Prechelt}},\ }\href {\doibase 10.1109/2.876288} {\bibfield  {journal}
  {\bibinfo  {journal} {Computer}\ }\textbf {\bibinfo {volume} {33}},\ \bibinfo
  {pages} {23 } (\bibinfo {year} {2000})}\BibitemShut {NoStop}%
\bibitem [{\citenamefont {Abbott}\ \emph {et~al.}(2017)\citenamefont {Abbott}
  \emph {et~al.}}]{GW170817}%
  \BibitemOpen
  \bibfield  {author} {\bibinfo {author} {\bibfnamefont {B.}~\bibnamefont
  {Abbott}} \emph {et~al.},\ }\href {\doibase 10.1103/physrevlett.119.161101}
  {\bibfield  {journal} {\bibinfo  {journal} {Physical Review Letters}\
  }\textbf {\bibinfo {volume} {119}} (\bibinfo {year} {2017}),\
  10.1103/physrevlett.119.161101}\BibitemShut {NoStop}%
\bibitem [{\citenamefont {Abbott}\ \emph
  {et~al.}(2021{\natexlab{a}})\citenamefont {Abbott} \emph {et~al.}}]{GWTC-3}%
  \BibitemOpen
  \bibfield  {author} {\bibinfo {author} {\bibfnamefont {R.}~\bibnamefont
  {Abbott}} \emph {et~al.} (\bibinfo {collaboration} {LIGO Scientific, VIRGO,
  KAGRA}),\ }\href@noop {} {\  (\bibinfo {year} {2021}{\natexlab{a}})},\
  \Eprint {http://arxiv.org/abs/2111.03606} {arXiv:2111.03606 [gr-qc]}
  \BibitemShut {NoStop}%
\bibitem [{\citenamefont {Abbott}\ \emph {et~al.}(2020)\citenamefont {Abbott}
  \emph {et~al.}}]{GW190814}%
  \BibitemOpen
  \bibfield  {author} {\bibinfo {author} {\bibfnamefont {R.}~\bibnamefont
  {Abbott}} \emph {et~al.} (\bibinfo {collaboration} {LIGO Scientific,
  Virgo}),\ }\href {\doibase 10.3847/2041-8213/ab960f} {\bibfield  {journal}
  {\bibinfo  {journal} {Astrophys. J. Lett.}\ }\textbf {\bibinfo {volume}
  {896}},\ \bibinfo {pages} {L44} (\bibinfo {year} {2020})},\ \Eprint
  {http://arxiv.org/abs/2006.12611} {arXiv:2006.12611 [astro-ph.HE]}
  \BibitemShut {NoStop}%
\bibitem [{\citenamefont {Cornish}\ and\ \citenamefont
  {Littenberg}(2015)}]{Cornish_2015}%
  \BibitemOpen
  \bibfield  {author} {\bibinfo {author} {\bibfnamefont {N.~J.}\ \bibnamefont
  {Cornish}}\ and\ \bibinfo {author} {\bibfnamefont {T.~B.}\ \bibnamefont
  {Littenberg}},\ }\href {\doibase 10.1088/0264-9381/32/13/135012} {\bibfield
  {journal} {\bibinfo  {journal} {Classical and Quantum Gravity}\ }\textbf
  {\bibinfo {volume} {32}},\ \bibinfo {pages} {135012} (\bibinfo {year}
  {2015})}\BibitemShut {NoStop}%
\bibitem [{\citenamefont {Littenberg}\ and\ \citenamefont
  {Cornish}(2015)}]{PhysRevD.91.084034}%
  \BibitemOpen
  \bibfield  {author} {\bibinfo {author} {\bibfnamefont {T.~B.}\ \bibnamefont
  {Littenberg}}\ and\ \bibinfo {author} {\bibfnamefont {N.~J.}\ \bibnamefont
  {Cornish}},\ }\href {\doibase 10.1103/PhysRevD.91.084034} {\bibfield
  {journal} {\bibinfo  {journal} {Phys. Rev. D}\ }\textbf {\bibinfo {volume}
  {91}},\ \bibinfo {pages} {084034} (\bibinfo {year} {2015})}\BibitemShut
  {NoStop}%
\bibitem [{\citenamefont {Cahillane}\ \emph {et~al.}(2017)\citenamefont
  {Cahillane}, \citenamefont {Betzwieser}, \citenamefont {Brown}, \citenamefont
  {Goetz}, \citenamefont {Hall}, \citenamefont {Izumi}, \citenamefont
  {Kandhasamy}, \citenamefont {Karki}, \citenamefont {Kissel}, \citenamefont
  {Mendell}, \citenamefont {Savage}, \citenamefont {Tuyenbayev}, \citenamefont
  {Urban}, \citenamefont {Viets}, \citenamefont {Wade},\ and\ \citenamefont
  {Weinstein}}]{Cahillane_2017}%
  \BibitemOpen
  \bibfield  {author} {\bibinfo {author} {\bibfnamefont {C.}~\bibnamefont
  {Cahillane}}, \bibinfo {author} {\bibfnamefont {J.}~\bibnamefont
  {Betzwieser}}, \bibinfo {author} {\bibfnamefont {D.~A.}\ \bibnamefont
  {Brown}}, \bibinfo {author} {\bibfnamefont {E.}~\bibnamefont {Goetz}},
  \bibinfo {author} {\bibfnamefont {E.~D.}\ \bibnamefont {Hall}}, \bibinfo
  {author} {\bibfnamefont {K.}~\bibnamefont {Izumi}}, \bibinfo {author}
  {\bibfnamefont {S.}~\bibnamefont {Kandhasamy}}, \bibinfo {author}
  {\bibfnamefont {S.}~\bibnamefont {Karki}}, \bibinfo {author} {\bibfnamefont
  {J.~S.}\ \bibnamefont {Kissel}}, \bibinfo {author} {\bibfnamefont
  {G.}~\bibnamefont {Mendell}}, \bibinfo {author} {\bibfnamefont {R.~L.}\
  \bibnamefont {Savage}}, \bibinfo {author} {\bibfnamefont {D.}~\bibnamefont
  {Tuyenbayev}}, \bibinfo {author} {\bibfnamefont {A.}~\bibnamefont {Urban}},
  \bibinfo {author} {\bibfnamefont {A.}~\bibnamefont {Viets}}, \bibinfo
  {author} {\bibfnamefont {M.}~\bibnamefont {Wade}}, \ and\ \bibinfo {author}
  {\bibfnamefont {A.~J.}\ \bibnamefont {Weinstein}},\ }\href {\doibase
  10.1103/physrevd.96.102001} {\bibfield  {journal} {\bibinfo  {journal}
  {Physical Review D}\ }\textbf {\bibinfo {volume} {96}} (\bibinfo {year}
  {2017}),\ 10.1103/physrevd.96.102001}\BibitemShut {NoStop}%
\bibitem [{\citenamefont {Acernese}\ \emph {et~al.}(2022)\citenamefont
  {Acernese} \emph {et~al.}}]{Acernese_2022}%
  \BibitemOpen
  \bibfield  {author} {\bibinfo {author} {\bibfnamefont {F.}~\bibnamefont
  {Acernese}} \emph {et~al.} (\bibinfo {collaboration} {{The} {Virgo}
  {Collaboration}}),\ }\href {\doibase 10.1088/1361-6382/ac3c8e} {\bibfield
  {journal} {\bibinfo  {journal} {Classical and Quantum Gravity}\ }\textbf
  {\bibinfo {volume} {39}},\ \bibinfo {pages} {045006} (\bibinfo {year}
  {2022})}\BibitemShut {NoStop}%
\bibitem [{\citenamefont {Sun}\ \emph {et~al.}(2020)\citenamefont {Sun},
  \citenamefont {Goetz}, \citenamefont {Kissel}, \citenamefont {Betzwieser},
  \citenamefont {Karki}, \citenamefont {Viets}, \citenamefont {Wade},
  \citenamefont {Bhattacharjee}, \citenamefont {Bossilkov}, \citenamefont
  {Covas}, \citenamefont {Datrier}, \citenamefont {Gray}, \citenamefont
  {Kandhasamy}, \citenamefont {Lecoeuche}, \citenamefont {Mendell},
  \citenamefont {Mistry}, \citenamefont {Payne}, \citenamefont {Savage},
  \citenamefont {Weinstein}, \citenamefont {Aston}, \citenamefont {Buikema},
  \citenamefont {Cahillane}, \citenamefont {Driggers}, \citenamefont {Dwyer},
  \citenamefont {Kumar},\ and\ \citenamefont {Urban}}]{Sun_2020}%
  \BibitemOpen
  \bibfield  {author} {\bibinfo {author} {\bibfnamefont {L.}~\bibnamefont
  {Sun}}, \bibinfo {author} {\bibfnamefont {E.}~\bibnamefont {Goetz}}, \bibinfo
  {author} {\bibfnamefont {J.~S.}\ \bibnamefont {Kissel}}, \bibinfo {author}
  {\bibfnamefont {J.}~\bibnamefont {Betzwieser}}, \bibinfo {author}
  {\bibfnamefont {S.}~\bibnamefont {Karki}}, \bibinfo {author} {\bibfnamefont
  {A.}~\bibnamefont {Viets}}, \bibinfo {author} {\bibfnamefont
  {M.}~\bibnamefont {Wade}}, \bibinfo {author} {\bibfnamefont {D.}~\bibnamefont
  {Bhattacharjee}}, \bibinfo {author} {\bibfnamefont {V.}~\bibnamefont
  {Bossilkov}}, \bibinfo {author} {\bibfnamefont {P.~B.}\ \bibnamefont
  {Covas}}, \bibinfo {author} {\bibfnamefont {L.~E.~H.}\ \bibnamefont
  {Datrier}}, \bibinfo {author} {\bibfnamefont {R.}~\bibnamefont {Gray}},
  \bibinfo {author} {\bibfnamefont {S.}~\bibnamefont {Kandhasamy}}, \bibinfo
  {author} {\bibfnamefont {Y.~K.}\ \bibnamefont {Lecoeuche}}, \bibinfo {author}
  {\bibfnamefont {G.}~\bibnamefont {Mendell}}, \bibinfo {author} {\bibfnamefont
  {T.}~\bibnamefont {Mistry}}, \bibinfo {author} {\bibfnamefont
  {E.}~\bibnamefont {Payne}}, \bibinfo {author} {\bibfnamefont {R.~L.}\
  \bibnamefont {Savage}}, \bibinfo {author} {\bibfnamefont {A.~J.}\
  \bibnamefont {Weinstein}}, \bibinfo {author} {\bibfnamefont {S.}~\bibnamefont
  {Aston}}, \bibinfo {author} {\bibfnamefont {A.}~\bibnamefont {Buikema}},
  \bibinfo {author} {\bibfnamefont {C.}~\bibnamefont {Cahillane}}, \bibinfo
  {author} {\bibfnamefont {J.~C.}\ \bibnamefont {Driggers}}, \bibinfo {author}
  {\bibfnamefont {S.~E.}\ \bibnamefont {Dwyer}}, \bibinfo {author}
  {\bibfnamefont {R.}~\bibnamefont {Kumar}}, \ and\ \bibinfo {author}
  {\bibfnamefont {A.}~\bibnamefont {Urban}},\ }\href {\doibase
  10.1088/1361-6382/abb14e} {\bibfield  {journal} {\bibinfo  {journal}
  {Classical and Quantum Gravity}\ }\textbf {\bibinfo {volume} {37}},\ \bibinfo
  {pages} {225008} (\bibinfo {year} {2020})}\BibitemShut {NoStop}%
\bibitem [{\citenamefont {Abbott}\ \emph {et~al.}(2023)\citenamefont {Abbott}
  \emph {et~al.}}]{LIGOScientific:2023vdi}%
  \BibitemOpen
  \bibfield  {author} {\bibinfo {author} {\bibfnamefont {R.}~\bibnamefont
  {Abbott}} \emph {et~al.} (\bibinfo {collaboration} {LIGO Scientific, VIRGO,
  KAGRA}),\ }\href@noop {} {\  (\bibinfo {year} {2023})},\ \Eprint
  {http://arxiv.org/abs/2302.03676} {arXiv:2302.03676 [gr-qc]} \BibitemShut
  {NoStop}%
\bibitem [{\citenamefont {Lin}(1991)}]{JSDivergence}%
  \BibitemOpen
  \bibfield  {author} {\bibinfo {author} {\bibfnamefont {J.}~\bibnamefont
  {Lin}},\ }\href {\doibase 10.1109/18.61115} {\bibfield  {journal} {\bibinfo
  {journal} {IEEE Transactions on Information Theory}\ }\textbf {\bibinfo
  {volume} {37}},\ \bibinfo {pages} {145} (\bibinfo {year} {1991})}\BibitemShut
  {NoStop}%
\bibitem [{\citenamefont {Abbott}\ \emph {et~al.}(2019)\citenamefont {Abbott}
  \emph {et~al.}}]{LIGOScientific:2018hze}%
  \BibitemOpen
  \bibfield  {author} {\bibinfo {author} {\bibfnamefont {B.~P.}\ \bibnamefont
  {Abbott}} \emph {et~al.} (\bibinfo {collaboration} {LIGO Scientific,
  Virgo}),\ }\href {\doibase 10.1103/PhysRevX.9.011001} {\bibfield  {journal}
  {\bibinfo  {journal} {Phys. Rev. X}\ }\textbf {\bibinfo {volume} {9}},\
  \bibinfo {pages} {011001} (\bibinfo {year} {2019})},\ \Eprint
  {http://arxiv.org/abs/1805.11579} {arXiv:1805.11579 [gr-qc]} \BibitemShut
  {NoStop}%
\bibitem [{\citenamefont {{LIGO Open Science Center
  (LOSC)}}(2017)}]{GW170817_data}%
  \BibitemOpen
  \bibfield  {author} {\bibinfo {author} {\bibnamefont {{LIGO Open Science
  Center (LOSC)}}},\ }\href {\doibase doi:10.7935/K5B8566F} {\enquote {\bibinfo
  {title} {{Data release for event GW170817}},}\ } (\bibinfo {year}
  {2017})\BibitemShut {NoStop}%
\bibitem [{\citenamefont {{LIGO Scientific
  Collaboration}}(2018{\natexlab{b}})}]{lalsuite_code}%
  \BibitemOpen
  \bibfield  {author} {\bibinfo {author} {\bibnamefont {{LIGO Scientific
  Collaboration}}},\ }\href {\doibase 10.7935/GT1W-FZ16} {\enquote {\bibinfo
  {title} {{LIGO} {A}lgorithm {L}ibrary - {LALS}uite},}\ }\bibinfo
  {howpublished} {free software (GPL)} (\bibinfo {year}
  {2018}{\natexlab{b}})\BibitemShut {NoStop}%
\bibitem [{\citenamefont {Abbott}\ \emph
  {et~al.}(2021{\natexlab{b}})\citenamefont {Abbott} \emph
  {et~al.}}]{LIGOScientific:2019lzm}%
  \BibitemOpen
  \bibfield  {author} {\bibinfo {author} {\bibfnamefont {R.}~\bibnamefont
  {Abbott}} \emph {et~al.} (\bibinfo {collaboration} {LIGO Scientific,
  Virgo}),\ }\href {\doibase 10.1016/j.softx.2021.100658} {\bibfield  {journal}
  {\bibinfo  {journal} {SoftwareX}\ }\textbf {\bibinfo {volume} {13}},\
  \bibinfo {pages} {100658} (\bibinfo {year} {2021}{\natexlab{b}})},\ \Eprint
  {http://arxiv.org/abs/1912.11716} {arXiv:1912.11716 [gr-qc]} \BibitemShut
  {NoStop}%
\end{thebibliography}%
